\documentclass[longauth]{aa}

\usepackage{graphicx}
\graphicspath{{resolved-KS-plots/}{resolved-tau-plots/}{observables/}}

\usepackage{txfonts}
\usepackage[breaklinks, colorlinks, citecolor=blue, linkcolor=blue]{hyperref}
\begin{document}

   \title{VERTICO III: The Kennicutt-Schmidt relation in Virgo cluster galaxies}

   \author{M. J. Jim\'enez-Donaire
          \inst{\ref{oan},\ref{yebes}} \fnmsep\thanks{\email{mdonaire@oan.es}}
          \and
          Toby Brown \inst{\ref{herzberg}}
          \and
          Christine D. Wilson \inst{\ref{mcmaster}}
          \and
          Ian D. Roberts \inst{\ref{leiden}}
          \and
          Nikki Zabel \inst{\ref{uct}}
          \and
          Sara L. Ellison \inst{\ref{victoria}}
          \and
          Mallory Thorp \inst{\ref{victoria}}
          \and
          Vicente Villanueva \inst{\ref{maryland}}
          \and
          Ryan Chown \inst{\ref{uwo},\ref{space}}          
          \and
          Dhruv Bisaria \inst{\ref{queens}}          
          \and
          Alberto D. Bolatto \inst{\ref{maryland}}          
          \and
          Alessandro Boselli \inst{\ref{marseille}}
          \and
          Barbara Catinella \inst{\ref{uwa},\ref{astro3d}}
          \and
          Aeree Chung \inst{\ref{yonsei}}
          \and
          Luca Cortese \inst{\ref{uwa},\ref{astro3d}}          
          \and
          Timothy A. Davis \inst{\ref{cardiff}}
          \and          
          Claudia D. P. Lagos \inst{\ref{uwa},\ref{astro3d}}
          \and
          Bumhyun Lee \inst{\ref{kasi}}
          \and
          Laura C. Parker \inst{\ref{mcmaster}}          
          \and
          Kristine Spekkens \inst{\ref{royal}}
          \and
          Adam R. H. Stevens \inst{\ref{uwa}}          
          \and
          Jiayi Sun \inst{\ref{mcmaster},\ref{cita}}
          }

   \institute{Observatorio Astronómico Nacional (IGN), C/Alfonso XII, 3, E-28014 Madrid, Spain \label{oan}\\
              \email{mdonaire@oan.es}
         \and
             Centro de Desarrollos Tecnológicos, Observatorio de Yebes (IGN), 19141 Yebes, Guadalajara, Spain \label{yebes}\\
         \and
             Herzberg Astronomy and Astrophysics Research Centre, National Research Council of Canada, 5071 West Saanich Rd, Victoria, BC, V9E 2E7, Canada \label{herzberg}\\
         \and
             Department of Physics \& Astronomy, McMaster University, 1280 Main St West, Hamilton, Ontario L8S 4M1, Canada \label{mcmaster}\\
         \and
             Leiden Observatory, Leiden University, PO Box 9513, 2300 RA Leiden, The Netherlands \label{leiden}\\
         \and
             Department of Astronomy, University of Cape Town, Private Bag X3, Rondebosch 7701, South Africa \label{uct}\\
         \and
             Department of Physics \& Astronomy, University of Victoria, Finnerty Road, Victoria, British Columbia, V8P 1A1, Canada \label{victoria}\\
         \and
             Department of Astronomy, University of Maryland, College Park, MD 20742, USA \label{maryland}\\
         \and
             Department of Physics \& Astronomy, The University of Western Ontario, London ON N6A 3K7, Canada \label{uwo}\\     
         \and
             Institute for Earth and Space Exploration, The University of Western Ontario, London ON N6A 3K7, Canada \label{space}\\
         \and
             Department of Physics, Engineering Physics, and Astronomy, Queen’s University, Kingston, ON K7L 3N6, Canada \label{queens}\\
         \and
             Aix-Marseille Universit\'{e}, CNRS, CNES, LAM, Marseille, France \label{marseille}\\
         \and
             International Centre for Radio Astronomy Research, The University of Western Australia, 35 Stirling Hwy, 6009 Crawley, WA, Australia \label{uwa}\\
         \and
             ARC Centre of Excellence for All Sky Astrophysics in 3 Dimensions (ASTRO 3D), Australia \label{astro3d}\\
         \and
             Department of Astronomy, Yonsei University, 50 Yonsei-ro, Seodamun-gu, Seoul, 03722, Republic of Korea \label{yonsei}\\
         \and
             School of Physics \& Astronomy, Cardiff University, Queens Buildings, The Parade, Cardiff CF24 3AA, UK \label{cardiff}\\
         \and
             Korea Astronomy and Space Science Institute, 776 Daedeokdae-ro, Yuseong-gu, Daejeon 34055, Republic of Korea \label{kasi}\\
         \and
             Royal Military College of Canada, PO Box 17000, Station Forces, Kingston, ON, Canada K7K 7B4 \label{royal}\\
         \and
             Canadian Institute for Theoretical Astrophysics (CITA), University of Toronto, 60 St George Street, Toronto, ON M5S 3H8, Canada \label{cita}\\        
             }

   \date{Received 09 August 2022; accepted 23 November 2022}

 
 \abstract
   {}
   {In this VERTICO (Virgo Environment Traced in CO) science paper we aim to study how the star formation process depends on galactic environment and gravitational interactions in the context of galaxy evolution. We explore the scaling relation between the star formation rate surface density $(\Sigma_\mathrm{SFR})$ and the molecular gas surface density $(\Sigma_\mathrm{mol})$, also known as the Kennicutt-Schmidt relation, in a subsample of Virgo cluster spiral galaxies.}
   {We use new Atacama Compact Array and Total Power (ACA+TP) observations from the VERTICO-ALMA (Atacama Large Millimeter/submillimeter Array) Large Program at $720\,\mathrm{pc}$ resolution to resolve the molecular gas content, as traced by the $^{12}\mathrm{CO}\,(2-1)$ transition, across the disks of 37 spiral galaxies in the Virgo cluster. In combination with archival UV and IR observations used to determine the star formation rate (SFR), we estimate the parameters of the Kennicutt-Schmidt (KS) relation for the entire ensemble of galaxies, and within individual galaxies.}
   {We find the KS slope for the entire population to be $N=0.97\pm0.07$, with a characteristic molecular gas depletion time of $1.86\,\mathrm{Gyr}$ for our full sample, in agreement with previous work in isolated, nearby star-forming galaxies. In individual galaxies, we find that the KS slope ranges between 0.69 and 1.40, and typical star formation efficiencies of molecular gas that can vary from galaxy to galaxy by a factor of $\sim 4$. These galaxy-to-galaxy variations account for $\sim 0.20\,\mathrm{dex}$ in scatter in the ensemble KS relation, which is characterized by a $0.42\,\mathrm{dex}$ scatter. In addition, we find that the HI-deficient galaxies in the Virgo cluster show a steeper resolved KS relation and lower molecular gas efficiencies than HI-normal cluster galaxies.}
   {While the molecular gas content in galaxies residing in the Virgo cluster appears, to first order, to behave similarly to that in isolated galaxies, our VERTICO sample of galaxies shows that cluster environments play a key role in regulating star formation. The environmental mechanisms affecting the HI galaxy content also have a direct impact in the star formation efficiency of molecular gas in cluster galaxies, leading to longer depletion times in HI-deficient members.}


   \maketitle
%

\section{Introduction}

In the context of galaxy evolution, a key goal of any star formation theory is to understand how efficiently galaxies can convert their interstellar gas content into stars. A common approach to tackle this question is to quantitatively measure the relationship between the gas density and star formation rate (SFR). As such, scaling relations constitute a fundamental tool to investigate star formation, since they arise as a natural consequence of the interplay between physics in star-forming regions and galactic scales. Following \citet{Schmidt1959}, who proposed that the rate of star formation is proportional to the square of the gas column density, the work developed by \citet{Kennicutt1989} and \citet{Kennicutt1998} on integrated galaxies supported this idea. The well-known Kennicutt-Schmidt (KS) relation connects the SFR and gas surface densities as:

\begin{equation}
    \Sigma_{\mathrm{SFR}}\propto\Sigma_{\mathrm{gas}}^N
\end{equation}

\noindent where $N$ is the power law index and $\Sigma_{\mathrm{gas}}=\Sigma_{\mathrm{HI}}+\Sigma_{\mathrm{H}_2}$, is the total gas column density, combining the contribution of the atomic $(\Sigma_{\mathrm{HI}})$ and molecular $(\Sigma_{\mathrm{H}_2})$ gas. While \citet{Kennicutt1989} and \citet{Kennicutt1998} found $N=1.40\pm0.15$, similar studies employing a range of SFR tracers and varied sampling methodologies derived $N=0.9-2.1$ \citep[e.g.,][]{Buat1992,Wong2002,Schuster2007,Kennicutt2007}. Scaling relations, such as the KS relation, depend on both the local physical conditions within the gas and the physics of the galaxy's star forming disk \citep[e.g.,][]{Kennicutt1998,Wong2002}. Therefore, characterizing these relations and their physical scatter has the potential to provide unique insight into the physics and scales that regulate star formation.

Intuitively, however, the connection between $\Sigma_{\mathrm{SFR}}$ and $\Sigma_{\mathrm{H}_2}$ is physically more appropriate, since star formation takes place within cold molecular gas. While this process takes places locally, inside molecular clouds, it is extremely complex to characterize since it involves physical mechanisms that affect a wide range of physical scales (from star-forming clumps to giant molecular cloud complexes) and densities ($\sim 10^2-10^6\,\mathrm{cm}^{-3}$). Tracing the total molecular component of the gas is key to understand star formation in detail \citep[e.g.,][]{McKee2007,Lada2010,Lada2012,Longmore2013,Klessen2016}); however, H$_2$ molecules are very difficult to observe directly. The main reason for this is that the H$_2$ molecule lacks a dipole moment, as well as the low temperatures found in molecular clouds, which prevents vibrational excitation. Observers usually trace the bulk molecular gas indirectly instead, via the carbon monoxide (CO) rotational transitions. Given its large abundance, CO is the next most reliable molecular tracer in the universe \citep[for an in-depth review, see][]{Bolatto2013}. 

Over the last decade, extensive CO mapping across nearby galaxies has been possible thanks to the great sensitivity and speed of single-dish radio telescopes and interferometers. These routine observations have tremendously improved our understanding of the physics regulating star formation in galaxies. High spatial resolution datasets (e.g., 700\,pc) have allowed studies such as the seminal work by \citet{Bigiel2008} to show that a KS-type power law with index $N=1.0\pm0.2$ holds down to sub-kpc scales between $\Sigma_{\mathrm{SFR}}$ and $\Sigma_{\mathrm{H2}}$  across a large sample of spiral, star-forming galaxies. Later studies have also found $N\sim 0.8-1.4$ for this KS relation \citep[e.g.,][]{Blanc2009,Schruba2011,Leroy2013,Usero2015,Lin2019,Ellison2021,Pessa2021} , and reinforced that $\Sigma_{\mathrm{SFR}}$ is strongly correlated with the local $\Sigma_{\mathrm{H_2}}$, and only tangentially connected to the atomic gas content.

While the power-law index $N$ can depend strongly on the methods used to quantify gas masses and SFRs, the variation in the scatter has also been related to distinct evolutionary life cycles in individual star-forming regions \citep[e.g.,][]{Schruba2010}. In addition, there is growing evidence that the galactic environment can potentially regulate the molecular gas and its connection to star formation. Within galaxies, it has been shown that galactic structures and galactic environment can alter the molecular gas probability distribution functions and the evolution of giant molecular clouds \citep[e.g.,][]{Hughes2013,Colombo2014,Sun2020,Meidt2021,Sun2022}. The GMC mass distribution function is steeper in inter-arm region than in the spiral arms, and more massive clouds, characterized by larger turbulence, tend to reside in environments at higher gas, star, and SFR surface densities, and stronger shear. Even at larger scales, the molecular efficiency of star formation has been shown to be directly affected in cluster environments due to enhanced H$_2$ masses as well as triggered star formation when compared to their field counterparts \citep[e.g.,][]{Mok2017}. These differences could in turn be reflected in changes in the KS slope and scatter, directly related to a variation in molecular gas densities and star formation properties across the disk of cluster members. The general dependence of the KS parameters on spatial scale is still unclear. While studies such as \citet{Williams2018} found the molecular KS index strongly decreases with increasing scales, becoming approximately linear at $\sim$kpc scales, \citet{Pessa2021} recently found no evidence for systematic dependences of the slope on the spatial resolution of the PHANGS (Physics at High Angular resolution in Nearby GalaxieS) data.

This KS star formation relation was initially studied in spiral, isolated galaxies. Later studies, however, showed that the KS relation also applies in samples of unresolved cluster spiral galaxies \citep{Fumagalli2008,Vollmer2012}. However, it is still unclear whether the KS relation is affected by the environment in which galaxies reside \citep[e.g.,][]{Boquien2011,Lizee2021,Vollmer2021}. There is plenty of evidence that the SFR of galaxies in clusters is significantly reduced \citep[see e.g.,][and references therein]{Boselli2022}, which could, in turn, modify the expected KS relation. The surrounding environment, indeed, might perturb the relation between gas column density and SFR both at galactic scales or within galactic disks. The large variation found for the power-law indices and the scatter at fixed molecular gas surface densities makes it unclear whether a unique star formation law could hold among and within galaxies. Several Virgo cluster galaxies which are affected by ram pressure stripping have been studied by \citep{Nehlig2016} and \citet{Lee2017}, finding no significant variations from the global KS relation from \citet{Kennicutt1998}. Nevertheless, these galaxies showed clear regions with varying depletion times where the gas is compressed due to ram pressure stripping, in the sense that the SFR are higher in regions where the intracluster medium exerts a higher pressure.

Resolving galaxy disks inside cluster environments constitutes a unique laboratory to study star formation. These gravitationally-bound environments can evolve in a different way than their galaxy counterparts in the field. Their diverse and often extreme environments, as well as interactions between members, could have a great impact on their molecular gas content \citep[][]{Violino2018,Pan2018,Thorp2022} that may affect the star-formation rates in a variety of fashions. In recent years, observations have shown that physical processes within galaxy clusters, such as ram-pressure stripping \citep[e.g.,][]{Gunn1972,Cortese2021}, or galaxy-to-galaxy interactions such as harassment or mergers have a great impact on their atomic gas content \citep{Boselli2006,Boselli2014}. In addition, these interactions have been shown to clearly affect the molecular gas content in the cluster members, as evidenced by the galaxy members in the Virgo cluster \citep{Fumagalli2009,Boselli2014b}. Given that molecular gas constitutes the immediate fuel for star formation, the cluster galaxies’ environment is directly shaped, both morphologically and kinematically, by their members interactions \citep[e.g.,][]{Bumhyun2017,Bumhyun2018,Zabel2019,Lizee2021}. Understanding how cluster environments are connected to the local-scale star formation rates and molecular gas star formation efficiency is therefore key to constrain theories of star formation and galaxy evolution \citep{Chung2014,Nehlig2016,Moretti2020,Cramer2020,Lee2020,Cramer2021,Boselli2021}.

\textit{Are there significant differences in the KS law between star forming regions in cluster galaxies and those of normal, star-forming disks? Does the cluster environment modify the star formation process at sub-kpc scales?} To address these questions, we employ a wide set of multi-wavelength data to trace star formation and molecular gas, and measure $\Sigma_\mathrm{SFR}$ and $\Sigma_\mathrm{H2}$ in a sample of 37 galaxies from the Virgo cluster of galaxies. The present study constitutes one of the first science papers from the Virgo Environment Traced in CO \citep[VERTICO,][]{Brown2021} program. While recent efforts have been made in this direction, VERTICO is the first survey probing the molecular gas content at sub-kpc scales in a complete and statistically significant number of galaxies of the Virgo cluster. By measuring CO in a homogeneous fashion, over a wide variety of regions, VERTICO is ideal for revealing the dependencies of local star formation efficiency (SFE) with gas density and location as a function of environmental influence for the first time. By analyzing the KS relation in cluster members, this study is in a unique position to unravel the mechanisms that regulate star formation at sub-kpc scales inside a galaxy cluster environment.

This paper is organized as follows: in Section \ref{sec:data} we describe the VERTICO observations and data reduction, while Section \ref{sec:ancillary} provides an overview of the ancillary data used to complement the VERTICO dataset. In Section \ref{sec:methods} we detail the main methods to convert the observable quantities into the main physical parameters. We present our main results in Section \ref{sec:results} and provide a discussion and an interpretation in Section \ref{sec:discussion}. Finally a short summary of our work can be found in Section \ref{sec:summary}.

\section{Observations and data reduction}
\label{sec:data}
\subsection{VERTICO galaxies}
We use a sub-sample of 37 galaxies included in the VERTICO\footnote{\url{https://www.verticosurvey.com/}} \citep{Brown2021} survey. The full VERTICO sample consists of a total of 51 Virgo Cluster galaxies, selected from the Very Large Array Imaging of Virgo in Atomic Gas survey \citep[VIVA,][]{Chung2009} because their environment seems to be actively affecting them, via ram pressure stripping, starvation, and tidal interaction. In addition, these sources cover a wide range of star formation properties and span two orders of magnitude in stellar mass. Our selected sub-sample only excludes highly-inclined ($i > 80^\mathrm{o}$) targets from the original dataset, in order to minimize any major projection effects. A summary with the basic properties of the working sample of galaxies can be found in Table \ref{tab:sample}, orientation parameters are drawn from \citet{Brown2021}, global SFR and stellar masses are adopted from \citet{Leroy2019}, and H\,I deficiencies are calculated using the predicted H\,I mass from field galaxies at fixed stellar mass, from \citet{Zabel2022}. Based on \citet{Mei2007}, we employ a common distance of 16.5\,Mpc to all Virgo galaxies throughout the paper.

The observations for VERTICO were carried out using the ALMA Atacama Compact Array (ACA) during Cycle 7 (2019.1.00763.L.). Out of the total 51 galaxies in the sample, 37 targets were newly observed, while 15 galaxies were already surveyed and the data were publicly available in the ALMA archive \citep{Cramer2020,Leroy2021a}. From our selected sub-sample, 13 galaxies had archival data (NGC\,4254, NGC\,4321, NGC\,4293, NGC\,4298, NGC\,4424, NGC\,4457, NGC\,4535, NGC\,4536, NGC\,4548, NGC\,4569, NGC\,4579, NGC\,4654, NGC\,4694). To recover the CO extended emission at large angular scales, Total Power (TP) observations were needed for 25 out of 37 targets. The VERTICO program targeted spectroscopic observations of the $^{12}$CO\,(2-1) transition, as well as the main isotopologues $^{13}$CO\,(2-1) and C$^{18}$O\,(2-1) and ALMA Band 6 continuum. For each galaxy, a Nyquist-sampled mosaic was performed in order to map their molecular gas disk. The nominal flux calibration uncertainty of ALMA in Band 6 during Cycle 7 was 5-10\% according to the ALMA Cycle 7 Technical Handbook. A detailed description of the data reduction and imaging process is available in \citet{Brown2021}.

After imaging, two working datasets were generated by convolving each datacube to 9” resolution and 15” resolution, which correspond to physical scales of 720\,pc and 1.2\,kpc respectively, at the distance of Virgo. While most of our work is carried out using the final products at 9” resolution, these two resolution sets allow us to beam-match the observations to ancillary infrared data and UV data, needed for the analysis (see Section \ref{sec:methods}). Integrated intensity maps are computed directly from these final data products, implementing a masking process described in \cite{Sun2018}. This method employs a spatially and spectrally varying noise, which we measure in every pixel and channel before primary beam correction. For that, a mask is generated by combining a core mask, selecting spaxels with $\mathrm{S/N}\geq 3.5$ in three consecutive channels or more, and a wing mask for spaxels with $\mathrm{S/N}\geq 2$ in two consecutive channels or more. We refer the reader to the VERTICO survey paper \cite{Brown2021} for a detailed description of the method implementation. Moment-$0$ maps are then calculated the integrated intensity along the spectral axis in K\,km\,s$^{-1}$. We further convert these line measurements into molecular gas surface densities, as described in Section \ref{sec:molgas}.

\begin{table*}
\centering
\caption{The VERTICO sub-sample of low-inclination ($i\leq80^o$) galaxies.} 
\begin{tabular}{lccccccc}
\hline
\hline
Galaxy & RA & DEC & $i$ & PA & log$_{10}\,{\textrm{SFR}}$ & log$_{10}\,{M_*}$ & def$_{\mathrm{HI}\,\,M_*}$\\
& hh:mm:ss & dd:mm:ss & $^\mathrm{o}$ & $^\mathrm{o}$ & $M_{\odot}\,\mathrm{yr}^{-1}$ & $M_{\odot}$ & dex \\
\hline 
IC3392 & 12:28:43 & +14:59:57 & 68 & 219 & $-$1.30 & 9.51 & 1.50 $^{+ 0.02 } _{- 0.02}$\\
NGC4064 & 12:04:11 & +18:26:39 & 70 & 150 & $-$1.07 & 9.47 & 1.50 $^{+ 0.01} _{- 0.02}$\\
NGC4189 & 12:13:47 & +13:25:35 & 42 & 70 & $-$0.33 & 9.75 & 0.63 $^{+ 0.07} _{- 0.03}$\\
NGC4254 & 12:18:50 & +14:25:06 & 39 & 243 & 0.70 & 10.52 & -0.03 $^{+ 0.02} _{- 0.06}$ \\
NGC4293 & 12:21:13 & +18:23:03 & 67 & 239 & $-$0.27 & 10.50 & 2.16 $^{+ 0.02} _{- 0.09}$\\
NGC4294 & 12:21:18 & +11:30:39 & 74 & 151 & $-$0.39 & 9.38 & 0.18 $^{+ 0.02} _{- 0.12}$\\
NGC4298 & 12:21:33 & +14:36:20 & 52 & 132 & $-$0.26 & 10.11 & 0.68 $^{+ 0.06} _{- 0.16}$\\
NGC4299 & 12:21:41 & +11:30:06 & 14 & 128 & $-$0.34 & 9.06 & -0.72 $^{+ 0.53 } _{- 0.30}$\\
NGC4321 & 12:22:55 & +15:49:33 & 32 & 280 & 0.54 & 10.71 & 0.36 $^{+ 0.03} _{- 0.01}$\\
NGC4351 & 12:24:01 & +12:12:15 & 48 & 251 & $-$0.91 & 9.37 & 0.55 $^{+ 0.02} _{- 0.13}$\\
NGC4380 & 12:25:22 & +10:01:00 & 61 & 158 & $-$0.77 & 10.11 & 1.27 $^{+ 0.06} _{- 0.16}$\\
NGC4383 & 12:25:25 & +16:28:12 & 56 & 17 & 0.01 & 9.44 & -0.39 $^{+ 0.01 } _{- 0.05}$\\
NGC4394 & 12:25:56 & +18:12:52 & 32 & 312 & $-$0.79 & 10.34 & 0.90 $^{+ 0.06} _{- 0.00}$\\
NGC4405 & 12:26:07 & +16:10:52 & 46 & 18 & $-$0.88 & 9.75 & 1.73 $^{+ 0.07} _{- 0.03}$\\
NGC4419 & 12:26:56 & +15:02:51 & 74 & 131 & $-$0.12 & 10.06 & 1.65 $^{+ 0.11 } _{- 0.13}$\\
NGC4424 & 12:27:12 & +09:25:14 & 61 & 274 & $-$0.52 & 9.89 & 1.14 $^{+ 0.07} _{- 0.07}$\\
NGC4450 & 12:28:29 & +17:05:05 & 51 & 170 & $-$0.55 & 10.70 & 1.35 $^{+ 0.03} _{- 0.00}$\\
NGC4457 & 12:28:59 & +03:34:14 & 37 & 256 & $-$0.49 & 10.42 & 1.29 $^{+ 0.06} _{- 0.04}$\\
NGC4501 & 12:31:59 & +14:25:11 & 65 & 320 & 0.43 & 11.00 & 0.80 $^{+ 0.03} _{- 0.02}$\\
NGC4532 & 12:34:19 & +06:28:06 & 64 & 159 & $-$0.16 & 9.25 & -0.25 $^{+ 0.12 } _{- 0.05}$\\
NGC4535 & 12:34:20 & +08:11:54 & 48 & 12 & 0.31 & 10.49 & 0.07 $^{+ 0.02 } _{- 0.10}$\\
NGC4536 & 12:34:27 & +02:11:16 & 74 & 118 & 0.47 & 10.19 & -0.26 $^{+ 0.02 } _{- 0.08}$\\
NGC4548 & 12:35:27 & +14:29:44 & 37 & 318 & $-$0.28 & 10.65 & 0.96 $^{+ 0.02 } _{- 0.02}$\\
NGC4561 & 12:36:08 & +19:19:22 & 28 & 60 & $-$0.64 & 9.09 & -0.64 $^{+ 0.41} _{- 0.46}$\\
NGC4567 & 12:36:33 & +11:15:29 & 49 & 251 & 0.03 & 10.25 & -0.27 $^{+ 0.00} _{- 0.04}$\\
NGC4568 & 12:36:34 & +11:14:22 & 70 & 211 & 0.29 & 10.47 & 0.40 $^{+ 0.01} _{- 0.09}$\\
NGC4569 & 12:36:50 & +13:09:55 & 69 & 203 & 0.16 & 10.86 & 1.12 $^{+ 0.03} _{- 0.04}$\\
NGC4579 & 12:37:43 & +11:49:06 & 40 & 273 & 0.08 & 10.92 & 1.20 $^{+ 0.02} _{- 0.04}$\\
NGC4580 & 12:37:48 & +05:22:06 & 46 & 337 & $-$0.90 & 9.94 & 1.99 $^{+ 0.11} _{- 0.07}$\\
NGC4606 & 12:40:58 & +11:54:44 & 69 & 38 & $-$1.33 & 9.61 & 1.86 $^{+ 0.01} _{- 0.02}$\\
NGC4651 & 12:43:43 & +16:23:38 & 53 & 75 & $-$0.10 & 10.31 & 0.09 $^{+ 0.03} _{- 0.00}$\\
NGC4654 & 12:43:57 & +13:07:33 & 61 & 300 & 0.31 & 10.26 & -0.01 $^{+ 0.00} _{- 0.03}$\\
NGC4694 & 12:48:15 & +10:59:01 & 62 & 323 & $-$0.85 & 9.94 & 1.03 $^{+ 0.11} _{- 0.07}$\\
NGC4698 & 12:48:23 & +08:29:15 & 66 & 347 & $-$0.83 & 10.49 & 0.37 $^{+ 0.02} _{- 0.10}$\\
NGC4713 & 12:49:58 & +05:18:40 & 45 & 89 & $-$0.20 & 9.31 & -0.42 $^{+ 0.10} _{- 0.00}$\\
NGC4772 & 12:53:29 & +02:10:06 & 60 & 325 & $-$1.08 & 10.18 & 0.48 $^{+ 0.02} _{- 0.09}$\\
NGC4808 & 12:55:49 & +04:18:15 & 72 & 127 & $-$0.19 & 9.63 & -0.27 $^{+ 0.01} _{- 0.03}$\\
\hline
\end{tabular}

\begin{minipage}{2.0\columnwidth}
    \vspace{1mm}
    {\bf Notes:} The right ascension and declination of the galaxies optical centers are taken from the NASA/IPAC Extragalactic Database\footnote{\url{https://ned.ipac.caltech.edu/}}. The orientation parameters (optical inclinations and position angles) are drawn from \citet{Brown2021}, and are calculated using fits to SDSS photometry. The average star formation rates and integrated stellar mass are adopted from \citet{Leroy2019}. H\,I deficiencies are adopted from \citet{Zabel2022}, and are calculated using the predicted H\,I mass from field galaxies at fixed stellar mass.
\end{minipage}

\label{tab:sample}
\end{table*}

\begin{table*}
\centering
\caption{{Literature compilation of slopes from resolved molecular KS relations.}}
\begin{tabular}{lcccr}
\hline
\hline
Reference & Index ($N$) & Scale & $\alpha_\mathrm{CO}$ & Fitting Method \\
&  &  & ($M_\odot\,\mathrm{pc}^{-2}/(\mathrm{K\,km\,s}^{-1})$) & \\
\hline
VERTICO & $0.97\pm0.07$ & 720\,pc & 4.35 & Least Trimmed Squares\\
\citet{Querejeta2021} & $0.97\pm0.06^a$ & $\sim 1-2$\,kpc & $4.35 \times Z'^{-1.6}$ & Bisector Ordinary Least Squares (OLS)\\
\citet{Pessa2021} & $1.06\pm0.01$ & 100\,pc & $4.35 \times Z'^{-1.6}$ & OLS\\
\citet{Sanchez2021} &  $0.95\pm0.21$ & $\sim$kpc & 4.3 & OLS\\
\citet{Ellison2021} &  $0.86\pm0.01$ & $\sim$kpc & 4.3 & Orthogonal Distance Regression (ODR)\\
\citet{Morselli2020} & $0.80\pm0.12$ & 500\,pc & 4.3 & OLS\\
\citet{Lin2019} & $1.05\pm0.01$ & $\sim$kpc & 4.3 & ODR\\
\cite{Usero2015}  & $1.01\pm0.08$ & $\sim 1-2$\,kpc & 4.4 &Total Least Squares\\
\citet{Leroy2013}  & $1.00\pm0.15$ & $\sim$kpc & 4.35 & Monte Carlo fitting\\
\citet{Schruba2011} & $1.0\pm0.1$ & $\sim$0.2-2\,kpc & 4.35 & Bisector OLS\\
\cite{Genzel2010} & $1.17\pm0.09$ & > kpc & $4.35^b$ & Not specified\\
\cite{Daddi2010} & $1.31\pm0.09$ & > kpc & $3.6^c$ & Not specified\\
\citet{Blanc2009}  & $0.82\pm0.05$ & $\sim$750\,pc & $6.0^d$ & Monte Carlo fitting\\
\citet{Bigiel2008} & $1.0\pm0.02$ & $\sim$750\,pc & 4.35 & Bisector OLS\\
\cite{Kennicutt2007} & $1.37\pm0.03$ & $\sim$500\,pc & $6.0^d$ & Bivariate Least Squares\\

\hline
\end{tabular}

    \begin{minipage}{2.0\columnwidth}
        \vspace{1mm}
        {\bf Notes:} The $\alpha_\mathrm{CO}$ factors correspond to the CO\,(1-0) transition. ($^a$) \citet{Querejeta2021} report a range of values between $0.90-1.43$ for the slopes, which is due to the choice of $\alpha_\textrm{CO}$ conversion factor. ($^b$) \citet{Genzel2010} employ a conversion factor of 4.35 for star-forming galaxies (SFGs) and 1.36 for submillimeter galaxies (SMGs). ($^c$) \citet{Daddi2010} use a conversion factor of 3.6 for $z=0.5-2.5$ normal galaxies, 4.6 for local spirals, and 0.8 for local ULIRGs and distant SMGs. ($^d$) \citet{Blanc2009} and \citet{Kennicutt2007} adopt the conversion factor derived in \citet{Bloemen1986}, with $X_{\mathrm{CO}}=2.8\times10^{20}\,\mathrm{cm}^{-2}\,\mathrm{(K\,km\,s^{-1})}^{-1}$.

    \end{minipage}

\label{tab:previous_results}
\end{table*}

\begin{table}
\centering
\caption{Best-fit values for the ensemble VERTICO samples.}
\begin{tabular}{cccc}
\hline
\hline
Sample & Index ($N$) & $\mathrm{log}_{10}\,A$ & $\sigma$ \\
 & & ($\mathrm{log}_{10}\, M_\odot\,\mathrm{yr}^{-1}\,\mathrm{pc}^{2}$) & (dex) \\
\hline
Galaxies: $720\mathrm{pc}$ & $0.97\pm0.07$ & $-3.20\pm0.08$ & $0.42$\\
Galaxies: $1.2\mathrm{kpc}$ & $0.91\pm0.08$ & $-3.15\pm0.11$ & $0.35$\\
HI-deficient & $0.93\pm0.04$ & -2.98 $\pm$ 0.09 & 0.30 \\
HI-normal & $0.84\pm0.04$ & -3.26 $\pm$ 0.10 & 0.41 \\
\hline
\end{tabular}

    \begin{minipage}{0.95\columnwidth}
        \vspace{1mm}
        {\bf Notes:} Best-fit slope ($N$), intercept (referenced to $\Sigma_\mathrm{mol}$ units of $1\,M_\odot\,\mathrm{pc}^{-2}$), and scatter ($\sigma$) results for the resolved KS relation in the VERTICO sample, comparing the ensembles at $720$~pc and $1.2$~kpc resolution, as well as the HI-normal and HI-deficient samples of galaxies.
    \end{minipage}
\label{tab:global_slopes}
\end{table}

\section{Ancillary data and control sample}
\subsection{Ancillary data}
\label{sec:ancillary}
All galaxies in the VERTICO sample are well-studied Virgo galaxies with existing datasets in a wide range of wavelengths, which allows us to characterize and compare the distributions of gas, stars and recent star formation.

We employ a combination of near-UV as well as near- and mid-IR photometry to derive star formation rates and stellar masses, following \citet{Leroy2019}. In particular, we use near ultra-violet  $\lambda=231$\,nm data (hereafter NUV) from the Galaxy Evolution Explorer \citep[GALEX,][]{Martin2005}. The near and mid-IR data were taken from the Wide-field Infrared Survey Explorer \citep[WISE,][]{Wright2010}, from which we use WISE bands 1 ($\lambda=3.4\,\mu\textrm{m}$), 3 ($\lambda=12\,\mu\textrm{m}$) and 4 ($\lambda=22\,\mu\textrm{m}$).

\subsection{Extragalactic CO sample: HERACLES}
In this study we make use of the HERA CO-Line Extragalactic Survey \citep[HERACLES,][]{Leroy2009} to compare the kpc scale properties of the VERTICO targets with a resolved sample of field galaxies. HERACLES is a large program that used the IRAM-30m single-dish telescope to provide wide-field CO\,(2-1) emission maps of molecular gas sampled at $\sim 13"$ resolution, which corresponds to $\sim500$\,pc at the median distance of the sample, inside the optical radius of nearby galaxies. The final sample consisted of 48 nearby, star-forming galaxy disks, which have good CO extent and sensitivity well beyond the H\,I-H$_2$ transition radius. We use a final sub-sample of 21 HERACLES galaxies detected in CO, after removing galaxy duplicates in both samples (NGC\,4254, NGC\,4321, NGC\,4536, NGC\,4569, and NGC\,4579) and interacting candidates (NGC\,2146, NGC\,2798, NGC\,3034, NGC\,3077, and NGC\,5713). The resolved working observables are calculated direcly from the HERACLES data products at native resolution, using the VERTICO pipeline \citep[see][for a detailed description]{Brown2021}. The HERACLES sample of galaxies is drawn from previous extragalactic surveys that provide data across different wavelengths such as The H\,I Nearby Galaxy Survey \citep[THINGS,][]{Walter2008} and the {\it Spitzer} Infrared Nearby Galaxies Survey \citep[SINGS,][]{Kennicutt2003}. As previously described in \citet{Zabel2022}, these samples were constructed with galaxies following an approximately flat far-infrared (FIR) luminosity distribution. As a result, the HERACLES sample under-represents FIR-faint galaxies and it is likely biased towards galaxies with large gas mass by design. 

\section{Physical parameters}
\label{sec:methods}

\subsection{Molecular gas}
\label{sec:molgas}
The molecular gas mass surface densities can be estimated using the velocity-integrated CO\,(1-0) observations, as a probe of the molecular gas content. In this paper, we derive the molecular surface densities directly from the VERTICO and HERACLES surveys CO\,(2-1) integrated intensity maps, assuming a fixed CO\,(2-1)-to-CO\,(1-0) line ratio, $R_{21}$, of 0.7 as derived by \citet{Brown2021}. This line ratio value agrees well with previous results for normal star-forming galaxies in the literature, such as HERACLES ($0.46-0.97$), EMPIRE ($0.51-0.87$) and PHANGS ($0.50-0.83$), using both transition lines at matched resolution in nearby, star-forming galaxies \citep[e.g.,][]{Leroy2013, JimenezDonaire2019,denBrok2021,Leroy2021c}. We derive the molecular surface density as
\begin{equation}
    \Sigma_{\textrm{mol}}=\alpha_{\textrm{CO}}\,\frac{I_{\textrm{CO\,(2-1)}}}{R_{21}}\,\textrm{cos}(i),
\end{equation}

\noindent where $i$ is the inclination of the galaxy disk. $\alpha_{\textrm{CO}}$ is the CO-to-molecular mass conversion factor for the the CO\,(1-0) transition. Throughout the paper we assume a constant Galactic value, $\alpha^{MW}_{\textrm{CO}}=4.35\,M_\odot\,\textrm{pc}^{-2}\textrm{(K km s}^{-1}\textrm{)}^{-1}$ \citep[see][for a detailed description of its derivation]{Bolatto2013}, for which the corresponding $X_\mathrm{CO}$ is $2\times10^{20}\,\textrm{cm}^{-2}\textrm{(K km s}^{-1}\textrm{)}^{-1}$. We use this value for consistency with previous VERTICO works \citep{Brown2021,Zabel2022}, as well as other CO nearby galaxy surveys \citep[e.g.,][but see Table \ref{tab:previous_results} for an overview]{Leroy2009,Wilson2009,Saintonge2011,Bolatto2017}. This value already includes a 1.36 factor to account for the presence of helium. It is well known that the CO-to-H$_2$ conversion factor depends on metallicity \citep{Wilson1995,Boselli2002,Bolatto2013,Sandstrom2013}.
The galaxies analyzed in this work are all massive spirals with a similar metallicity range. We thus decided to use a similar conversion factor for all of them. 
Second order effects related to the observed metallicity gradients in galaxies might induce systematic effects in the radial variation of the deduced molecular
gas column density. The recent analysis of PHANGS galaxies done by \citet{Pessa2021}, however, suggests that these effects, if present, are negligible and thus justify our 
choice of a constant conversion factor. An example of molecular mass surface density map can be found on the right panel of Figure \ref{fig:products}, for the VERTICO galaxy NGC\,4501.

\subsection{Star Formation Rates}
Our resolved star formation rate measurements are obtained from GALEX and WISE photometry following the procedure laid out in \citet{Leroy2019}. All images are convolved from their native resolution to a common working resolution of 9", using the \citet{Aniano2011} convolution kernels. All Gaia Data Release (DR) 2 stars within the image area were masked. Image backgrounds are estimated and subtracted with the Astropy Background2D function.

\begin{figure*}
\centering
\includegraphics[scale=0.8]{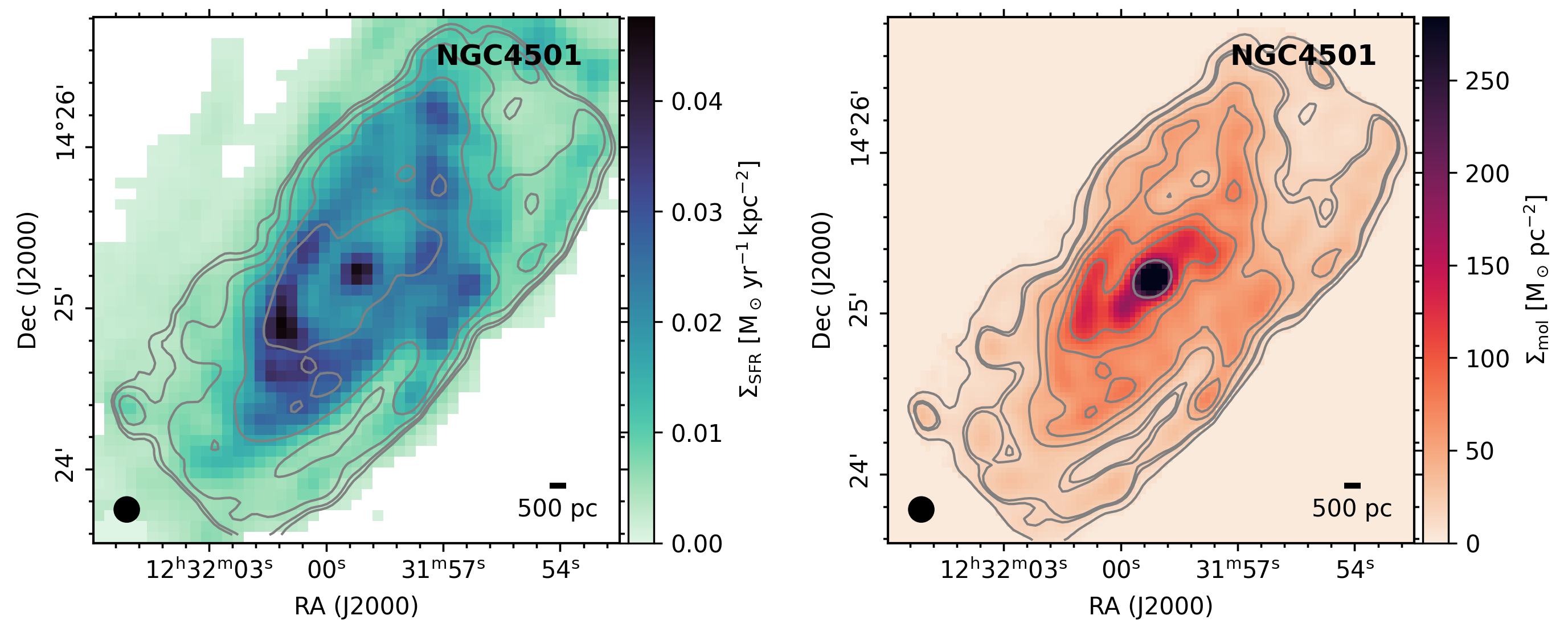}

\caption{Example of the data products used for each galaxy in the VERTICO survey. Left: SFR surface density measured for NGC\,4501 using GALEX NUV and WISE3 photometry. Right: Molecular gas surface density map derived from the VERTICO CO\,(2-1) data products. Molecular gas surface brightness contours at 3, 5, 10, 20 and 40-$\sigma$ detection are overlaid on both panels. The illustrated beam in the lower left corner of each panel is 9" in diameter and represents the working angular resolution. This beam corresponds to 720\,pc at the distance of Virgo (16.5 Mpc). The complete figure set for the VERTICO subsample used in this paper is available online.
}
\label{fig:products}
\end{figure*}

SFR maps were constructed from a combination of GALEX NUV and WISE3 photometry. The left panel of Figure \ref{fig:products} shows an example of SFR map for the VERTICO galaxy NGC\,4501. We use WISE3 images to quantify the obscured SFR at an angular resolution of 9'' similar to the one reached by ALMA. The WISE3 band is preferred to the 
WISE4 band at $22\mu$m for its better angular resolution, although the latter is believed to be a better SFR tracer due to less contamination from polycyclic aromatic hydrocarbons (PAH) emission \citep{Boselli2004}. That said, \citet{Leroy2019} show that WISE4 is a much more robust SFR tracer than WISE3, despite the higher sensitivity and resolution of WISE3 imaging. To address this we apply a pixel-by-pixel WISE3-to-WISE4 color correction to the WISE3 images. Specifically, for each galaxy we fit a linear relation between WISE3/WISE4 color and galactocentric radius and then modify the WISE3 image on a pixel-by-pixel basis according to the galactocentric radius of each pixel and the expected WISE3/WISE4 ratio from the linear fit. A detailed derivation of these maps can be found in Roberts et al. (in preparation). This approach is also suggested in the Appendix of \citet{Leroy2019} for cases where it is necessary to use WISE3 over WISE4 as the mid-IR SFR tracer. With GALEX NUV and color-corrected WISE3 images for each galaxy, we then apply the NUV+WISE4 SFR calibration from \citet{Leroy2019} to derive spatially resolved SFR maps in units $M_\odot$/yr/kpc$^2$. This employs the stellar initial mass function of \citet{Kroupa2003} with bounds of 0.1 and 100$M_\odot$, a slope -2.35
between 1 and 100$M_\odot$, and a slope of -1.3 between 0.1 and 1$M_\odot$. The same procedure was followed to construct resolved SFR maps for all of the HERACLES galaxies, therefore ensuring a homogeneous comparison between galaxy samples. We compared our estimated SFRs for the subsample of galaxies that are also included in HERACLES, to the SFRs calculated in \citet{Leroy2013} and \citet{Bigiel2008} using a combination of GALEX FUV and $24\mu$m images. For that subsample of galaxies, we find that variations are on the order of $\sim7\%$.

\section{Results}
\label{sec:results}
In this section we present our main results for the resolved Kennicutt-Schmidt relation in the VERTICO galaxies, and we explore how this relation changes with angular resolution. We show the resolved relation for the full sample of galaxies included in Table \ref{tab:sample} as well as for each individual galaxy to analyze galaxy-to-galaxy variations. In addition, we explore how HI-deficiency may impact the shape of the KS relation and the depletion times observed in the full sample.

\subsection{Ensemble Star Formation Relations in VERTICO}
\label{sec:ensemble}

\begin{figure*}
\centering
\includegraphics[scale=0.51]{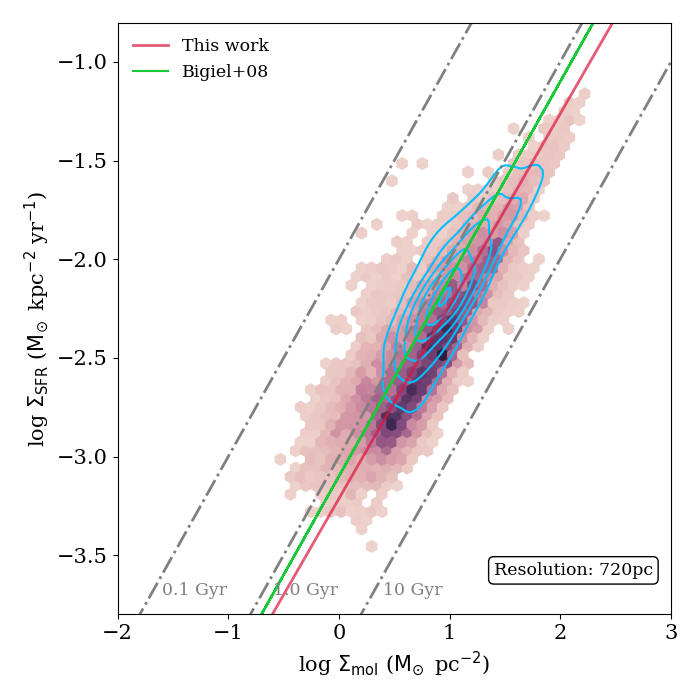}\,
\includegraphics[scale=0.51]{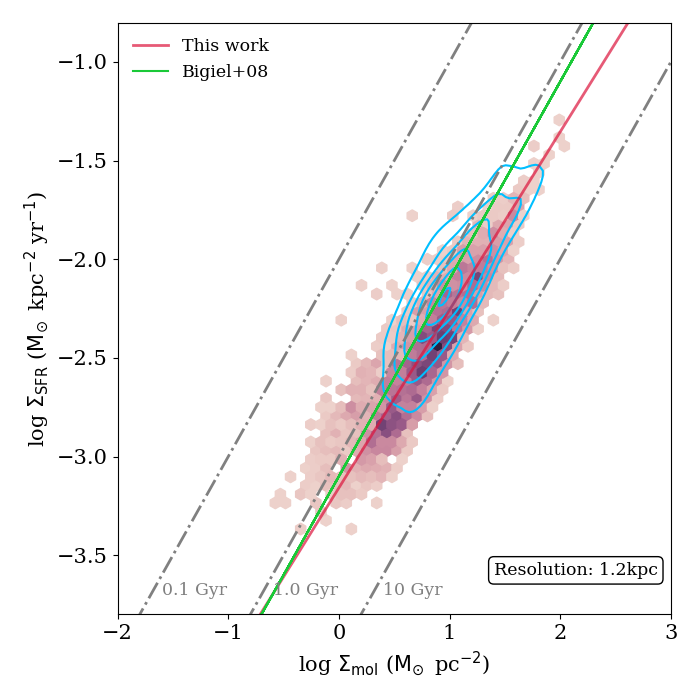}\,
\caption{Left: Resolved KS relation for the ensemble of VERTICO galaxies at a resolution of 720\,pc. Right: Resolved KS relation for the ensemble of VERTICO galaxies with data convolved to a linear resolution of 1.2\,kpc. Only galaxies with $i<80^o$ have been selected, excluding edge-on targets, in order to avoid absorption and strong projection effects. Color code shows increasing point density. Contours show the corresponding KS relation using the sub-sample of field galaxies extracted from the HERACLES survey. Note that values for $\Sigma_\textrm{SFR}$ and $\Sigma_\textrm{mol}$ have been corrected to face-on via cos\,$i$.
}
\label{fig:global-ks-inc}
\end{figure*}

Figure \ref{fig:global-ks-inc} shows the relation between the star formation rate surface density, $\Sigma_\textrm{SFR}$, and the molecular gas surface density, $\Sigma_\textrm{mol}$, for the ensemble sample of galaxies selected from VERTICO. The left panel shows the VERTICO data at 9" (720\,pc) resolution, while the right panel shows the same dataset convolved to 15", corresponding to physical scales of 1.2\,kpc. In both panels, the individual bins are color-coded by point density, where brighter colors show higher density of data points. The panels also include levels of constant depletion times of 0.1, 1.0 and 10 Gyr, respectively, as dashed grey lines. These diagonal lines represent an estimation of the time that the rate of star formation would need in order to deplete the molecular gas reservoir to produce stars.

The blue contours shown on top of the VERTICO data correspond to sub-samples of HERACLES field galaxies, chosen to match the fixed-physical scales samples in each panel. Therefore, these data are also plotted at 720\,pc and 1.2\,kpc resolution in order to compare two datasets that are intrinsically very different, between cluster and field galaxies. The plot shows that the KS relation between SFR surface density and the molecular gas surface density for the high-density points of VERTICO galaxies is very similar to that found for the HERACLES field galaxies. The scatter in the data shown in Figure \ref{fig:global-ks-inc} is approximately symmetric around depletion times of 1-3\,Gyr. Figure \ref{fig:global-ks-inc} therefore demonstrates visually the first main conclusion of this paper: when considered as an ensemble, the KS relation for Virgo cluster disk galaxies is consistent with field galaxies.

In order to better quantify this relation, we perform a robust linear regression following the method presented in \citet{Cappellari2013}, which implements the least trimmed squares (LTS) regression approach of \citet{Rousseeuw1984}. This method takes into account the uncertainties in all variables, the presence of outliers in the data, as well as unknown intrinsic scatter in excess of uncertainties. Expressing the power-law relation $\Sigma_\textrm{SFR} = a\,(\Sigma_\textrm{mol})^N$, in log-space as:
\begin{equation}
    \textrm{log}\,\Sigma_\textrm{SFR} = A + N\,\textrm{log}\,\Sigma_\textrm{mol},
\end{equation}
we can identify $N$ as the slope of the KS relation, and $A=\textrm{log}_{10}\,a$ as the intercept. We treat both $\Sigma_\textrm{SFR}$ and $\Sigma_\textrm{mol}$ as independent variables with uncertainties, and quote the fitting results in Table \ref{tab:global_slopes}. We derive a slope of $0.97 \pm 0.07$ for the resolved KS relation for the VERTICO data at 720\,pc, and an intrinsic scatter about the fit of 0.42\,dex.

Since molecular cloud formation, stellar feedback, and the significance of SFR tracers can depend strongly on spatial scales \citep[e.g.,][]{Orr2018,Hani2020}, we investigate whether the resolved KS relation experiences any changes when working at a significantly lower resolution. We perform the same robust fit to the data shown in the right panel of Figure \ref{fig:global-ks-inc}, and derive a slope of $0.91 \pm 0.09$ with a RMS scatter of $\Sigma_\mathrm{SFR}$ about the fit of 0.35\,dex, at 1.2\,kpc scales. These results indicate that there are no significant differences in the KS parameters as a consequence of working at different resolutions, these agree well within the uncertainties of the data. As a result of averaging data at lower resolution, we see an expected decrease in the number of points that enter the fitting procedure, which increases the uncertainty of the fit. Consequently, this also effectively narrows the distribution, in good agreement with findings in previous studies of isolated galaxies \citep[e.g.,][]{Bigiel2008,Schruba2010,Pessa2021}.

The star formation efficiency of the molecular gas can be defined as the ratio between the traced rate of star formation and the amount of molecular gas that is available for star formation. In terms of our measured surface densities:
\begin{equation}
    \textrm{SFE}_{\mathrm{mol}}=\frac{\Sigma_\mathrm{SFR}}{\Sigma_\mathrm{mol}},
    \label{eqn:sfe}
\end{equation}
Inversely, the molecular gas depletion time gives a quantitative estimate of the time necessary for a given region to consume the available gas if star formation remains constant at the present rate. It can be defined as
\begin{equation}
    \tau^{\mathrm{CO}}_\textrm{dep} = \Sigma_\textrm{mol}/\Sigma_\textrm{SFR},
\end{equation}

\noindent in the particular case in which the molecular gas is traced by CO. Conversely,
\begin{equation}
    \tau^{\mathrm{CO}}_\textrm{dep} = 1/\textrm{SFE}_\textrm{mol}.
\end{equation}

\noindent In comparison with the mean depletion times of $\sim2$\,Gyr found in \citet{Bigiel2008} and \citet{Leroy2013} for HERACLES, we derive a median H$_2$ depletion time of 1.8\,Gyr for our full sample of VERTICO galaxies. These values are within the range of median depletion times in the PHANGS galaxies \citep[1.18 to 2.10 Gyr,][]{Utomo2018,Querejeta2021}, but overall larger than the 0.6\,Gyr derived from integrated values for the Herschel Reference Survey sample \citep{Boselli2014b}. Our results are in reasonable agreement within uncertainties with the range of KS slopes found in previous studies \citep[e.g.,][see Table \ref{tab:previous_results}]{Bigiel2008,Usero2015,Lin2019,Ellison2021,Pessa2021,Querejeta2021}. The derived slopes at both resolutions are fully compatible within uncertainties to the slope of 1.0 found for the HERACLES samples used in \citet{Bigiel2008} and \citet{Leroy2013}. It is important to note that the slope of the KS relation has been reported to vary significantly depending on the choice of CO-to-H$_2$ conversion factor, resolution, and the details of the fitting methodology employed \citep[e.g.,][]{Leroy2013,Ellison2021,Querejeta2021}, as reflected in Table \ref{tab:previous_results}.

\subsection{The KS relation in HI-deficient environments}

It is well known that atomic gas in galaxies, which ultimately constitutes the fuel for molecular cloud formation, can be strongly affected by environmental
processes. In particular, galaxies residing in clusters can often be HI-deficient, showing asymmetric or even truncated HI disks \citep{Koopmann2004,Chung2009,Taylor2013,Yoon2017}. HI-deficiency is thus a useful parameter that allows us to quantify the effects of environmental influence on cluster galaxies. These differences could directly affect the way star formation proceeds in those galaxy disks. In fact, over the past decade recent studies have shown that molecular gas can also be influenced by dynamical processes in the cluster \citep{Vollmer2008,Fumagalli2009,Boselli2014,Zabel2019,Moretti2020}. Furthermore, the recent VERTICO study described in \citet{Zabel2022} shows that the environmental mechanisms observed in the atomic gas can also affect the molecular gas reservoirs of cluster galaxies. In particular, galaxies with large HI deficiencies have steeper and more centrally concentrated molecular gas radial profiles.

In this study we differentiate between HI-normal galaxies (i.e., not affected by the environment) and HI-deficient galaxies. We do so by establishing a cut based on global HI-deficiencies calculated for VERTICO in \citet{Zabel2022} (see their Table 1, column 10), using the predicted HI mass from field galaxies at fixed stellar mass. In this way, we consider HI-deficient galaxies as those with $\mathrm{def}_{\mathrm{HI},\,M_*}>0.3$.

\begin{figure*}
\centering
\includegraphics[scale=0.51]{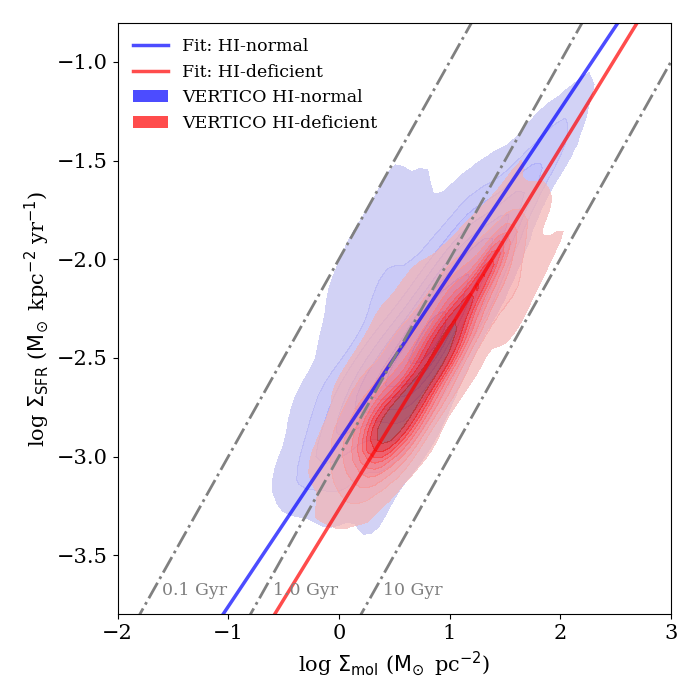}\,
\includegraphics[scale=0.51]{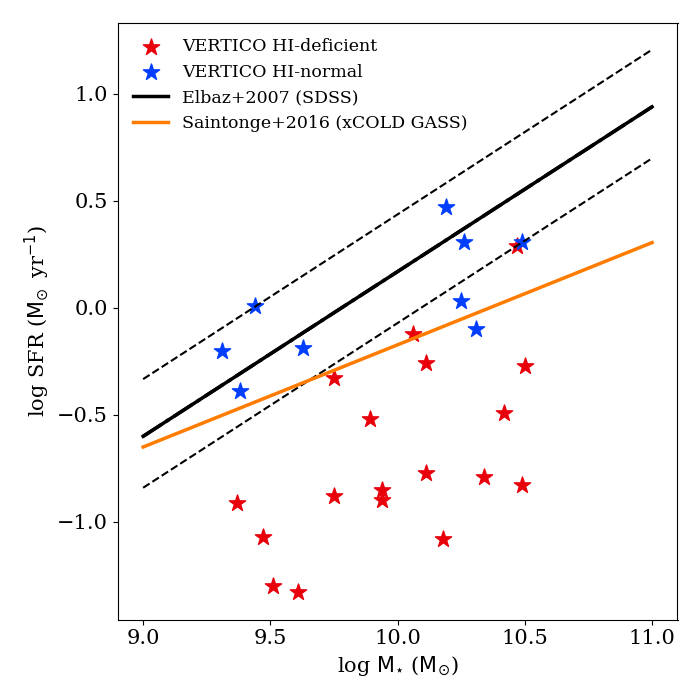}
\caption{Left: resolved KS relation for the VERTICO ensemble sample at a resolution of 720\,pc. The red contours show the relation obtained for HI-deficient galaxies, according to \citet{Zabel2022}, whereas the blue contours indicate the relation obtained for normal galaxies. Solid lines indicate LTS fits to the data. Similarly to Figure \ref{fig:global-ks-inc} only galaxies with $i < 80^o$ have been considered and values for $\Sigma_{\mathrm{SFR}}$ and $\Sigma_{\mathrm{mol}}$ have been corrected to face-on via $\mathrm{cos} i$. The plot shows longer gas depletion times for HI-deficient VERTICO galaxies. Right: $\textrm{SFR}-M_\star$ relation for the VERTICO subsample, limited to HI-deficient and HI-normal galaxies covering the same range in stellar mass. The different colors indicate HI-deficient (red) and HI-normal (blue) galaxies. Overlaid are the star formation main sequence (SFMS) relation obtained in \citet{Elbaz2007} in black (solid line), as well as that presented in xCOLD GASS in orange \citep{Saintonge2016}. While VERTICO HI-normal galaxies lie on the SFMS, HI-deficient galaxies are mostly scattered below the SFMS.}

\label{fig:global-ks-hidef}
\end{figure*}

The left panel of Figure \ref{fig:global-ks-hidef} shows the ensemble resolved KS relation for HI-deficient galaxies (red contours) and normal galaxies (blue contours), allowing us to compare between the two families of galaxies in the sample. By using the same fitting methodology described in Section \ref{sec:ensemble}, we obtain a resolved KS slope of $N=0.93\pm0.04$, a $< \mathrm{log}_{10} \Sigma_\mathrm{SFR}/\Sigma_\mathrm{mol} > =-9.34\pm0.04$ and a 1-$\sigma$ scatter of 0.30\,dex for those galaxies classified as HI-deficient. In comparison, the resulting quantities for the HI-normal galaxies show a characteristic slope of $N=0.84\pm0.04$, a $< \mathrm{log}_{10} \Sigma_\mathrm{SFR}/\Sigma_\mathrm{mol} > =-9.10\pm0.06$ and a 1-$\sigma$ scatter of 0.41\,dex. These two statistically different KS relations indicate that HI-deficient galaxies are offset towards more quiescent SFR surface densities by 0.23\,dex. These differences clearly show that the resolved KS slope and star formation efficiency are moderately affected by the galaxy environment, in the sense that more HI-deficient galaxies tend to have steeper slopes and show lower molecular gas efficiency than HI-rich galaxies.

We compare these two samples of galaxies in the right panel of Figure \ref{fig:global-ks-hidef}. This plot represents the $\textrm{SFR}-M_\star$ relation in VERTICO, limiting our sample to those HI-deficient and HI-normal galaxies covering the same range in stellar mass. This approach allows us to compare their properties more homogeneously. The blue data points indicate HI-normal (non-deficient) galaxies, whereas the red data markers represent the HI-deficient subsample of galaxies. The solid lines indicate the star formation main sequence (SFMS) relation obtained in \citet{Elbaz2007} in black, as well as that presented in xCOLD GASS in orange \citep{Saintonge2016}. The plot shows that VERTICO HI-normal galaxies lie on the SFMS, while HI-deficient galaxies are mostly scattered below the SFMS. HI-deficient galaxies also have global SFRs that are, on average, 0.60\,dex lower than those of galaxies on the SFMS, for a fixed stellar mass.

\begin{figure*}
\centering
\includegraphics[scale=0.30]{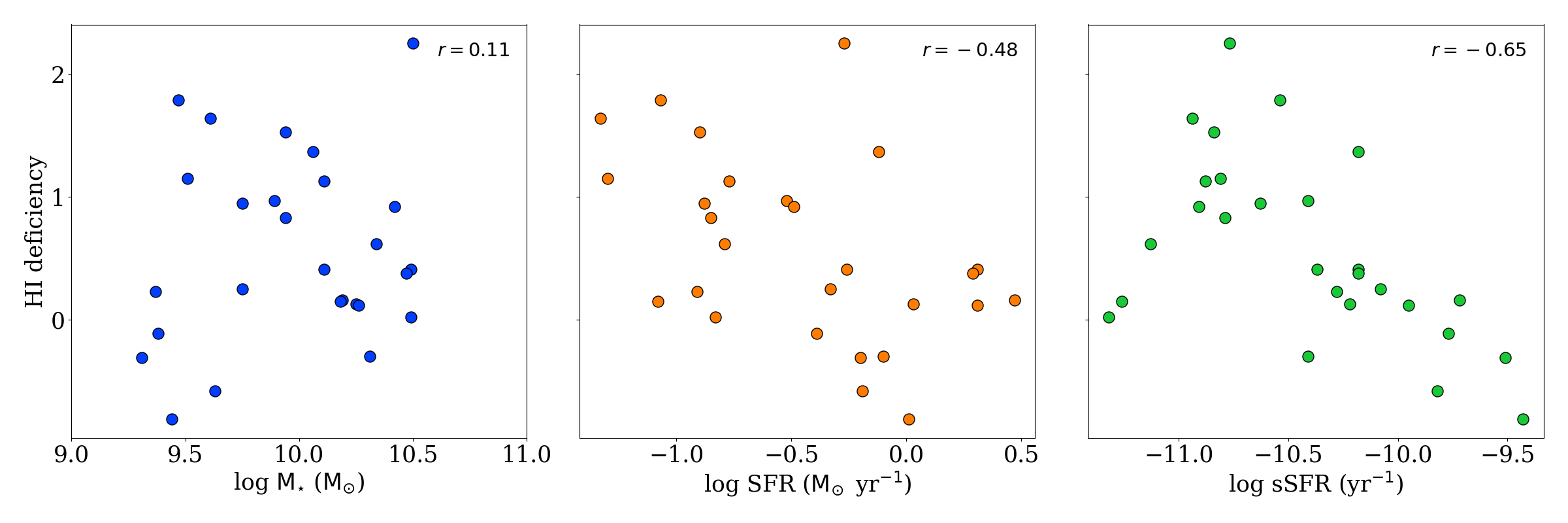}
\caption{HI-deficiency parameter as a function of the total stellar mass (left), global star formation rate (middle) and specific SFR (right) for each galaxy. The Pearson's correlation coefficients are shown on the different panels for each distribution. These quantities are calculated for the VERTICO sample of galaxies shown on the right panel of Figure \ref{fig:global-ks-hidef}.
}
\label{fig:pearson}
\end{figure*}

We explore the potential correlation between the total stellar mass, the global SFR and the specific SFR (sSFR) as a function of HI-deficiency in Figure \ref{fig:pearson}. We measure the strength of the correlation between each pair of variables with the Pearson correlation coefficient. As a reference, coefficients ranging between 0.7 and 1.0 (-0.7 and -1.0) indicate a strong positive (negative) linear correlation. Values between 0.3 and 0.7 (-0.3 and -0.7) show a moderate positive (negative) correlation. Finally, Pearson coefficients between 0 and 0.3 (0 and -0.3) indicate a weak positive (negative) linear relationship. The left panel shows that there is a weak correlation between HI-deficiency and stellar mass. However, we find that there is a moderate anti-correlation between HI-deficiency and the global and specific SFRs. These results indicate that the secular evolution, related to the galaxies' mass, is not the only factor driving the galaxies towards systematically lower values of SFR. On the contrary, the Virgo galaxies presented here are clearly being perturbed by environment-driven processes \citep[e.g.,][]{Chung2009,Yoon2017,Brown2021,Boselli2022,Zabel2022}. Our results show that these processes constitute an important factor in quenching galaxies by systematically lowering the star formation efficiency of the molecular gas.
 Our findings are directly related to the conclusions reported in \citet{Villanueva2022}. This study finds a clear trend of lower molecular SFE for increasingly disturbed HI galaxy disks in Virgo. These findings are also in line with those found by \citet{Zabel2022}. Quantities driving the HI morphology are also affecting the SFE in the molecular gas, without necessarily affecting the amount of molecular gas inside galaxy disks.

\subsection{The KS relation for individual galaxies}
\label{sec:individual}

Recent works \citep[e.g.,][]{Lin2019,Ellison2021,Pessa2021,Querejeta2021} have shown that there is a large galaxy-to-galaxy diversity in the observed KS scaling relations. In this section we explore the variety of KS relations among different galaxies, in particular in Figure \ref{fig:indiv_ks}. These plots are analogous to Figure \ref{fig:global-ks-inc}, but each panel shows the resolved KS relation for each of the 37 galaxies in our VERTICO sub-sample. In all panels, each marker represents an individual measurement at a common resolution of 720\,pc, and they are color-coded by the galactocentric radius.

Similarly to what we describe in Section \ref{sec:ensemble}, we perform a robust LTS fit to each individual galaxy, shown as a solid blue line in each panel of Figure \ref{fig:indiv_ks}. Table \ref{tab:indiv_slopes} shows that the slopes obtained by fitting individual $\Sigma_\textrm{SFR}$-to-$\Sigma_\textrm{mol}$ relations range between $N=0.69$ and $N=1.40$, with a mean typical scatter of $\pm0.21$dex. We note again that these uncertainties on the fitted variables do not reflect important systematic effects such as the choice of SFR tracer or a variable $\alpha_\textrm{CO}$ conversion factor. 

\begin{table*}
\centering
\caption{Fits to the individual molecular KS relations ($\Sigma_\textrm{SFR}=a\,\Sigma_\textrm{mol}^N$) in VERTICO.}
\begin{tabular}{lccccc}
\hline
\hline
Galaxy & Index ($N$) $\mathrm{log}_{10}\,A$ & Pearson coeff. & $\langle\mathrm{log}_{10} \Sigma_\mathrm{SFR}/\Sigma_\mathrm{mol}\rangle$ & Scatter ($\sigma$) \\
 & & ($\mathrm{log}_{10}\, M_\odot\,\mathrm{yr}^{-1}\,\mathrm{pc}^{2}$) & ($p-$value) & ($\mathrm{log}_{10}\, \mathrm{yr}^{-1}$) & (dex) \\
\hline
IC3392 & 0.91 $\pm$ 0.04 & -3.31 $\pm$ 0.09 & 0.9 (<0.001) & -9.40$\pm$0.09 & $\pm$0.20 \\ 
NGC4064 & 1.01 $\pm$ 0.05 & -3.02 $\pm$ 0.10 & 0.9 (<0.001) & -9.00$\pm$0.10 & $\pm$0.27\\ 
NGC4189 & 0.90 $\pm$ 0.02 & -2.98 $\pm$ 0.05 & 0.8 (<0.001) & -9.12$\pm$0.12 & $\pm$0.15\\ 
NGC4254 & 0.97 $\pm$ 0.01 & -3.34 $\pm$ 0.04 & 0.9 (<0.01) & -9.25$\pm$0.11 & $\pm$0.29\\ 
NGC4293 & 1.06 $\pm$ 0.04 & -2.55 $\pm$ 0.09 & 0.9 (<0.001) & -9.28$\pm$0.10 & $\pm$0.30\\ 
NGC4294 & 0.70 $\pm$ 0.05 & -2.55 $\pm$ 0.04 & 0.8 (<0.001) & -8.41$\pm$0.10 & $\pm$0.27\\ 
NGC4298 & 0.96 $\pm$ 0.01 & -3.32 $\pm$ 0.03 & 0.9 (<0.01) & -9.38$\pm$0.11 & $\pm$0.12\\ 
NGC4299 & 1.05 $\pm$ 0.10 & -2.83 $\pm$ 0.31 & 0.6 (<0.001) & -8.08$\pm$0.15 & $\pm$0.26\\ 
NGC4351 & 1.15 $\pm$ 0.06 & -2.93 $\pm$ 0.09 & 0.8 (<0.001) & -8.81$\pm$0.14 & $\pm$0.19\\ 
NGC4380 & 0.72 $\pm$ 0.03 & -3.17 $\pm$ 0.03 & 0.7 (<0.001) & -9.35$\pm$0.12 & $\pm$0.16\\ 
NGC4383 & 1.20 $\pm$ 0.03 & -3.12 $\pm$ 0.21 & 0.8 (<0.001) & -8.92$\pm$0.13 & $\pm$0.23\\ 
NGC4394 & 0.97 $\pm$ 0.06 & -3.06 $\pm$ 0.11 & 0.6 (<0.001) & -9.10$\pm$0.14 & $\pm$0.30\\ 
NGC4405 & 0.96 $\pm$ 0.04 & -3.14 $\pm$ 0.12 & 0.9 (<0.001) & -9.18$\pm$0.10 & $\pm$0.18\\ 
NGC4419 & 0.94 $\pm$ 0.02 & -3.17 $\pm$ 0.06 & 0.9 (<0.001) & -9.24$\pm$0.10 & $\pm$0.21\\ 
NGC4424 & 1.17 $\pm$ 0.06 & -3.09 $\pm$ 0.20 & 0.8 (<0.001) & -9.03$\pm$0.11 & $\pm$0.31\\ 
NGC4450 & 0.69 $\pm$ 0.05 & -3.15 $\pm$ 0.03 & 0.8 (0.001) & -9.32$\pm$0.14 & $\pm$0.25\\ 
NGC4457 & 0.97 $\pm$ 0.02 & -3.46 $\pm$ 0.08 & 0.9 (<0.001) & -9.57$\pm$0.10 & $\pm$0.20\\ 
NGC4501 & 0.92 $\pm$ 0.02 & -3.30 $\pm$ 0.03 & 0.9 (<0.01) & -9.41$\pm$0.09 & $\pm$0.18\\ 
NGC4532 & 0.98 $\pm$ 0.07 & -2.25 $\pm$ 0.19 & 0.7 (<0.001) & -8.33$\pm$0.13 & $\pm$0.34\\ 
NGC4535 & 0.90 $\pm$ 0.03 & -3.13 $\pm$ 0.02 & 0.8 (<0.01) & -9.24$\pm$0.14 & $\pm$0.28\\ 
NGC4536 & 0.88 $\pm$ 0.02 & -2.88 $\pm$ 0.02 & 0.9 (<0.01) & -8.90$\pm$0.13 & $\pm$0.25\\ 
NGC4548 & 0.79 $\pm$ 0.03 & -3.19 $\pm$ 0.03 & 0.8 (<0.001) & -9.32$\pm$0.12 & $\pm$0.21\\ 
NGC4561$^{*}$ & -- & -- & -- & -- & --\\ 
NGC4567 & 0.70 $\pm$ 0.03 & -2.78 $\pm$ 0.05 & 0.8 (<0.001) & -9.05$\pm$0.11 & $\pm$0.26\\ 
NGC4568 & 0.84 $\pm$ 0.02 & -3.12 $\pm$ 0.03 & 0.9 (<0.001) & -9.34$\pm$0.10 & $\pm$0.11\\ 
NGC4569 & 0.82 $\pm$ 0.02 & -3.32 $\pm$ 0.02 & 0.9 (<0.01) & -9.47$\pm$0.12 & $\pm$0.21\\ 
NGC4579 & 0.83 $\pm$ 0.03 & -3.21 $\pm$ 0.03 & 0.8 (<0.01) & -9.35$\pm$0.12 & $\pm$0.17\\ 
NGC4580 & 0.82 $\pm$ 0.02 & -3.18 $\pm$ 0.08 & 0.9 (<0.001) & -9.40$\pm$0.10 & $\pm$0.12\\ 
NGC4606 & 0.93 $\pm$ 0.04 & -3.24 $\pm$ 0.08 & 0.9 (<0.001) & -9.29$\pm$0.11 & $\pm$0.12\\ 
NGC4651 & 0.79 $\pm$ 0.03 & -2.80 $\pm$ 0.03 & 0.8 (<0.001) & -9.14$\pm$0.11 & $\pm$0.20\\
NGC4654 & 0.81 $\pm$ 0.02 & -2.93 $\pm$ 0.02 & 0.9 (<0.01) & -9.08$\pm$0.12 & $\pm$0.25\\ 
NGC4694 & 1.40 $\pm$ 0.05 & -3.31 $\pm$ 0.11 & 0.8 (<0.001) & -9.04$\pm$0.13 & $\pm$0.12\\ 
NGC4698 & 1.04 $\pm$ 0.07 & -3.09 $\pm$ 0.03 & 0.4 (<0.001) & -9.12$\pm$0.17 & $\pm$0.27\\ 
NGC4713 & 0.82 $\pm$ 0.04 & -2.51 $\pm$ 0.08 & 0.6 (<0.001) & -8.82$\pm$0.14 & $\pm$0.23\\ 
NGC4772$^{*}$ & -- & -- & -- & -- & --\\ 
NGC4808 & 0.74 $\pm$ 0.03 & -2.73 $\pm$ 0.02 & 0.9 (<0.001) & -8.87$\pm$0.12 & $\pm$0.15\\
\hline
Whole sample & $0.91\pm0.04$ & -3.21 $\pm$ 0.12 & 0.8 (<0.01) & -9.25$\pm$0.12 & 0.23 \\
\hline
\end{tabular}
\begin{minipage}{2.0\columnwidth}
    \vspace{1mm}
    {\bf Notes:} Best-fit slope, intercept (referenced to $\Sigma_\mathrm{mol}$ units of $1\,M_\odot\,\mathrm{pc}^{-2}$), Pearson correlation coefficient, median value of $\mathrm{log}_{10} \Sigma_\mathrm{SFR}/\Sigma_\mathrm{mol}$, and RMS scatter (1-$\sigma$) measured for the resolved KS relation in the individual VERTICO galaxies at 720\,pc resolution. ($^{*}$) Galaxies NGC\,4561 and NGC\,4772 could not be fitted using a power-law relation. The last row, corresponding to the whole sample, represents the mean values obtained by averaging the previous results for all the individual galaxies. These overall agree with our best-fit results presented in Table \ref{tab:global_slopes}, within uncertainties.
\end{minipage}
\label{tab:indiv_slopes}
\end{table*}

When visually inspecting the panels, it is clear that there is significant galaxy-to-galaxy variation (take galaxies NGC\,4457 and NGC\,4713 as examples), confirming that there is no single resolved KS relation that can accurately describe every galaxy. Figure \ref{fig:indiv_ks} visually demonstrates the second main conclusion of our work, that disk galaxies in VERTICO exhibit significant galaxy-to-galaxy variation in their KS relations. This result agrees with previous findings by studies using resolved molecular gas maps in isolated and cluster galaxies, such as \citet{Bigiel2008,Shetty2013,Shetty2014,Ellison2020,Zabel2020,Ellison2021}. However, while we see that the slope and normalization vary from galaxy to galaxy, almost all of them show a strong, direct proportionality between $\Sigma_\textrm{SFR}$ and $\Sigma_\textrm{mol}$. Both quantities appear to be related for most galaxies through an index which is close to the $N=1$ molecular KS law. In other words, the galaxies selected from the VERTICO survey contain molecular gas that appears to be forming stars at a nearly constant rate (or constant depletion times), although that rate can vary from galaxy to galaxy by a factor of $\sim 4$. This is one of the main sources of variations in the overall scatter we see in the resolved KS relation for the ensemble of galaxies, since galaxy-to-galaxy variations can account for a factor of $\sim 0.20$\,dex scatter in the ensemble (with a total of $0.42$\,dex) sample \citep{Ellison2021}.

Figure \ref{fig:slopes} shows the distribution of best-fit KS index for individual galaxies in VERTICO, as well as a comparison to other isolated galaxies in HERACLES \citep{Leroy2013} and PHANGS \citep{Pessa2021}. We stress the message described above: while there is a clear variety of KS indexes from galaxy to galaxy, 76\% of our targets' slopes are within the range of $0.8-1.2$. This agrees with previous results showing that, to first order, the reservoir of molecular gas is the key quantity regulating star formation. While the slope index $N$ is very close or compatible with unity, we note that the majority of our sample (about 67\%) is characterized by sub-linear slopes. This could be the result of a possible dependence on the assumed constant $\alpha_{\mathrm{CO}}$. Given that both metallicity and gas column density can decrease radially across galaxy disks, the slope of this KS relation can be modified as a result. On the other hand, this could also indicate that the star formation efficiency of the molecular gas tends to decrease at high molecular gas densities for most of the galaxies. A variable molecular gas star formation efficiency is also evident in the significant scatter we see for each individual KS relation, indicating that the $\Sigma_\textrm{SFR}$-to-$\Sigma_\textrm{mol}$ relation varies not only from galaxy to galaxy, but also within galaxies. This reflects the broad variety of physical conditions presented in Virgo galaxies. Some cases are particularly interesting, such as NGC\,4561 and NGC\,4772. Figure \ref{fig:indiv_ks} shows that the KS relation seems to break for these galaxies at face value, and therefore a power-law description is not a good approximation. It is important to note that these two targets are not representative of the full sample, since they have been marginally detected in CO in VERTICO \citep[see][]{Brown2021}. Their integrated intensity maps show a very small dynamic range in $\Sigma_\mathrm{mol}$ up to $\sim 5-7\,M_\odot\,\mathrm{pc}^{-2}$. This is also evident by the limited radial profiles shown in \citet{Brown2021}, where only a few detections are available in the inner regions of the galaxies disks ($r < 6\,\mathrm{kpc}$). 

There are two additional galaxies, NGC\,4501 and NGC\,4579, which present a clear bimodal behavior for different galactic environments. Figure \ref{fig:indiv_ks} indicates that their galaxy centers, within approximately 1.5 kpc, are characterized by much larger depletion times than those seen in their overall disks ($r>1.5\,$kpc). Their centers, therefore, appear to follow a parallel KS relation to their disks. We note that these galaxies' central dynamics are particularly striking, as displayed in the velocity fields and position-velocity (PV) diagrams available in \citet{Brown2021}. 

There are several possible explanations behind the central deviation of $\tau_\mathrm{dep}$ in NGC\,4501. First, its central kinematics show signatures of outflowing material \citep{Wong2004,Repetto2017}, likely due to
elliptical streaming, and this galaxy hosts a central AGN \citep{Brum2017,Ruschel2021}. The presence of central AGNs has been shown to affect the molecular gas reservoirs in the central regions of galaxies \citep{Ellison2021b,GarciaBurillo2021}. If the central molecular gas is outflowing, the standard Galactic conversion factor may not be applicable in this case, since the gas is not likely to be self-gravitating. Previous attempts to constrain molecular outflow conversion factors suggest that they could be two to four times below our constant Galactic value \citep{Weiss2005,Cicone2012,Cicone2018,Leroy2015,Zschaechner2018,Lutz2020}. Such a decrease in the conversion factor would also drive the central molecular gas surface densities to values two to four times lower, which could explain the difference in $\tau_\mathrm{dep}$ observed. In addition to that, this system has a clear bar which can drive non-circular motions in the molecular gas, acting to stabilize against star formation. Both the outflow and the bar can contain gas that may not be participating in the formation of new stars, as indeed is expected given that the large energies injected prevents it to collapse and form molecular clouds. Second, the center of NGC\,4501 is characterized by abundant older stellar populations which suggest a weaker star-forming activity. An inferred lower $\Sigma_\mathrm{SFR}$ in its central region would cause a visible vertical drop in the KS relation towards longer depletion times. Both of these effects will cause an observed shift towards lower $\Sigma_\mathrm{SFR}$ and larger $\Sigma_\mathrm{mol}$. Finally, H\,I emission maps and simulations \citep{Vollmer2008,Chung2009} have recently shown that NGC\,4501 is also undergoing early phases of ram pressure stripping. This could potentially drive the central galaxy away from the KS relation in its disk. Similarly to NGC\,4501, NGC\,4579 has a low-luminosity AGN in its center \citep{Kuno2007} and a compact inner nuclear ring that could help increase the molecular gas densities in its central region. However, it is challenging to estimate how the presence of a central AGN would affect the calculated SFRs, since it could directly affect both WISE3 and WISE4 filters differently, and therefore the color correction. In the presence of an AGN, the SFR derived using a combination of UV and IR tracers is likely inaccurate.

\smallskip

It is also interesting to point out the pair of interacting galaxies NGC\,4567 and NGC\,4568. While both galaxies seem to be well characterized by a sublinear KS slope, this pair of spiral galaxies overlap in the sky. In this case it is difficult to differentiate the components of each disk, since their line-of-sight velocities match in the overlapping region. Their H\,I emission maps \citep{Chung2009} show that they are physically connected and interacting gravitationally, which could potentially explain the suppression of the SFR seen in the low molecular gas surface density regions of NGC\,4568. \citet{Nehlig2016} studied in detail the influence of large-scale compression ($\sim$kpc scales) on density, molecular fraction and molecular SFE based on IRAM-30m CO\,(2-1) observations. In agreement with our higher resolution data, they also report a $0.9-1.0$ KS index. They also find lower SFE of the gas in the northwestern interacting part of the galaxies, as well as in the southwestern edge of NGC 4568. They conclude that gravitational ISM compression can justify an increase in gas surface densities as well as molecular fractions, which could potentially explain the observed changes in SFE. In that context, a recent study by \citet{Thorp2022} analyses the resolved molecular gas and star formation properties in a large set of galaxy mergers at different stages of interaction. While they find that scaling relations in post-mergers, pairs and isolated galaxies are similar, there is significant variation in individual mergers.

As described above, the entire VERTICO survey comprises a total of 51 Virgo Cluster galaxies, with a significant fraction of highly inclined disks. While these galaxies have not been taken into account in our analysis, we provide an overview of all KS relations in Appendix \ref{ap:observed} with Figure \ref{fig:obs_ks}, using direct observables (e.g., $I_\mathrm{CO}$).

\begin{figure*}
\includegraphics[scale=0.30]{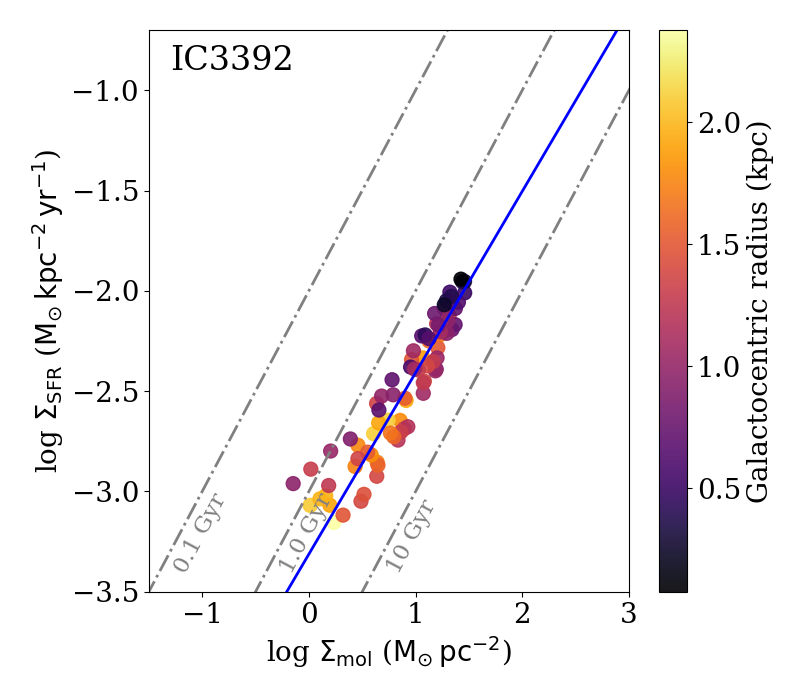}\,
\includegraphics[scale=0.30]{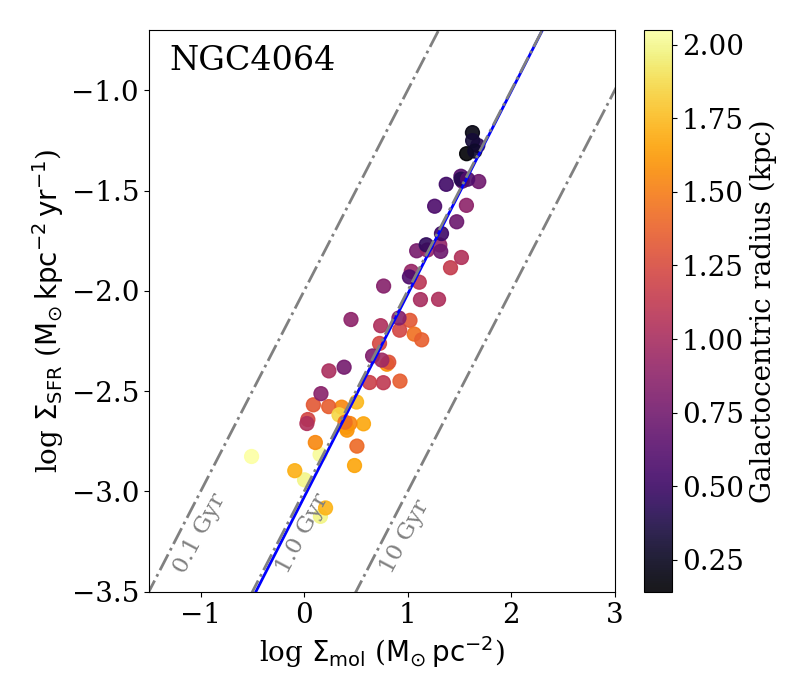}\,
\includegraphics[scale=0.30]{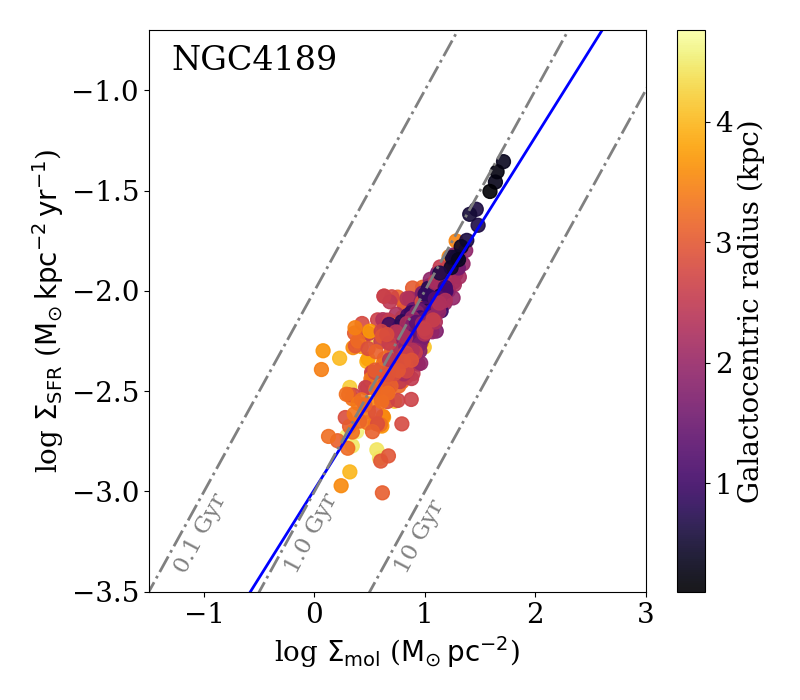}\\

\includegraphics[scale=0.30]{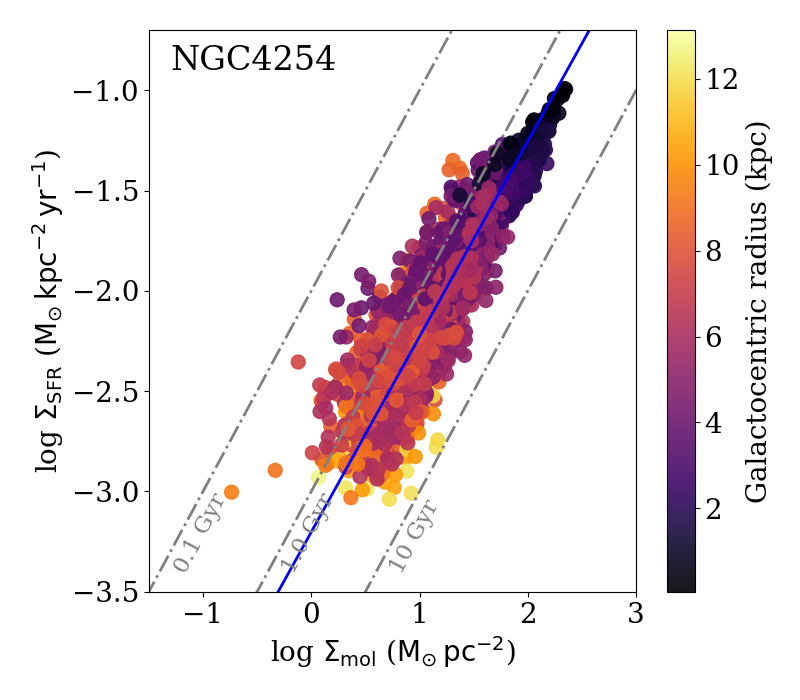}\,
\includegraphics[scale=0.30]{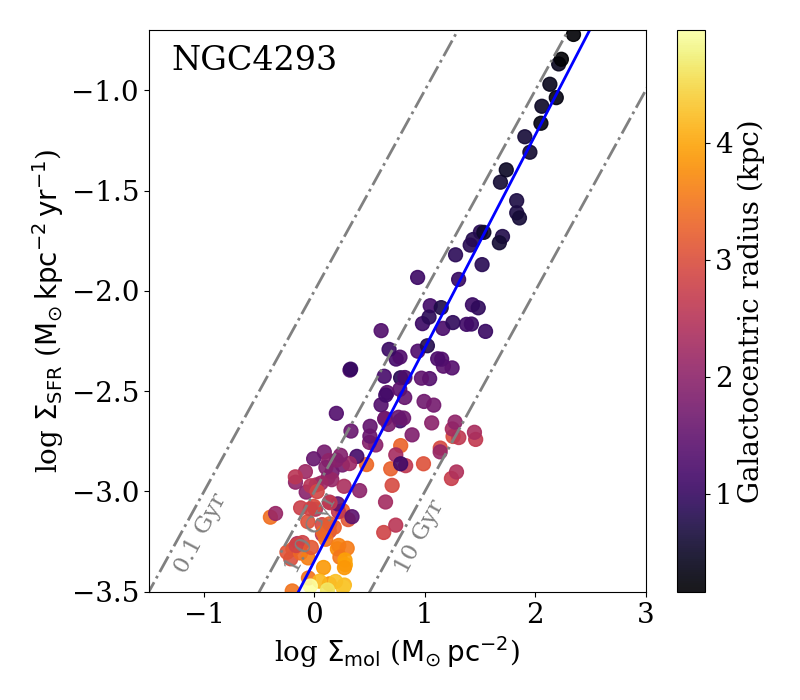}\,
\includegraphics[scale=0.30]{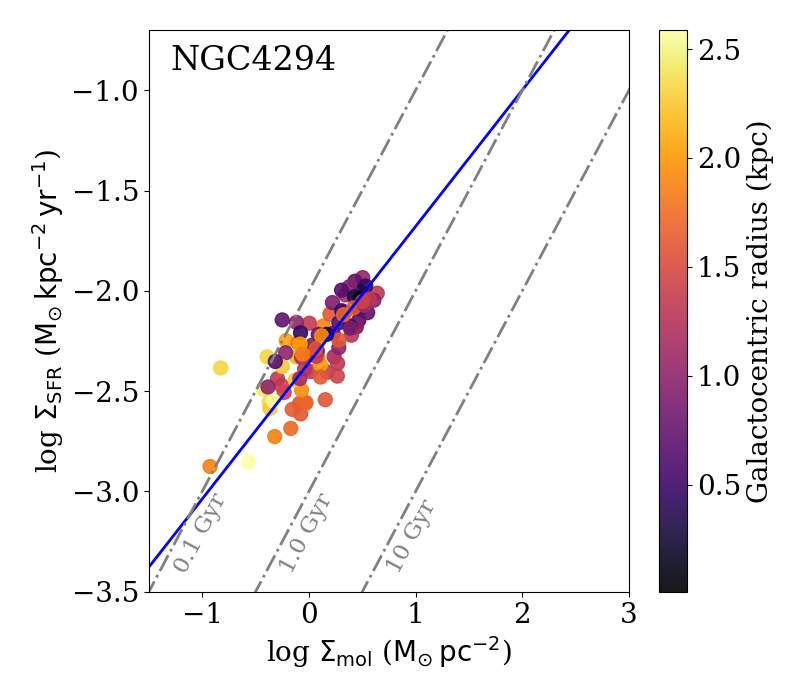}\\

\includegraphics[scale=0.30]{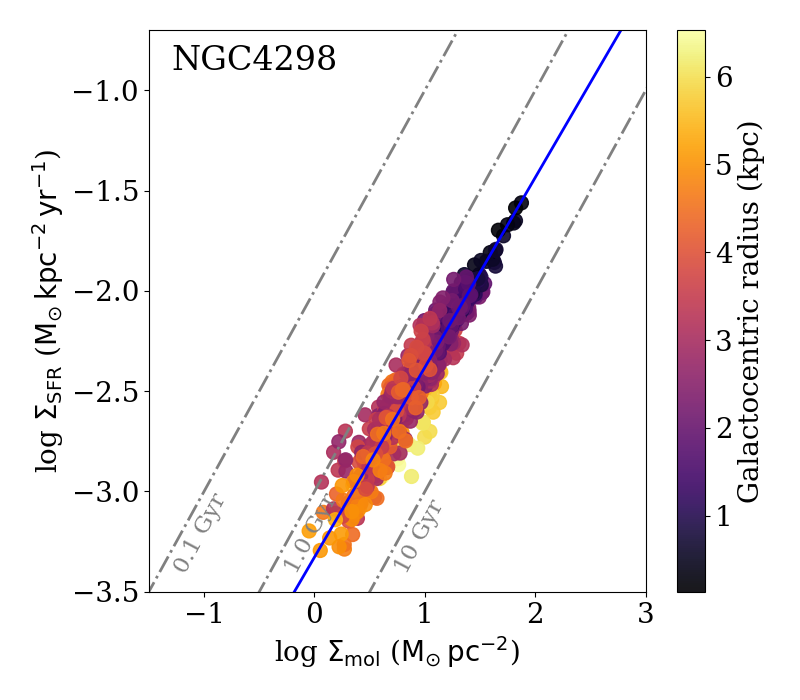}\,
\includegraphics[scale=0.30]{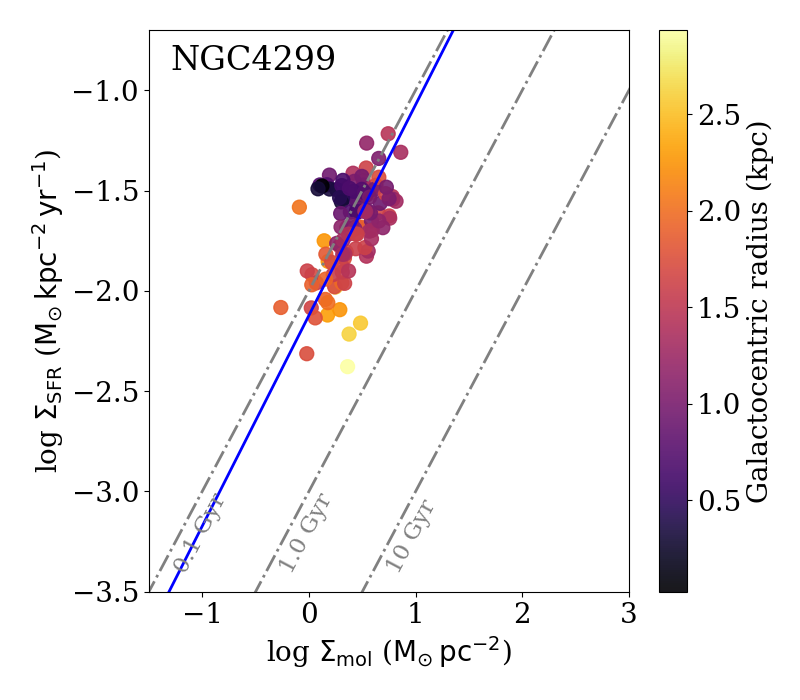}\,
\includegraphics[scale=0.30]{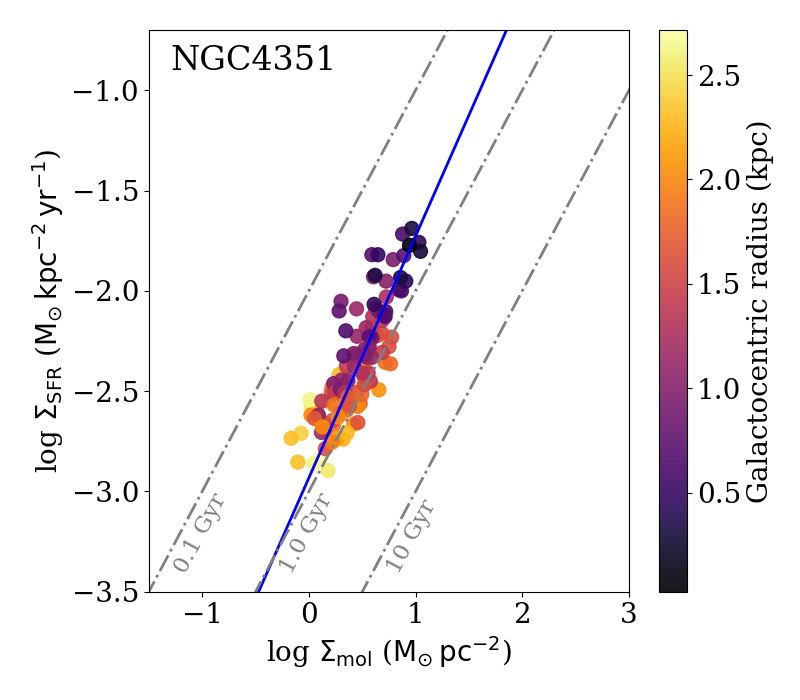}\\

\includegraphics[scale=0.30]{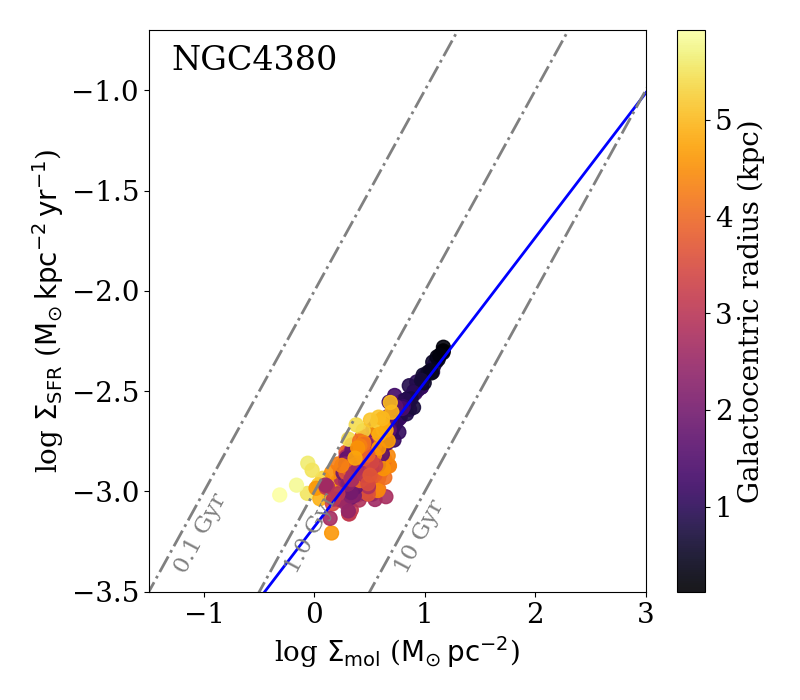}\,
\includegraphics[scale=0.30]{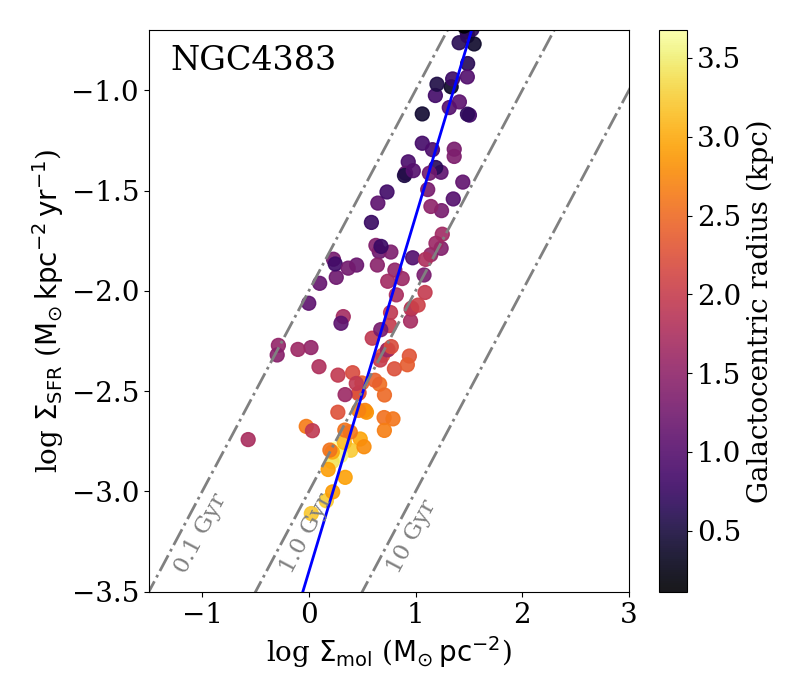}\,
\includegraphics[scale=0.30]{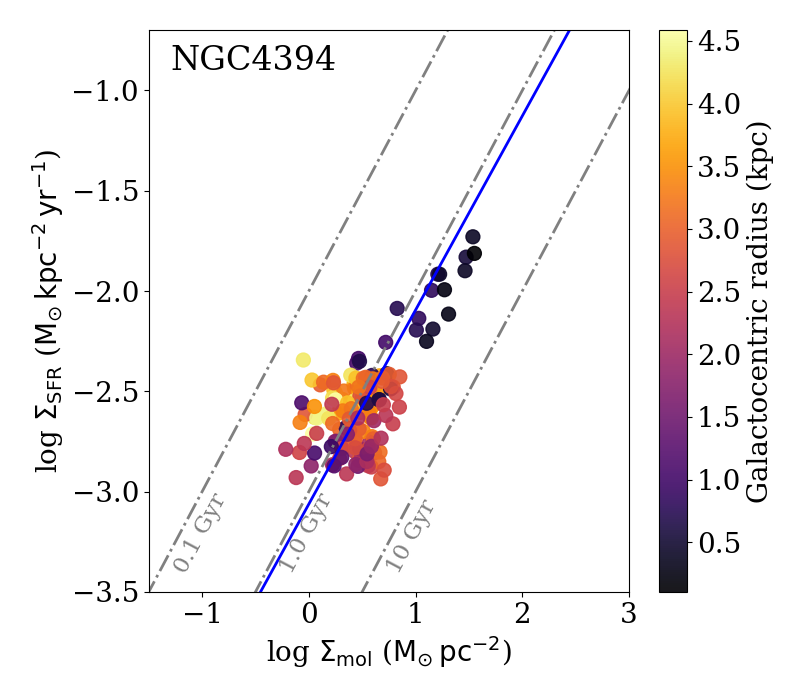}\\
\caption{The resolved Kennicutt-Schmidt relation for the sub-sample of galaxies in the VERTICO sample with inclinations  $i\leq 80^o$, represented as $\Sigma_\textrm{SFR}$ as a function of $\Sigma_\textrm{mol}$. All data points are convolved to a common working resolution of 720\,pc. Each data point is color-coded by the distance to the galaxy center. The diagonal dashed, grey lines show constant depletion times of 0.1, 1 and 10\,Gyr, respectively. For a fixed molecular gas mass, there are clear galaxy-to-galaxy variations in the KS relation. OLS weighted fits to the line-of-sight measurements are the solid blue lines shown in Table \ref{tab:indiv_slopes}. Note that values for $\Sigma_\textrm{SFR}$ and $\Sigma_\textrm{mol}$ have been corrected via cos\,$i$.}
\label{fig:indiv_ks}
\end{figure*}

\renewcommand\thefigure{\arabic{figure}}
\setcounter{figure}{4}

\begin{figure*}

\includegraphics[scale=0.30]{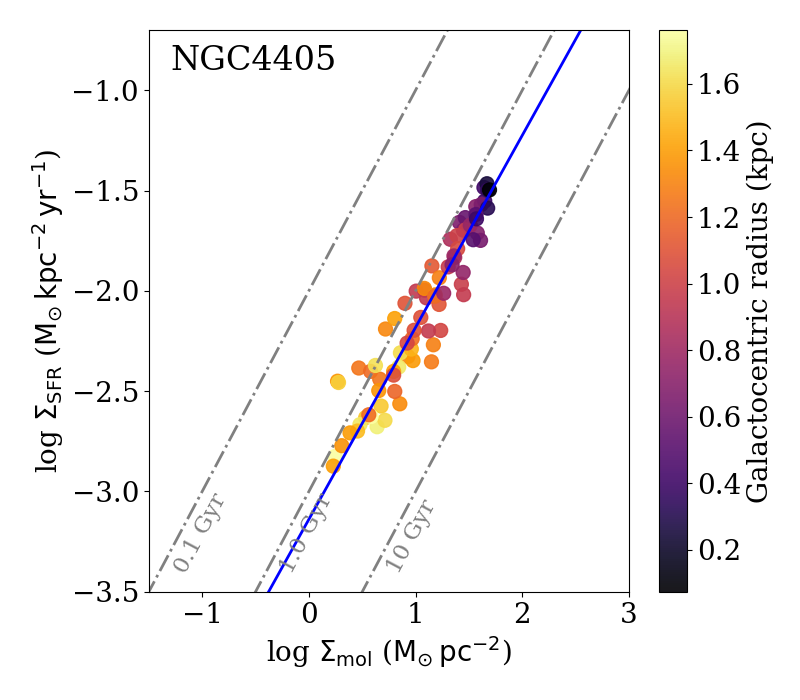}\, \includegraphics[scale=0.30]{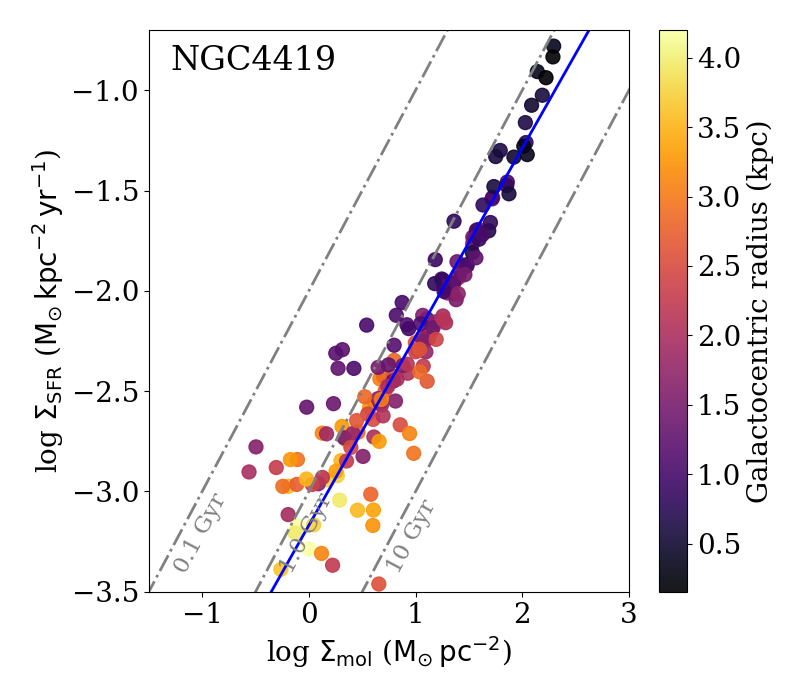}\,
\includegraphics[scale=0.30]{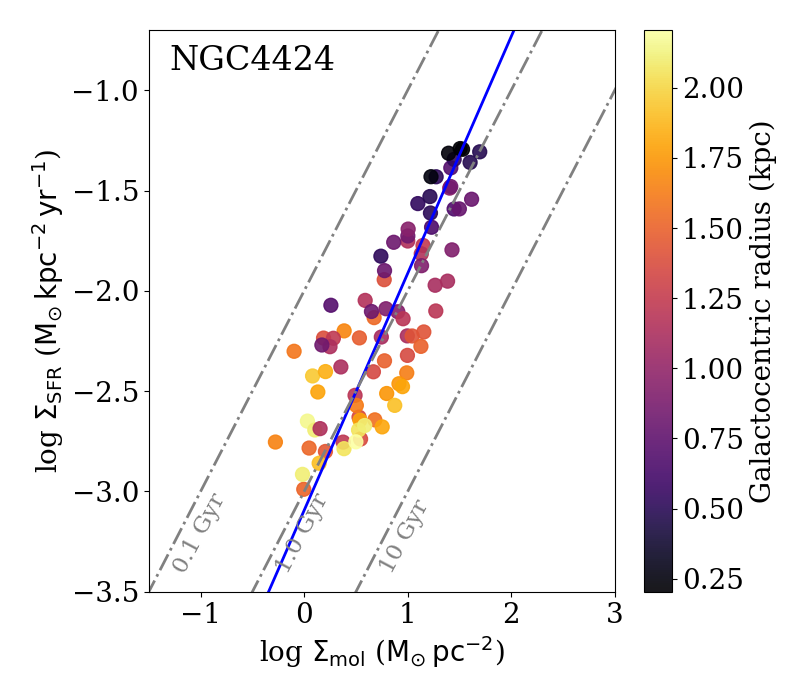}\\

\includegraphics[scale=0.30]{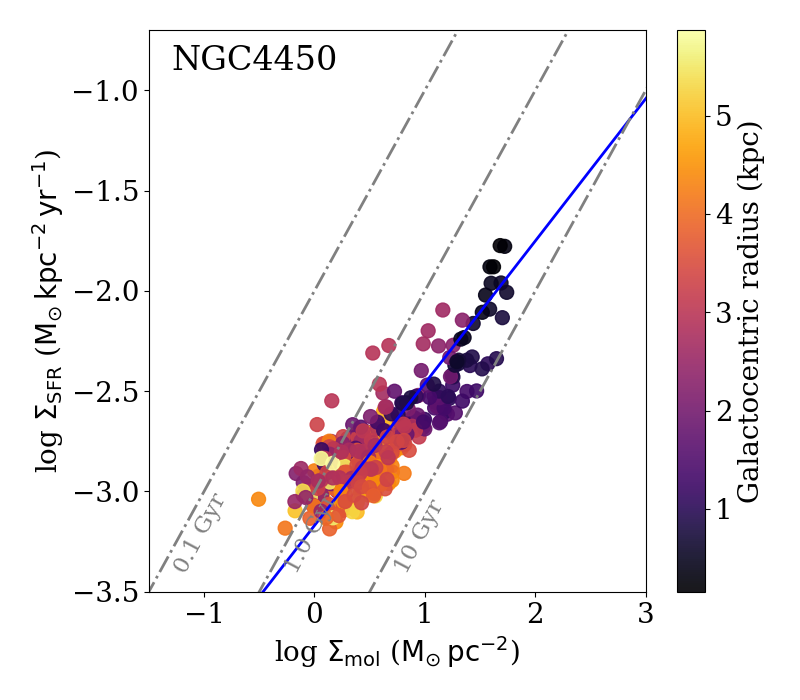}\, \includegraphics[scale=0.30]{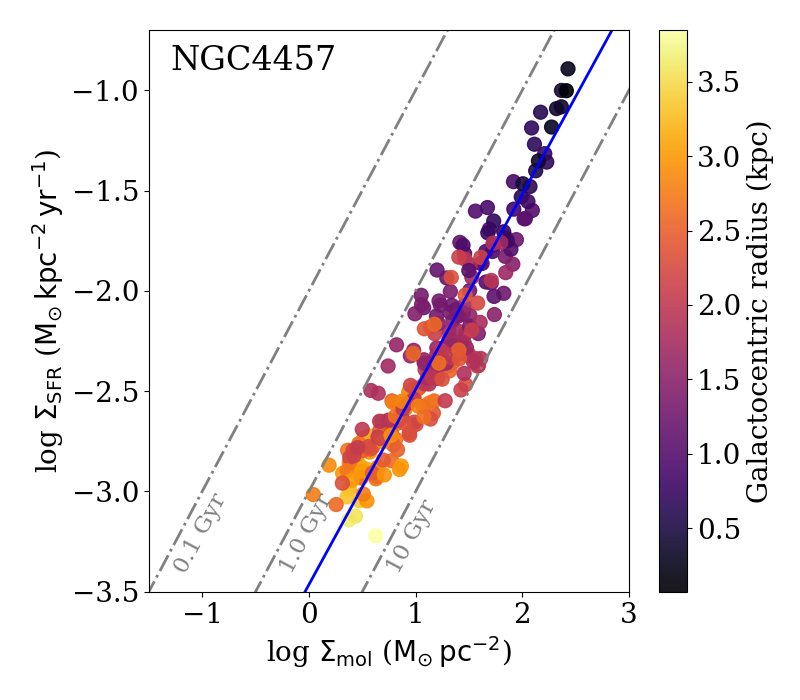}\,
\includegraphics[scale=0.30]{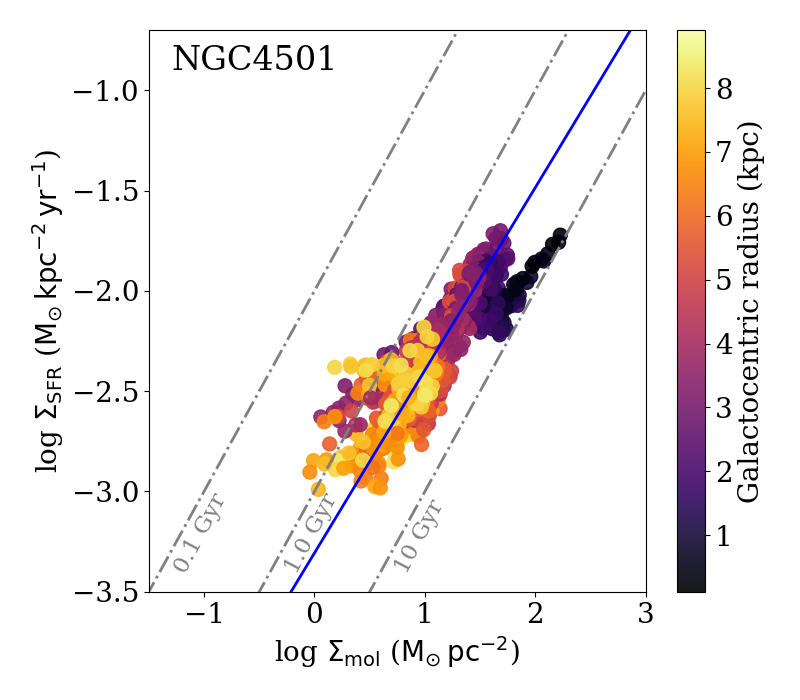}\\ 

\includegraphics[scale=0.30]{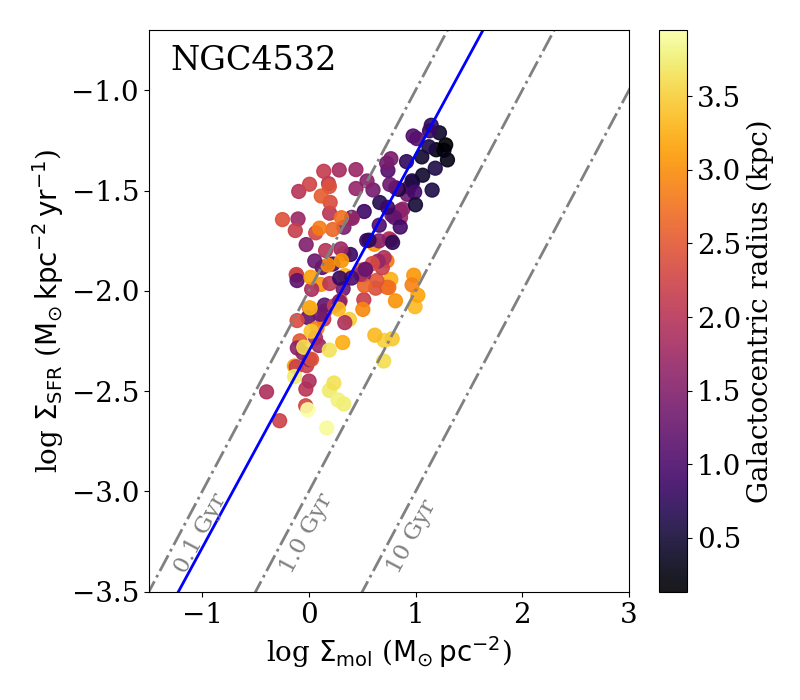}\, \includegraphics[scale=0.30]{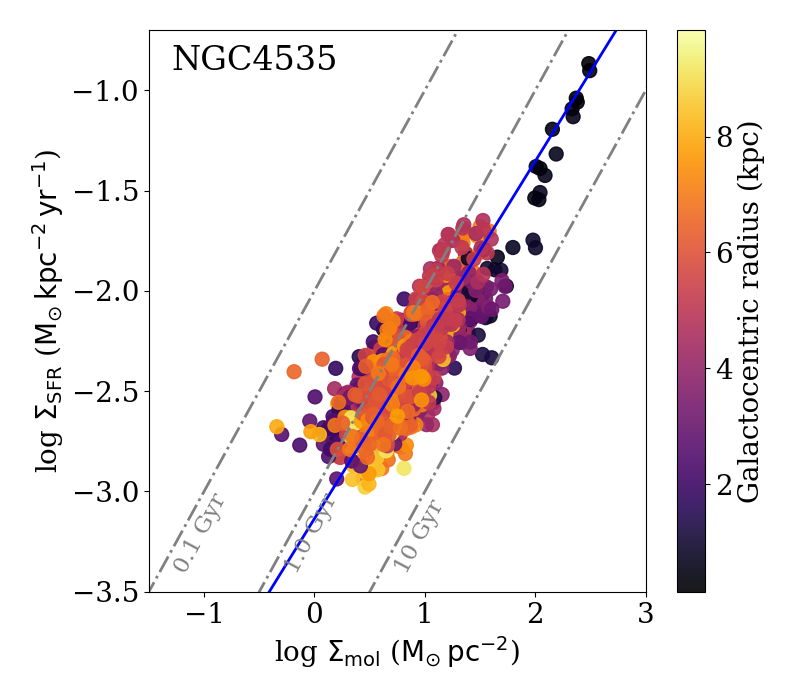}\,
\includegraphics[scale=0.30]{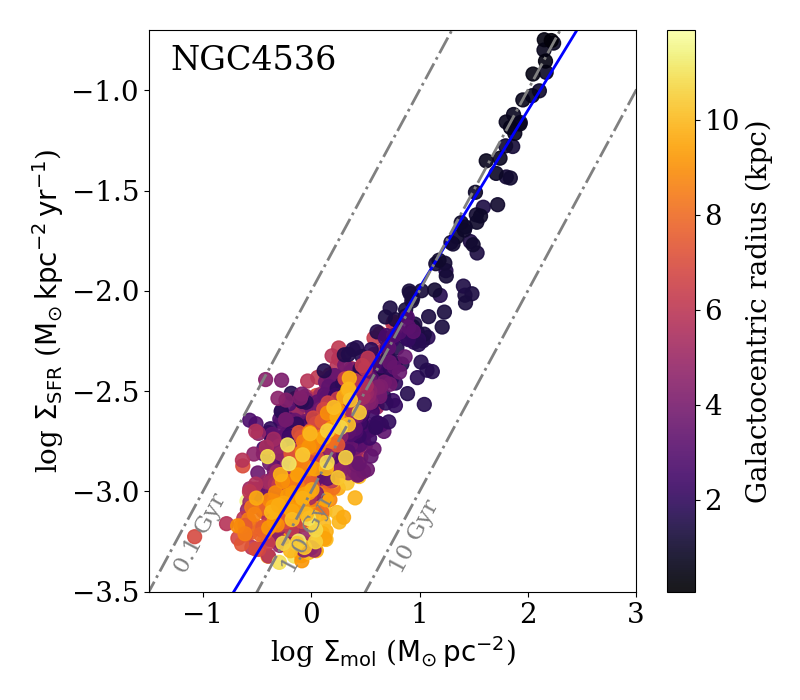}\\

\includegraphics[scale=0.30]{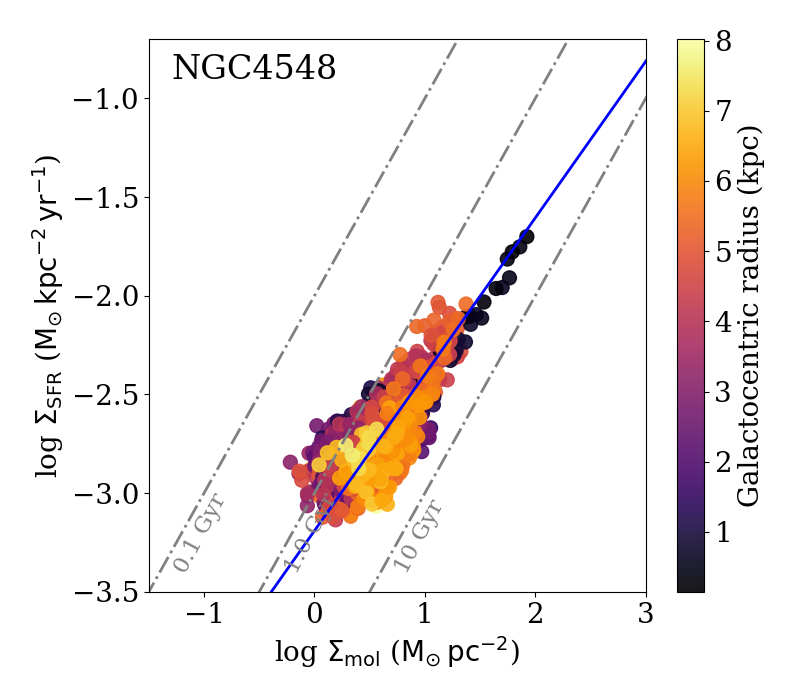}\, \includegraphics[scale=0.30]{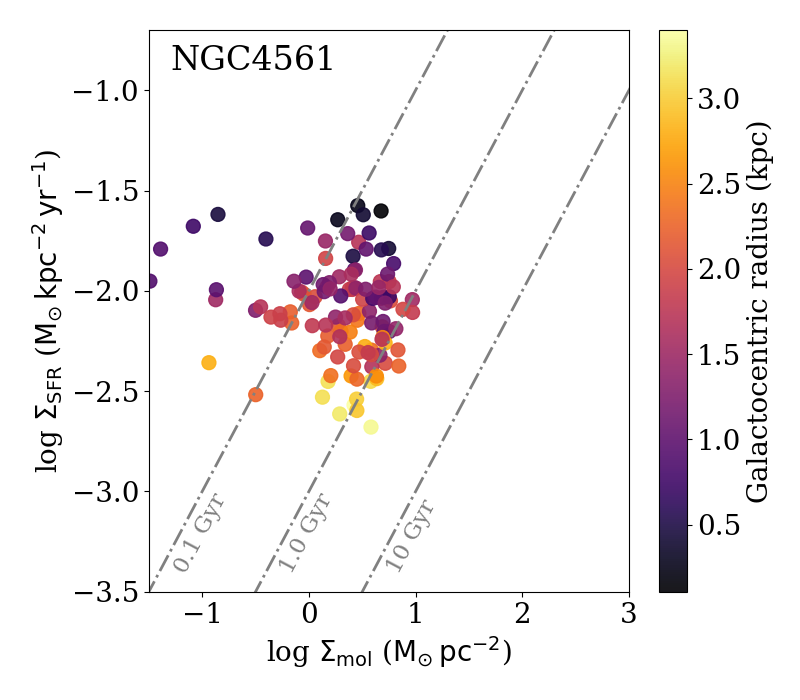}\,
\includegraphics[scale=0.30]{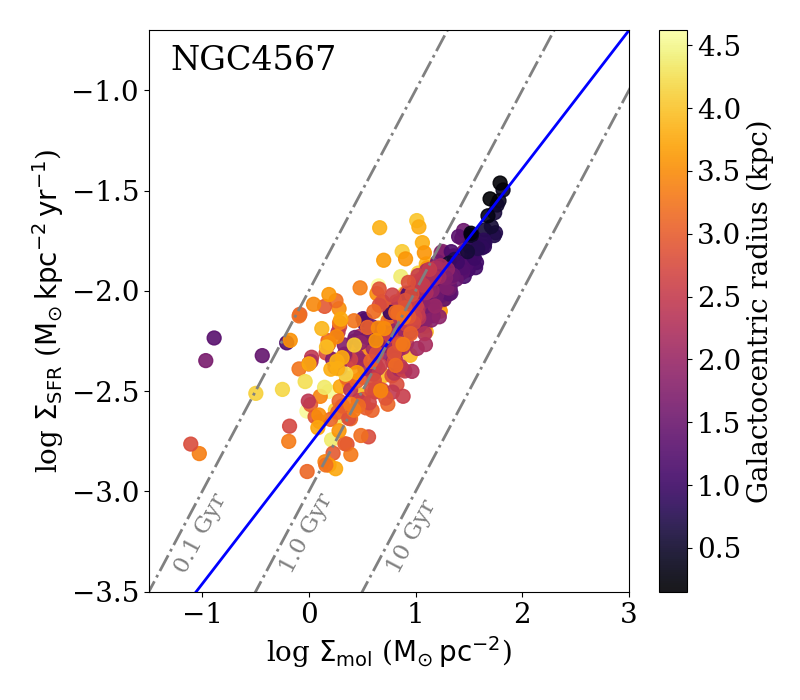}\\
\caption{continued. Note that we do not provide a robust LTS regression for NGC\,4561 since a KS power-law does not appear to be a good fit of the distribution of the data for that galaxy.}
\end{figure*}

\renewcommand\thefigure{\arabic{figure}}
\setcounter{figure}{4}

\begin{figure*}

\includegraphics[scale=0.30]{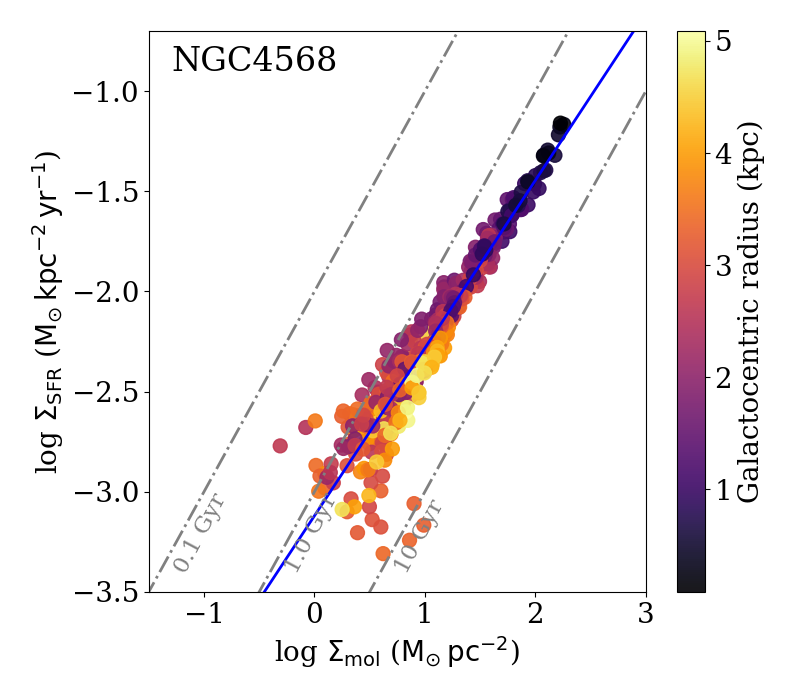}\, \includegraphics[scale=0.30]{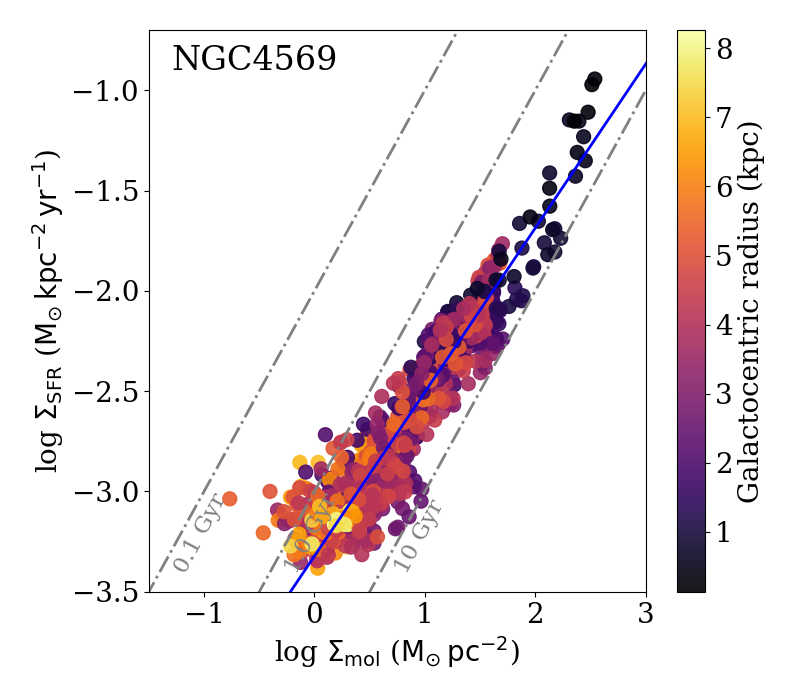}\,
\includegraphics[scale=0.30]{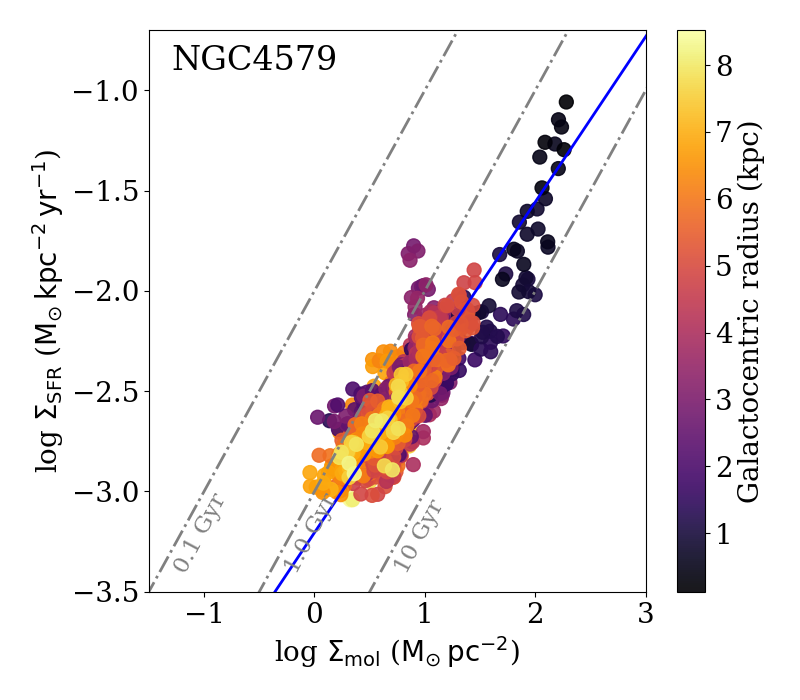}\\ 

\includegraphics[scale=0.30]{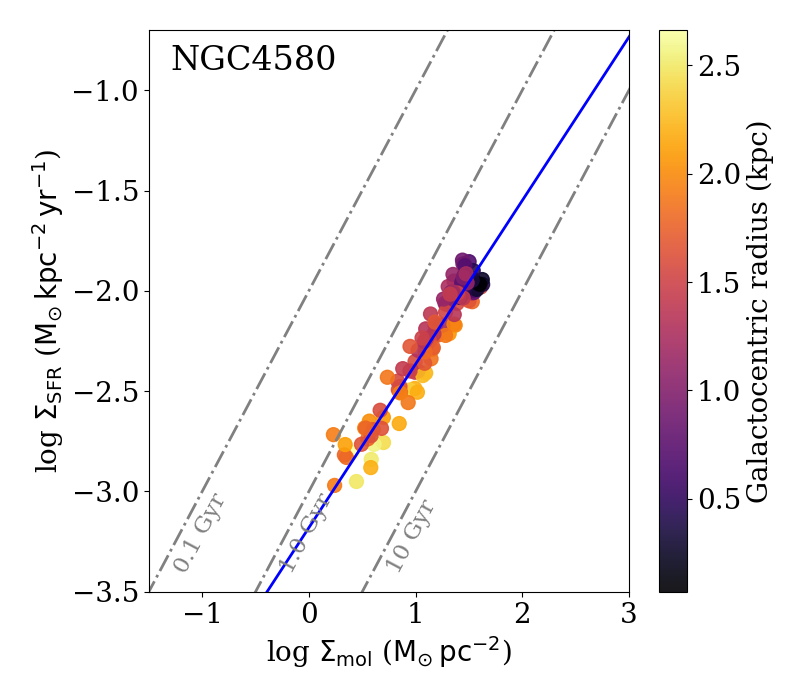}\, \includegraphics[scale=0.30]{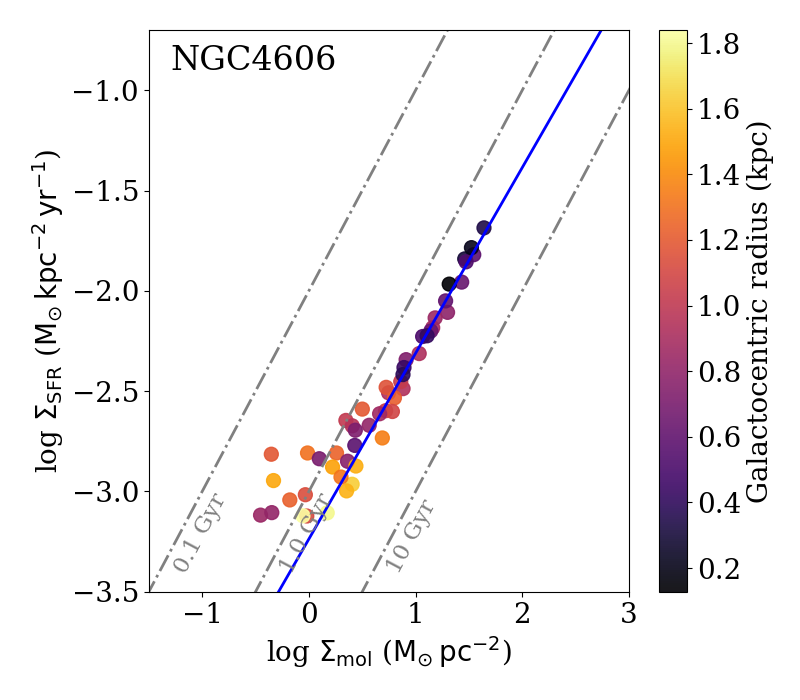}\,
\includegraphics[scale=0.30]{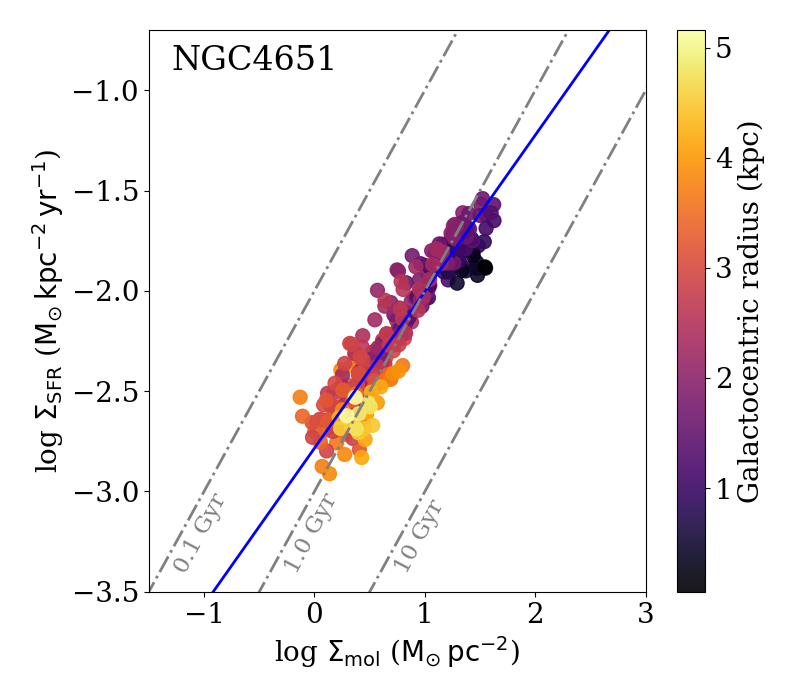}\\ 

\includegraphics[scale=0.30]{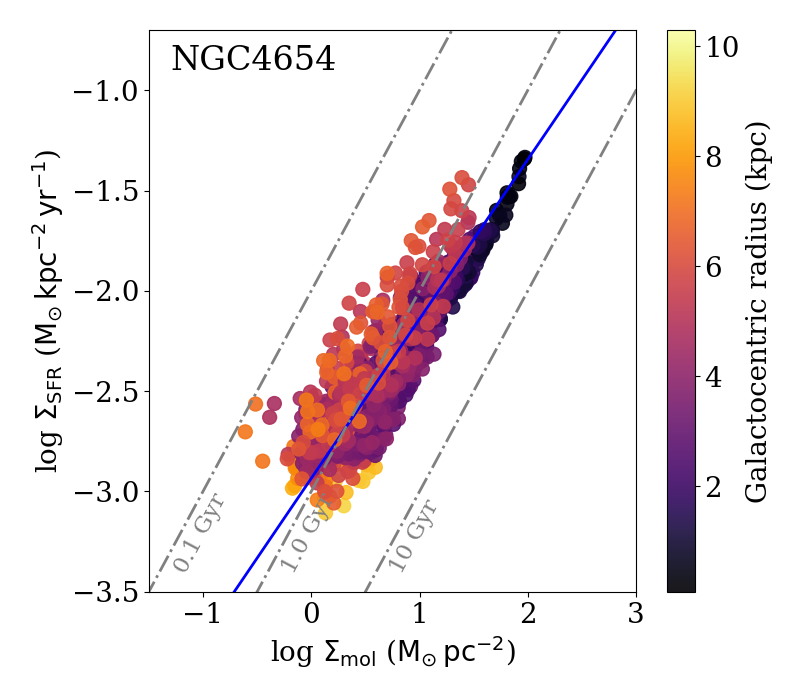}\, \includegraphics[scale=0.30]{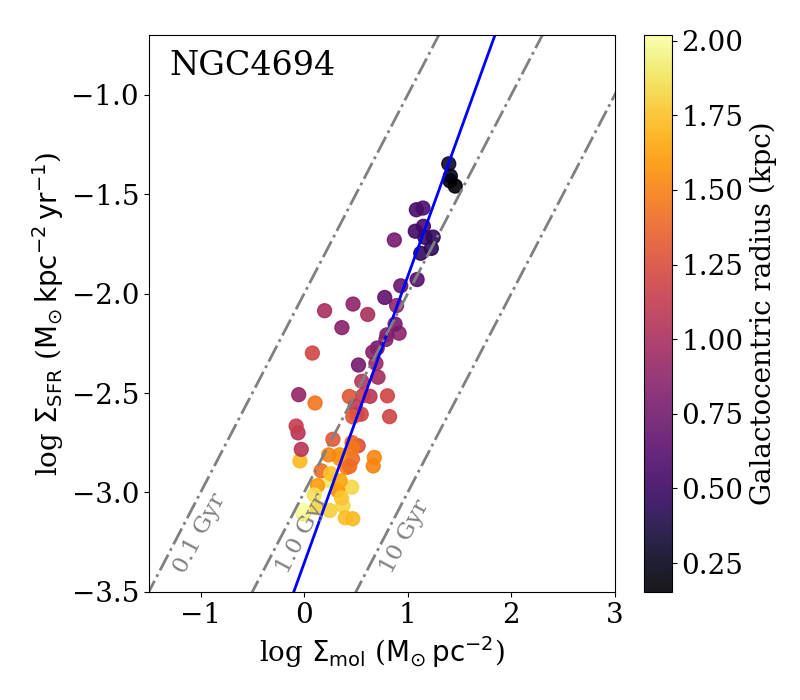}\,
\includegraphics[scale=0.30]{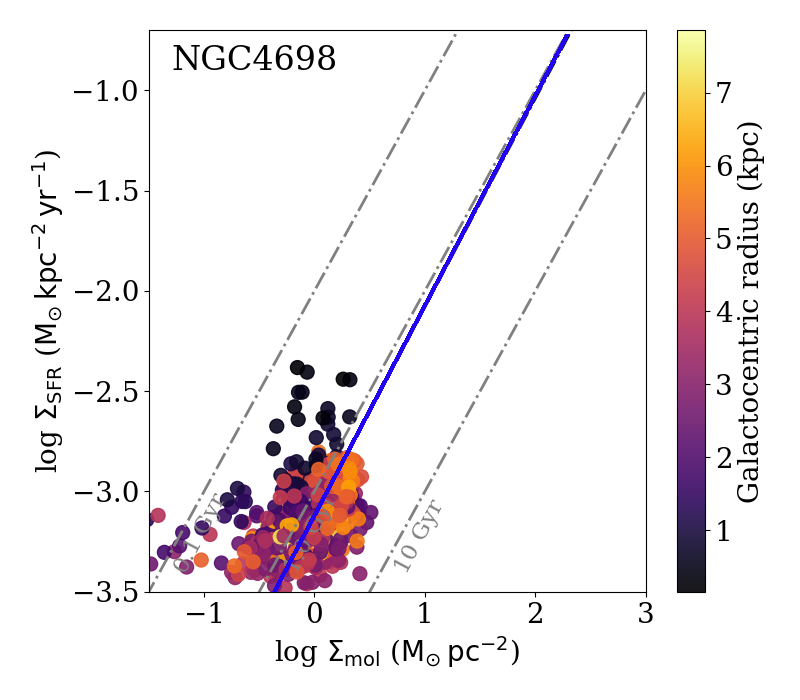}\\ 

\includegraphics[scale=0.30]{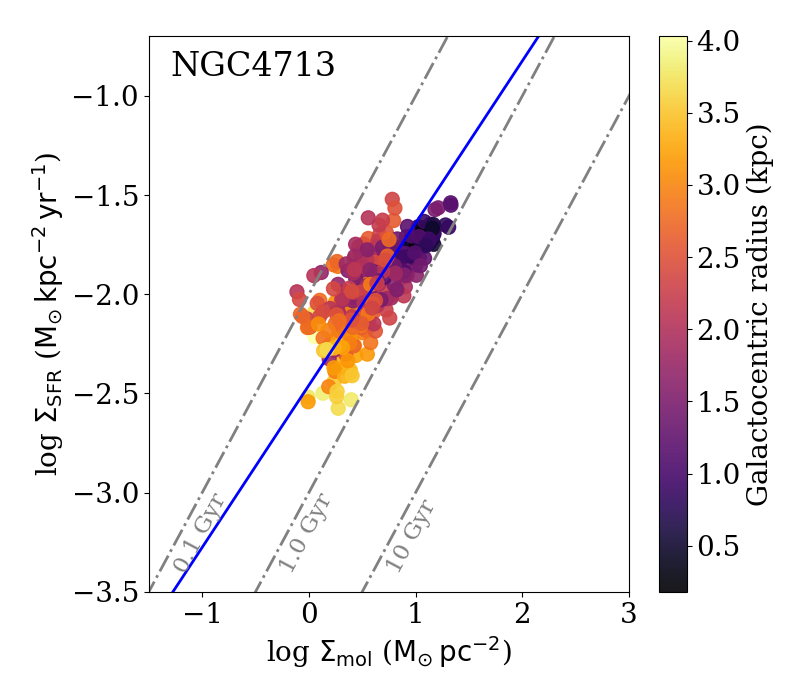}\, \includegraphics[scale=0.30]{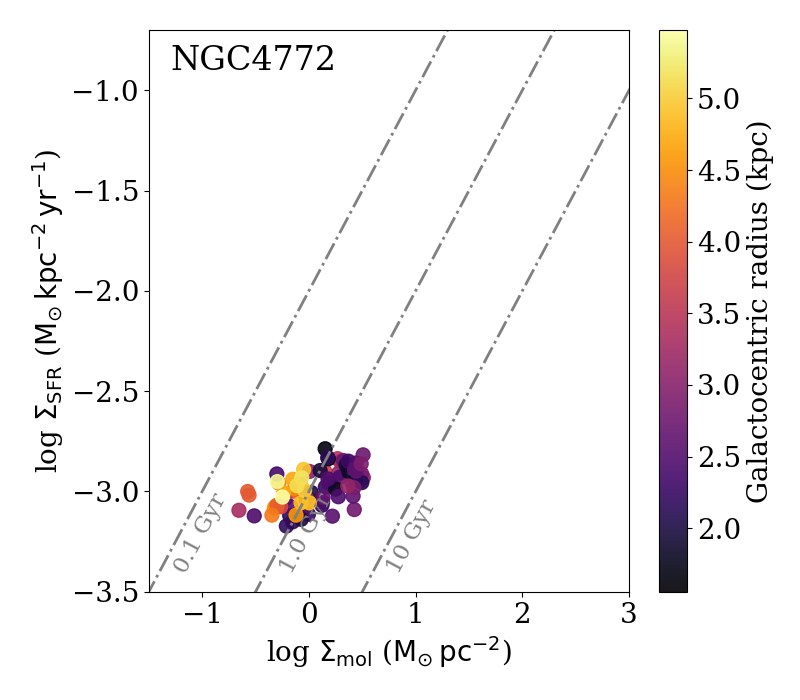}\,
\includegraphics[scale=0.30]{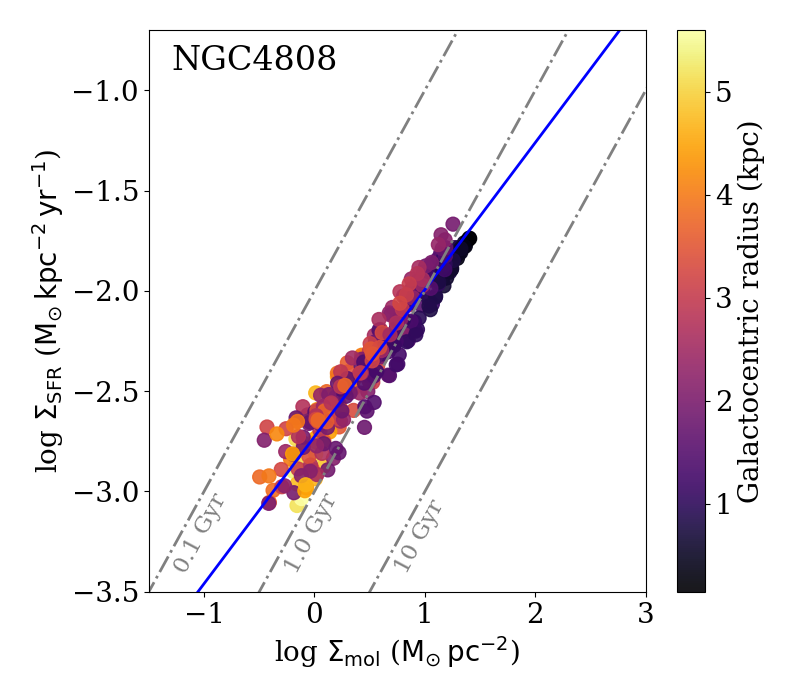}\\
\caption{continued. Note that we do not provide a robust LTS regression for NGC\,4772 since a KS power-law does not appear to be a good fit of the distribution of the data for that galaxy.}
\end{figure*}

\subsection{Star Formation Efficiencies in VERTICO}
\label{sec:sfe}
Our results can additionally be analyzed in terms of the star formation efficiency of the molecular gas, SFE$_\mathrm{mol}$ or, conversely, in terms of the depletion time $\tau_{\mathrm{dep}}$ as defined in section \ref{sec:ensemble}. VERTICO allows us to resolve different regions within galaxy disks, so that we can explore dependencies of the depletion times as a function of the local environment.

While combining our measurements into a single dataset is helpful to understand the ensemble tendencies of our galaxies, it can obscure real trends and differences among them. Figure \ref{fig:indiv_ks}, together with our measurements in Table \ref{tab:indiv_slopes} show systematic variations in the power-law index among galaxies, as well as in their median $\tau_\textrm{dep}$ values from galaxy-to-galaxy. In addition, we also encounter large internal variations in the depletion times for the majority of the sources, as reflected by the individual mean scatter of $\pm0.21$dex. This is also evident by the large scatter we see at fixed molecular gas surface densities in Figure \ref{fig:indiv_ks}, which spans almost an order of magnitude. Clear examples of a varying $\tau_\textrm{dep}$ are NGC\,4383, NGC\,4501, NGC\,4579 and NGC\,4654. Two of them, NGC\,4501 and NGC\,4579, have particularly active and AGN-dominated centers.

\begin{figure}[ht]
\centering
\includegraphics[scale=0.45]{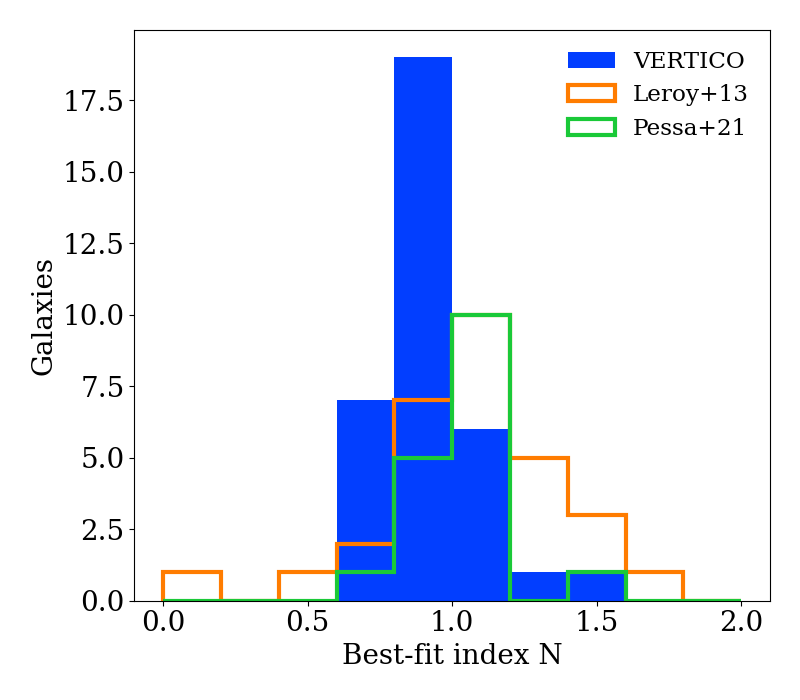}
\caption{Histogram of best-fit power-law KS index, $N$, for individual galaxies in VERTICO (blue color, see Table \ref{tab:indiv_slopes}) using a fixed CO-to-H$_2$ conversion factor. For a comparison, orange bars represent the sub-sample of HERACLES galaxies used in \citet{Leroy2013} and green bars show the best-fit slopes found for the sub-sample of PHANGS galaxies presented in \citet{Pessa2021}. The distribution of best-fit $N$ for individual galaxies in VERTICO spans the range $N\sim0.7-1.5$ and peaks near $N=0.9$, while this value is slightly larger for the HERACLES and PHANGS distributions.}
\label{fig:slopes}
\end{figure}

\begin{figure*}
\centering
\includegraphics[scale=0.372]{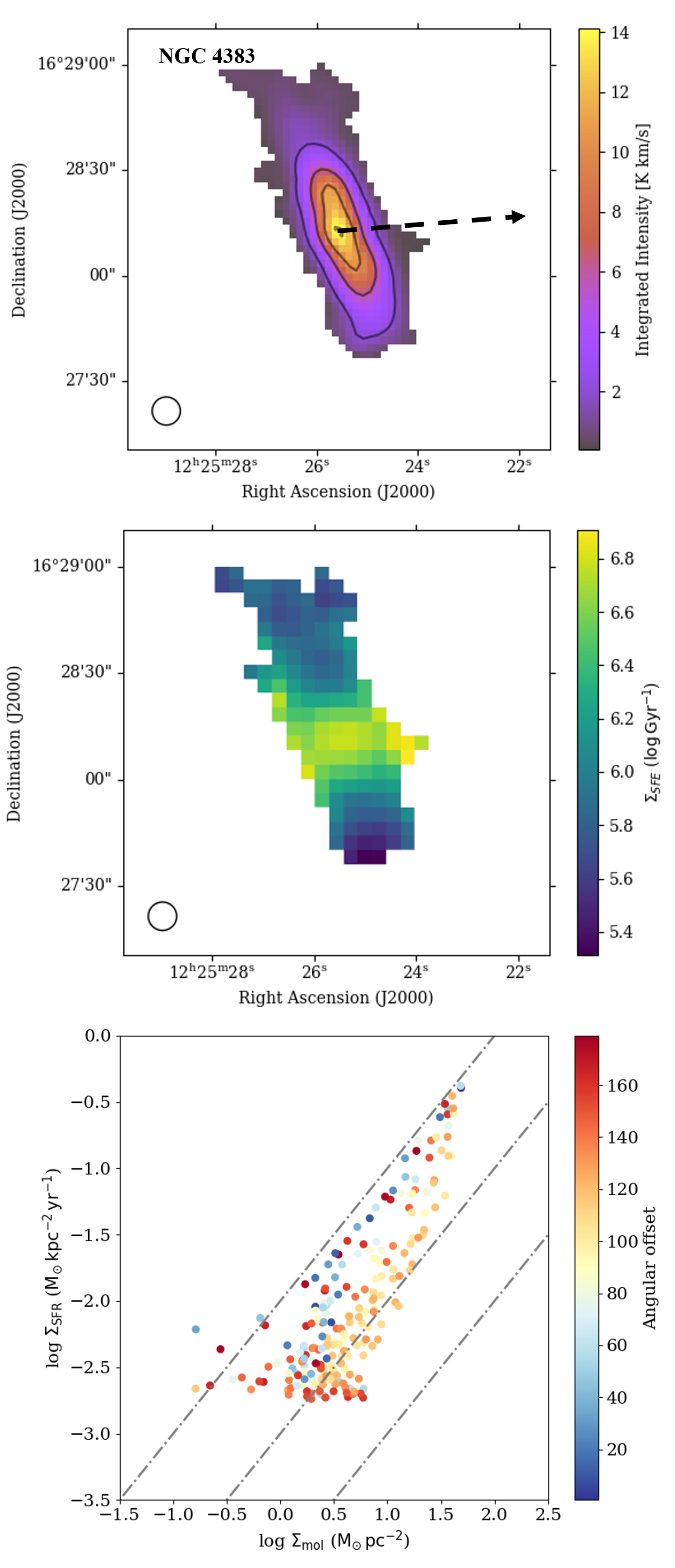}\,\includegraphics[scale=0.86]{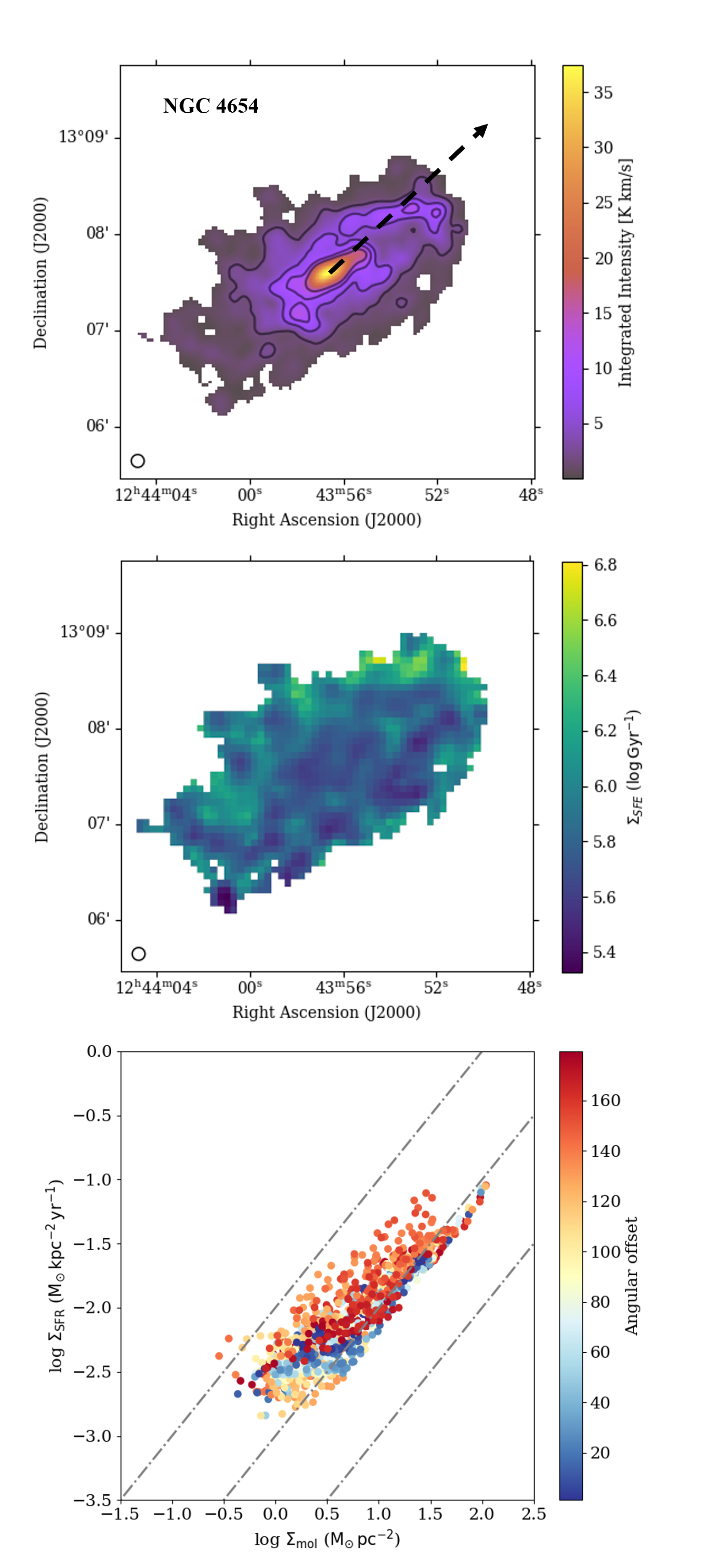}
\caption{Systematic azimuthal variation of SFE. Top: Integrated intensity map of CO\,(2-1) emission for NGC\,4383 (left panel) and NGC\,4654 (right panel). The black dashed arrow represents the direction where the largest azimuthal variations of SFE are found. Angular offsets are measured from this direction to each different line-of-sight. Middle: Map of resolved star formation efficiency of molecular gas, calculated as shown in Eq. \ref{eqn:sfe}. Bottom: Resolved Kennicutt-Schmidt relations, color coded by the angular separation from the preferential direction shown on the integrated intensity map.}
\label{fig:angle_az}
\end{figure*}

In addition to systematically quenching the star formation rates in highly perturbed systems, as seen in Figure \ref{fig:global-ks-hidef}, the cluster environment in which VERTICO galaxies live can also influence the resolved physical properties of the atomic and molecular gas within its disk through gravitational and/or hydrodynamic interactions. This could therefore favor local regions of gas compression through external interactions which may, in turn, create gradients or enhanced regions of star formation. We explore whether such regions exist in the disks of our VERTICO sub-sample by studying azimuthal variations of the measured SFE in the molecular gas. We start by taking the center of each galaxy to be its optical center. We sample each galaxy with 18 radial vectors in the RA-DEC plane. For each vector we calculate the angular separation to each pixel in the computed SFE map (Figure \ref{fig:angle_az}, middle). We select the angle which by eye most cleanly creates a divergent distribution in SFE by azimuthal angle, as shown in our resolved KS plot, color coded by the different angular offsets (Figure \ref{fig:angle_az}, bottom).

Figure \ref{fig:angle_az} shows example plots for two galaxies, NGC\,4383 and NGC\,4654, where these azimuthal variations are found to be the strongest in VERTICO. The top panels show integrated intensity maps of the $^{12}$CO\,(2-1) emission in both galaxies. In each intensity map, the black dashed arrow shows the reference direction from which angular offsets are measured for each different line-of-sight in the galaxy disk. Angular offsets are always measured starting from the arrow to each different point in the galaxy disk, and therefore range from 0$^o$ to 180$^o$. The middle panels show a resolved map of the SFE of the molecular gas in each galaxy. The bottom panels in Figure \ref{fig:angle_az} present the individually resolved KS relation, $\Sigma_\mathrm{SFR}$ as a function of $\Sigma_\mathrm{mol}$, where each data point is color coded by the angular separation computed from the reference direction marked in the left panel.

As seen in the bottom panels of Figure \ref{fig:angle_az}, the colors help us visualize that regions located at different angular distances from a preferred direction follow slightly different KS relations. For NGC\,4654, the split in colors is more evident for the direction as drawn in the bottom right panel of Figure \ref{fig:angle_az}. This direction coincides with the direction of interaction with the companion galaxy NGC\,4639 \citep[see][Fig. 1]{Vollmer2003,Lizee2021}, which caused the molecular gas to be asymmetrically distributed along the major axis of the galaxy. Figure \ref{fig:angle_az} shows that disk regions within $\sim 70^o$ from the direction of interaction tend to follow a KS relation characterized by slightly longer depletion times ($\sim0.9$ Gyr) than those regions in the opposite side ($\sim0.5$ Gyr). These results appear to be in agreement with previous results reported by \citet{Lizee2021}. Using new CO\,(2-1) data from the IRAM-30m telescope, the authors find an enhanced SFE, and therefore shorter depletion times, in the NW region of the galaxy, coincident with an arm of high molecular gas surface density and stellar surface densities. Through their observations and consequent modeling of the data, they hypothesize that this molecular gas surface density must have been enhanced after a period of time where the gas was being compressed through the interaction with NGC\,4639. This gas compression then led to increasing the SFR and the turbulent velocity of the region, giving rise to elevated efficiencies of the molecular gas.

In the case of NGC\,4383 we see both radial and azimuthal variations in the depletion time. The left panels of Figure \ref{fig:angle_az} show that regions with angular separations within $\sim 80-120^o$ (therefore perpendicular to the preferred direction) tend to follow a super linear KS relation characterized by longer depletion times. The region along the semi-minor axis of the galaxy shows much higher star formation efficiencies, with median depletion times of $\sim 0.4$\,Gyr. As previous studies pointed out \citep[see e.g.,][]{Koopmann2004,Chung2009}, NGC\,4383 is an amorphous starburst galaxy characterized by powerful outflow and bright H$\alpha$ and UV emission in its central region, with global SFR that are typically 2-3 times higher than any other isolated field galaxies of similar luminosity. The origin of this starburst-related outflow is still unclear. However, these elevated central SFRs have been argued to be a consequence of cluster interactions, likely to be the result of tidal interactions and gas accretion \citep{Chung2009}.

\section{Discussion}
\label{sec:discussion}
\subsection{Kennicutt-Schmidt in Virgo}
The relation between molecular gas surface densities and SFR surface densities is one of the best studied galaxy scaling relations. It helps us quantify the relative changes in star formation rate, as a consequence of variations in the resolved molecular gas mass. In particular, it is critical to constrain the slope of this relation since it is often used in the typical star formation prescriptions adopted in semi-analytic models or in resolved 2D models of galaxy evolution. A one-to-one correlation between these quantities is indicative of the molecular gas being the necessary fuel for star formation to occur, where gas is converted into stars at a constant efficiency of molecular gas. Steeper slopes indicate enhanced star formation for high molecular gas surface densities. A sub-linear slope, on the contrary, indicates that the observed depletion times for molecular gas increase at higher molecular gas surface densities.

The results we present in this work are in-line with previous studies (see Table \ref{tab:previous_results} for a brief summary) where a KS relation with a slope of $N\sim1$ has been determined \citep[e.g.,][]{Bigiel2008,Schruba2011,Leroy2013,Usero2015,Ellison2021}. For the ensemble VERTICO sample of galaxies we derive a slightly sub-linear slope of $0.97\pm0.07$, compatible with a linear KS relation within uncertainties. This implies that, globally, the VERTICO galaxies convert molecular gas into stars at a roughly constant efficiency across their disks, and agrees with previous findings in that the KS relation is more linear than its atomic version \citep{Bigiel2008,Leroy2013}. We also find a median $\tau_{\mathrm{dep}}=1.86\,\mathrm{Gyr}$ for our full sample, which is well within the range of typical values found in nearby, star-forming galaxies \citep[$\sim 1-3\,\mathrm{Gyr}$, see][]{Bigiel2008,Leroy2008,Leroy2013,Rahman2012,Querejeta2021}. It is important to note that we observe significant scatter, both within and among galaxies. 

While the VERTICO galaxies selected for this study live in a cluster of galaxies, their KS slopes and median depletion times largely overlap with those observed by \citet{Bigiel2008} and \citet{Leroy2013} in isolated galaxies, as seen by the HERACLES survey data. While such a direct comparison might suggest that the denser cluster environment does not alter the molecular gas reservoirs in a way that alters the global star formation properties of galaxy disks, molecular gas can be affected by hydrodynamic pressure exerted by the hot intracluster medium \citep{Tonnesen2009,Lee2020}. A clear example is the direct observational evidence of molecular gas found to be stripping via ram pressure in galaxy clusters. Some of the first observations to confirm that are in the Norma cluster \citep{Jachym2014}, in Virgo \citep{Verdugo2015} and the Coma cluster \citep{Jachym2017}. The work by \citet{Zabel2019} has recently presented clearer evidence of molecular gas stripping in the Fornax cluster, where CO is detected in extended tails from several late-type galaxies. It is important to note that the Virgo cluster, however, is much sparser, spiral-rich, and X-ray faint than some clusters. In comparison, the Coma cluster only contains $\sim 15\%$ of spiral galaxies.

Our VERTICO results also agree with previous observations by \citet{Zabel2020} in Fornax cluster galaxies, where the authors find that the overall KS relation is similar to those found by \citet{Kennicutt1998} and \citet{Bigiel2008}, but show slightly shorter depletion times due to the presence of dwarf galaxies with disturbed molecular gas. Numerical simulations \citep[e.g.,][]{diMatteo2007} already pointed out that galaxy interactions do not necessarily translate into a more efficient conversion of molecular gas into new stars. Overall, they find that all stages of interaction can follow the same general KS sequence. On the other hand \citet{Roediger2007,Roediger2008} also simulated ram pressure in cluster galaxies, which could lead to enhanced episodes of star formation on the leading side of some galaxies experiencing mild ram pressure, as observed in NGC\,4654.

The KS slopes for the vast majority of our sample, are lower than unity. Sublinear slopes indicate that the molecular gas depletion time increases with increasing gas surface density. Such sublinear slopes have been actively debated in previous works \citep[e.g.,][]{Shetty2013,Shetty2014} and could simply indicate that, on average, not all the CO-traced molecular gas is associated with star formation. The CO measurements could instead include a more diffuse phase \citep{Rahman2011,Pety2013}, while star formation is only taking place in the densest regions \citep[e.g.,][]{Gao2004}, such that CO emission does not correlate with star formation activity on a one-to-one relation. 

The VERTICO galaxies analyzed here also exhibit significant scatter in their resolved relations between $\Sigma_\mathrm{SFR}$ and $\Sigma_\mathrm{mol}$. This suggests that there is no single characteristic $\tau_{\mathrm{dep}}$ for all the galaxies, but rather a systematic variation of individual depletion times on sub-galactic scales. This result is also in agreement with findings in local galaxies \citep{Bigiel2008,Leroy2013,Shetty2014,Utomo2017,Ellison2021} and cluster members \citep{Zabel2020}. Therefore $\Sigma_\mathrm{SFR}$ must depend on other internal environmental properties besides molecular gas surface densities such as turbulence, metallicity, or gas fractions, among others. In some of our galaxies, we are able to attribute individual variations in the depletion times to a likely distinct behavior in the galaxy centers of NGC\,4501 and NGC\,4579, as well as tidal interactions and ram-pressure stripping in galaxies NGC\,4383 and NGC\,4654, respectively. 

\subsection{Fitting methodology and sample}
Several caveats and assumptions may have an impact in our derived slopes, and we warn the readers that comparisons among different observations and studies should be done with caution. One of the factors introducing large discrepancies in the KS slopes revolves around the fitting techniques employed. Various fitting methodologies can effectively yield very different KS parameters \citep{Leroy2013,Reyes2019}. In particular, \citet{Ellison2021} find significant differences between the ODR and OLS fitting procedures in the ALMaQUEST (ALMA MaNGA Quenching \& Star-Formation) survey, resulting in super-linear and sub-linear KS relations respectively, as well as in different mean depletion times. These differences are also evident from Table \ref{tab:previous_results}, where the fitting approach varies significantly. It is worth noting that interpreting the results from simple bisector and OLS regressions is statistically hard, since a linear slope could even result from a scenario where there is no correlation between a predictor and its response \citep[see e.g.,][]{Isobe1990,Shetty2014}.

The slope of the KS relation can also be affected by the choice of $\alpha_\mathrm{CO}$. While we have adopted a constant value in this work, it is well known that the CO-to-H$_2$ conversion factor varies within and among galaxies. Not only it has been found to decrease in regions with very high values of SFR \citep{Downes1998,Bryant1999}, but also galaxies tend to have lower CO-to-H$_2$ conversion factors in their central regions \citep{Sandstrom2013} when compared to their outskirts. In addition, it is inversely correlated with metallicity \citep[e.g.,][]{Wilson1995,Genzel2012,Bolatto2013,Accurso2017,Sun2020}. As an example, \citet{Querejeta2021} show that there is a significant variation (about $\sim 30\%$) in the slopes and intercepts derived for the KS relation in the PHANGS sample, depending on the adopted $\alpha_\mathrm{CO}$ conversion factor, with galaxy centers being the most sensitive environment. In particular, they quote KS slopes ranging between $0.90\pm0.05$ and $1.43\pm0.27$ for the overall disks. Additionally, \citet{Pessa2021} explores the impact of assuming a constant conversion factor, rather than a metallicity dependent one, on their measured KS relationship in PHANGS, finding a decrease of $\sim 5\%$ in the KS slope. In addition, they point out that the timescales probed by the SFR prescription strongly impacts the resulting KS parameters. In particular, they find that a longer timescale SFR tracer pushes previously low SFRs to higher values, which results into a significant flattening of the KS slope (from $1.06\pm0.01$ to $0.60\pm0.01$).

An additional concern to the fitting methodology employed is that it uses the same weighting scheme for every sampled region in our galaxies. Therefore it could be expected that galaxies with more extended H$_2$ discs (i.e., those less affected by the environment) and a larger number of resolved regions could be dominating the ensemble fit, driving the slope of the KS relation. As an example, NGC\,4254 and NGC\,4321 are two of the more extended VERTICO galaxies, typically considered as normal galaxies \citep[see e.g., ][]{Pessa2021}, and would be expected to dominate the KS slope and scatter. We address this issue by weighting each line-of-sight measurement by the number of independent measurements that a given galaxy contributes to the overall sample, as also described in \citet{Zabel2020}. Following this procedure, we obtain a resolved KS slope of $N=0.99\pm0.02$, with a median $\mathrm{log}_{10} \Sigma_\mathrm{SFR}/\Sigma_\mathrm{mol}=-9.25\pm0.06$ and a 1-$\sigma$ scatter of 0.40\,dex. While this result provides a slope that is more representative of every galaxy, it also shows that the slope is effectively the same as that derived in Figure \ref{fig:global-ks-inc} within the measured uncertainty range.

\subsection{Environmental influence on the resolved KS relation}

It is clear that environment is a key element shaping galaxy evolution. Clusters of galaxies are particularly interesting environments to study extreme and unusual ISM properties in galaxy disks. These rich environments can perturb both the atomic and molecular gas content, especially through ram pressure stripping of the gas, and thus drastically affecting their evolution. A large number of observational studies have shown that cluster galaxies tend to have lower atomic \citep[e.g.,][]{Haynes1984,Cayatte1990,Catinella2013} and molecular \citep[e.g.,][]{Fumagalli2009,Boselli2014} gas content than field galaxies. In this context, VERTICO aims to systematically uncover the links between resolved molecular gas and star formation in a large sample of late-type cluster galaxies.

RPS is one of the most important perturbations in cluster environments, and generally considered as the dominant process in rich galaxy clusters \citep[e.g.,][]{Vollmer2001,Boselli2006,Boselli2014b}. During this process, the ISM of galaxies, especially in the outer parts of their disks, is removed by the external pressure or due to its friction with the ICM. Both the atomic gas and the molecular gas (less severely) can be stripped during the interaction. As a consequence, the denser molecular gas component, within giant molecular clouds, also decreases since it cannot longer be replenished by the diffuse gas component \citep{Tonnesen2009}. In these cases, the star formation activity in those galaxies is also affected as they slowly consume the available gas reservoir \citep{Boselli2014}. In this context, the HI-deficiency parameter can be used as an indirect tracer of perturbed environments in statistical studies \citep[e.g.,][]{Haynes1984,Solanes2001,Gavazzi2005,Gavazzi2006,Catinella2013,Loni2021}. The works by \citet{Fumagalli2009} and \citet{Boselli2014} have additionally shown that the molecular gas content tends to decrease in perturbed, HI-deficient galaxies. This indicates that hydrodynamical processes like RPS can directly affect the cold, dense molecular gas component \citep{Wilson2009,Mok2016,Chung2017} and ultimately reduce the star formation activity, as previously observed in nearby clusters \citep[e.g.,][]{Kennicutt1983,Moss1993,Gavazzi2006,Haines2007,Vulcani2010}. 

The results presented in Section \ref{sec:results} show the same trend for Virgo cluster galaxies in a systematic way: HI-deficient galaxies tend to have a reduced global star formation activity and lower star formation efficiency of their molecular gas. This overall decrease of star formation situates the HI-deficient galaxies below the main sequence relation shown in the right panel of Figure \ref{fig:global-ks-hidef}, compared to unperturbed galaxies which are scattered around it. The HI-deficiency parameter is a good proxy for environmental influence, therefore our results suggest that cluster environment is largely responsible for lowering the SFE in HI-deficient galaxies. These results are comparable to those found in \citet{Davis2014} for the ATLAS 3D Project galaxies, where early type galaxies (ETGs) are systematically found offset from the KS relation for normal and starburst galaxies. ETGs, which are typically far more H-I deficient than late-type galaxies \citep[e.g.,][]{Huchtmeier1989}, show lower star formation efficiencies of the gas by a factor of $\sim2.5$. While both results are not directly comparable, since \citet{Davis2014} employ total gas surface densities rather than just molecular gas surface densities, they point towards a key role of the gas dynamics to regulate star formation in galaxies.

In this context, our results are in agreement with recent efforts to characterize the atomic and molecular gas in Virgo cluster galaxies. A first piece of evidence is presented in \citet{Zabel2022}, showing the effects of cluster environments in the VERTICO sample of galaxies. They find that VERTICO HI-deficient galaxies appear to have more compact molecular gas profiles, and indicate that these are a consequence of the same processes that act on the atomic gas removal via ram pressure stripping. In addition, they find a weak correlation between global HI and H$_2$ deficiencies, with galaxies that are HI-deficient because of their lower HI surface densities also being the most H$_2$ deficient (see Fig. 2 from \citet{Zabel2022} for a detailed summary). This weak correlation between HI and H$_2$ deficiencies indicates that environmental effects removing HI from galaxies also affect the molecular gas, albeit to a lesser extent. Similarly, \citet{Villanueva2022} note that VERTICO galaxies show a less extended molecular gas emission compared to their stellar distribution. In comparison to isolated galaxies, they find that these cluster galaxies have elevated molecular-to-atomic gas ratio and lower SFE of their molecular gas. Our results contribute to these major VERTICO efforts, providing additional evidence that the Virgo cluster environment affects the distribution and physical properties of the molecular gas, and its efficiency to form stars in such disturbed environments. We have shown the importance of environment in reducing the star formation efficiency of molecular gas. Future work will explore which mechanisms are most responsible and where, as well as how efficiency interacts with gas fueling to regulate the star formation cycle.

\bigskip

\section{Summary and conclusions}
\label{sec:summary}
In this work we employ new data for 37 spiral galaxies included in the ALMA large program VERTICO: the Virgo Environment Traced in CO. These observations were used to investigate the resolved properties of the molecular gas, and combined with tracers of star formation activity allowed us to explore the resolved scaling relation between $\Sigma_\mathrm{SFR}$ and $\Sigma_\mathrm{mol}$, known as the resolved Kennicutt-Schmidt relation, for the first time in a sample of star-forming galaxies within a cluster. In order to assess the universality of the KS relation, we explore the ensemble and individual scaling relations, and compared our results to a controlled sample of field, isolated galaxies from the HERACLES survey. Our main results are the following:

\begin{enumerate}
    \item We confirm a strong correlation between $\Sigma_\mathrm{SFR}$ and $\Sigma_\mathrm{mol}$ at a linear resolution of 720\,pc for the ensemble of galaxies, with a mean slope $N=0.97\pm0.07$, and a characteristic median depletion time of $\tau_{\mathrm{dep}}=1.86\,\textrm{Gyr}$ for our full sample. Our measurements agree well with those seen in previous work in isolated, star-forming galaxies in HERACLES \citep{Bigiel2008,Leroy2013} and other surveys of resolved molecular gas emission \citep{Schruba2011,Usero2015,Lin2019,Pessa2021,Querejeta2021}. This suggests that, overall, star formation in the Virgo cluster operates in a similar way to that seen in isolated galaxies.
    
    \item In individual galaxies, the derived KS parameters vary from galaxy to galaxy. We find slopes that range between $0.69$ and $1.40$, with a mean typical scatter of $\pm0.21$\,dex. and typical star formation efficiencies of molecular gas that can vary from galaxy to galaxy by a factor of $\sim 4$. This confirms that there is no single KS relation that can accurately describe every galaxy, as seen in previous studies \citep{Bigiel2008,Saintonge2011,Shetty2013,Zabel2020,Ellison2021}. 
    
    \item While the ensemble KS relation strongly correlates $\Sigma_\mathrm{SFR}$ and $\Sigma_\mathrm{mol}$ for the full sample, there is almost an order of magnitude variation in the observed scatter for the entire sample. Compared to the scatter in the individual KS relations, galaxy-to-galaxy variations appear to account for approximately half of the scatter seen in the ensemble KS relation. In addition, the large scatter seen in depletion times inside individual galaxies ($\sim 0.23$\,dex) indicates that there are systematic variations at sub-galactic scales, suggesting that $\Sigma_\mathrm{SFR}$ may also depend on other local environmental properties besides molecular gas surface densities. 
    
    \item We observe clear azimuthal variations of the molecular gas depletion times in two of our galaxies, NGC\,4383 and NGC\,4654. These can be directly linked to the presence of gravitational tidal interactions and ram-pressure stripping of the gas, respectively, and provide additional evidence that the cluster environment directly affects the resolved physical properties of the molecular gas.
    
    \item Using the HI-deficiency parameter as a proxy for cluster environment influence on the VERTICO sample, we find that the HI-deficient galaxies in the Virgo cluster show a steeper resolved KS relation and lower molecular gas efficiencies than HI-normal cluster galaxies. When comparing these subsamples of galaxies in the SFR-$M_*$ plane, we find that HI-normal galaxies tend to lie over the SFMS, while HI-deficient galaxies have average global SFRs that are systematically lower by $\sim$0.60\,dex, for a fixed stellar mass. This implies that the environmental mechanisms affecting the HI galaxy content also have a direct impact in the local molecular gas content and its star formation efficiency, which are reduced in HI-deficient cluster galaxies. This suppression results in the systematic quenching of the SFR, eventually leading to longer depletion times in HI-deficient members.
    
\end{enumerate}


\begin{acknowledgements}
    MJJD thanks Miguel Querejeta, Eric Pellegrini and Ivana Be\v{s}li\'c for helpful discussions during the development of this work. CDW acknowledges support from the Natural Sciences and Engineering Research Council of Canada and the Canada Research Chairs program. IDR acknowledges support from the ERC Starting Grant Cluster Web 804208. AC acknowledges support by the National Research Foundation of Korea (NRF), grant No. 2018R1D1A1B07048314. Parts of this research were supported by the Australian Research Council Centre of Excellence for All Sky Astrophysics in 3 Dimensions (ASTRO 3D), through project number CE170100013. LCP acknowledges support from the Natural Science and Engineering Council of Canada. The work of JS is partially supported by the Natural Sciences and Engineering Research Council of Canada (NSERC) through the CITA National Fellowship. V. V. acknowledges support from the scholarship ANID-FULBRIGHT BIO 2016 - 56160020 and funding from NRAO Student Observing Support (SOS) - SOSPA7-014. V. V. acknowledges partial support from NSF-AST1615960. 
    
    This paper makes use of the following ALMA data: 
    
    ADS/JAO.ALMA\href{https://almascience.nrao.edu/asax/?result_view=observation&projectCode=\%222019.1.00763.L\%22}{\#2019.1.00763.L},\\ ADS/JAO.ALMA\href{https://almascience.nrao.edu/asax/?result_view=observation&projectCode=\%222017.1.00886.L\%22}{\#2017.1.00886.L},\\ ADS/JAO.ALMA\href{https://almascience.nrao.edu/asax/?result_view=observation&projectCode=\%222016.1.00912.S\%22}{\#2016.1.00912.S},\\ ADS/JAO.ALMA\href{https://almascience.nrao.edu/asax/?result_view=observation&projectCode=\%222015.1.00956.S\%22}{\#2015.1.00956.S}. \\
    
ALMA is a partnership of ESO (representing its member states), NSF (USA) and NINS (Japan), together with NRC (Canada), MOST and ASIAA (Taiwan), and KASI (Republic of Korea), in cooperation with the Republic of Chile. The Joint ALMA Observatory is operated by ESO, AUI/NRAO and NAOJ. The National Radio Astronomy Observatory is a facility of the National Science Foundation operated under cooperative agreement by Associated Universities, Inc.
      
This research has made use of the NASA/IPAC Extragalactic Database (NED), which is operated by the Jet Propulsion Laboratory, California Institute of Technology, under contract with the National Aeronautics and Space Administration.

\end{acknowledgements}

\begin{appendix}
\section{Observed relations}
\label{ap:observed}

The full VERTICO survey comprises a total of 51 Virgo Cluster galaxies, which are included in  the  VIVA survey \citep{Chung2009}. However, a significant fraction of the VERTICO sample is composed of galaxies with highly inclined disks \citep[see Table 1 in][for a complete summary of the basic properties of the full VERTICO sample]{Brown2021}. In order to minimize the effects of extinction, our sub-sample includes only galaxies with moderate inclinations ($i\leq80^\mathrm{o}$).

As described in Section \ref{sec:methods}, the conversion between the observed intensities and the physical quantities, such as molecular gas surface densities, is subject to assumptions and therefore uncertainty. In order to provide an overview of all galaxies belonging to the VERTICO survey, we show our resulting resolved KS relations in Figure \ref{fig:obs_ks} using direct observables (e.g., $I_\mathrm{CO}$), without inclination corrections.

\begin{figure*}
\includegraphics[scale=0.30]{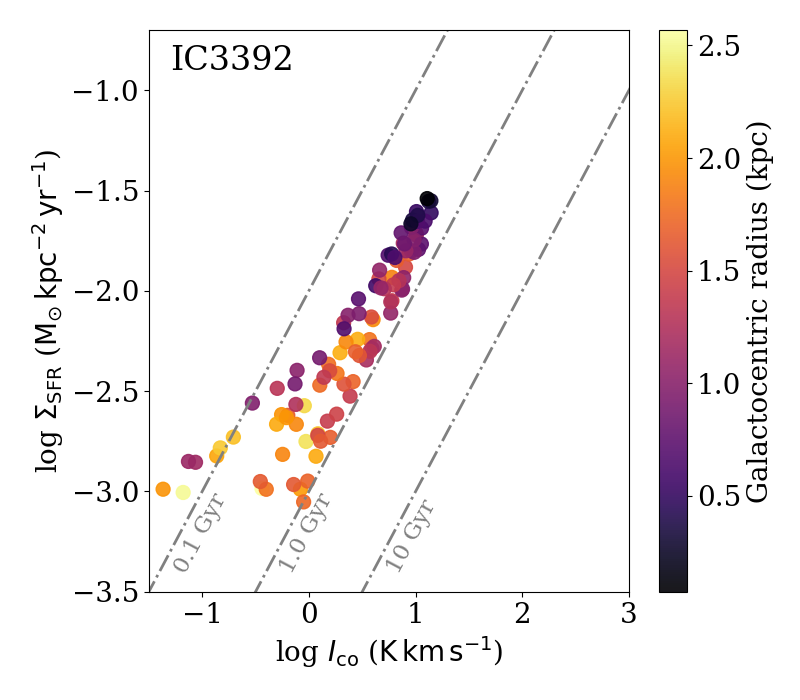}\,
\includegraphics[scale=0.30]{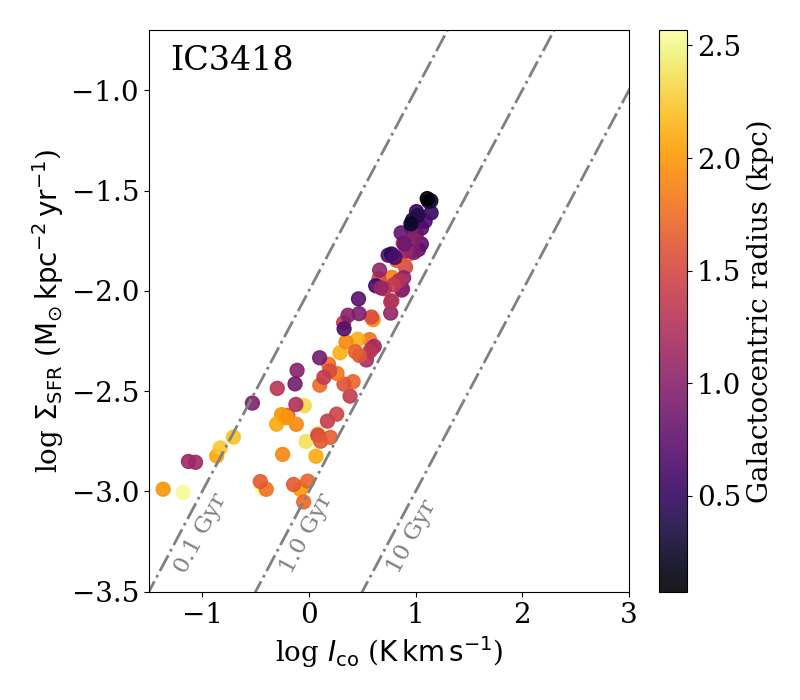}\,
\includegraphics[scale=0.30]{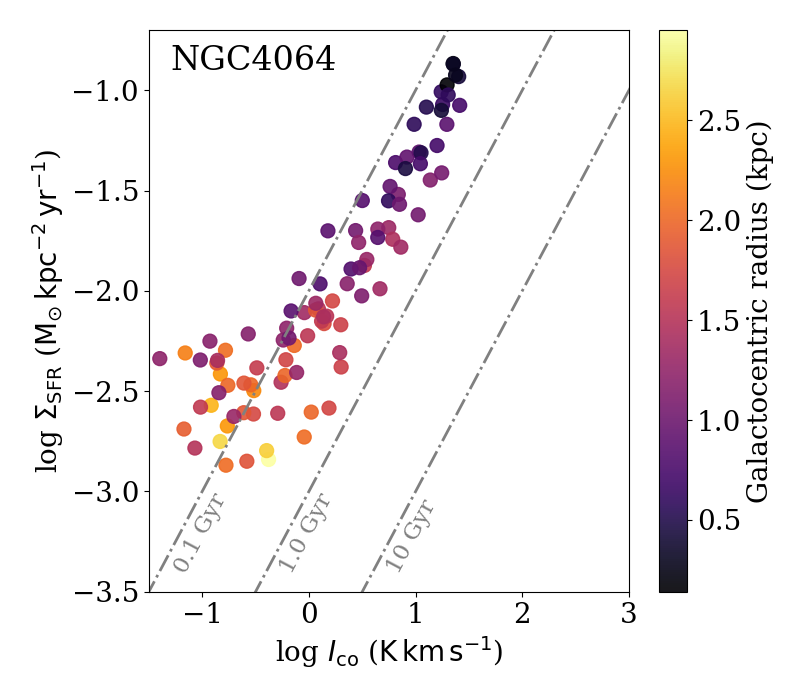}\\

\includegraphics[scale=0.30]{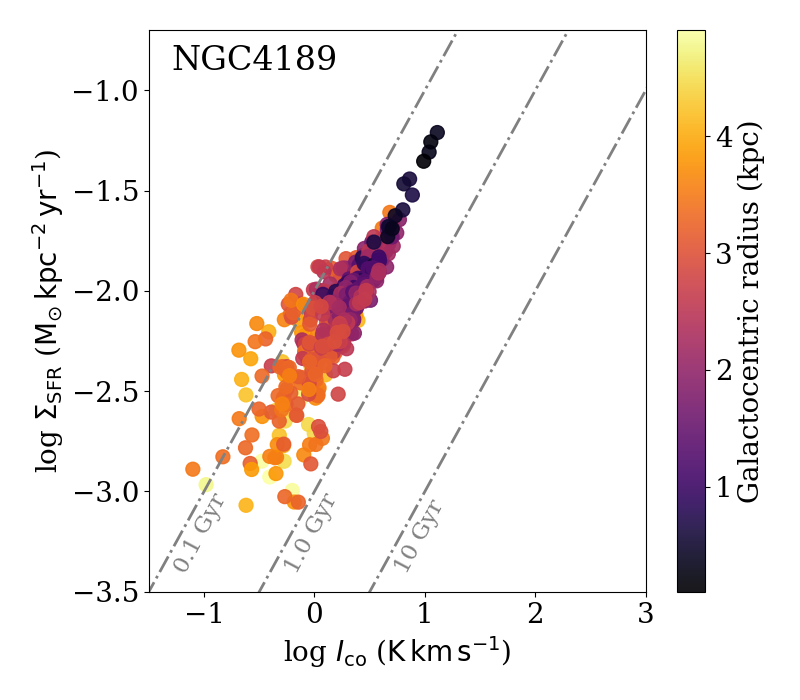}\,
\includegraphics[scale=0.30]{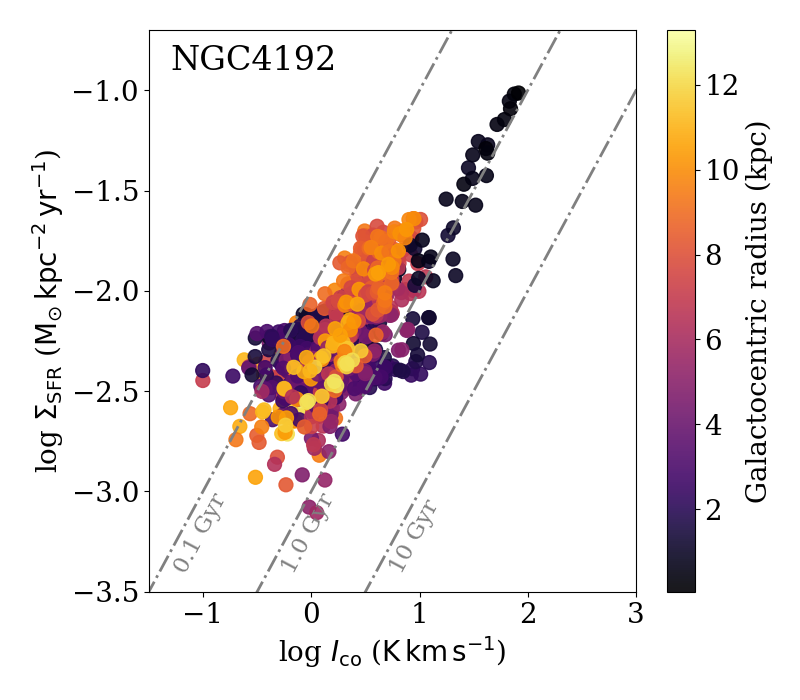}\,
\includegraphics[scale=0.30]{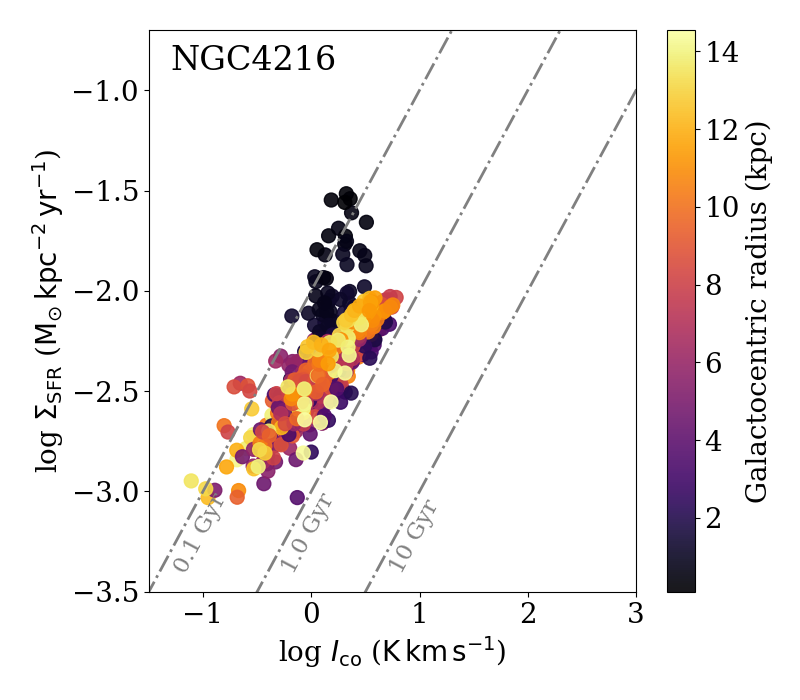}\\

\includegraphics[scale=0.30]{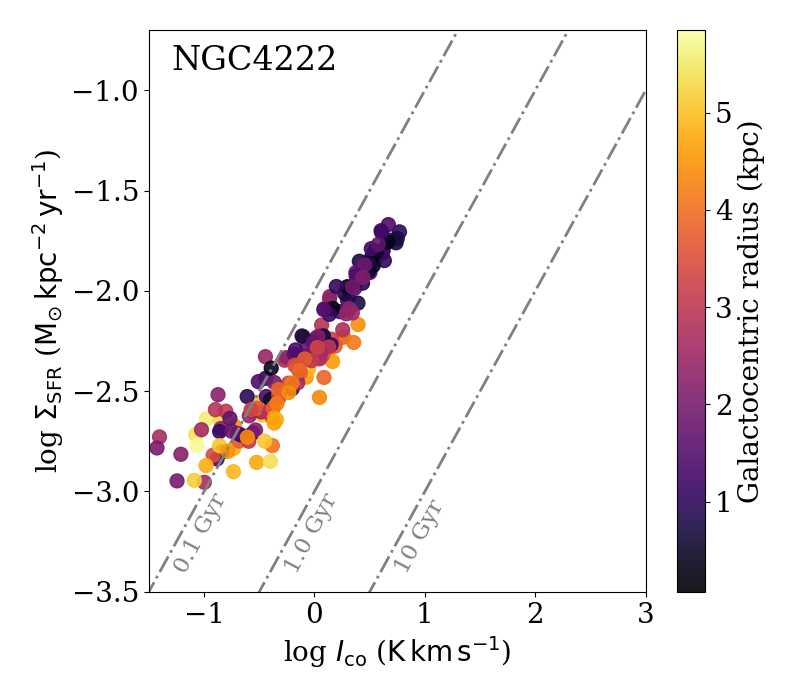}\,
\includegraphics[scale=0.30]{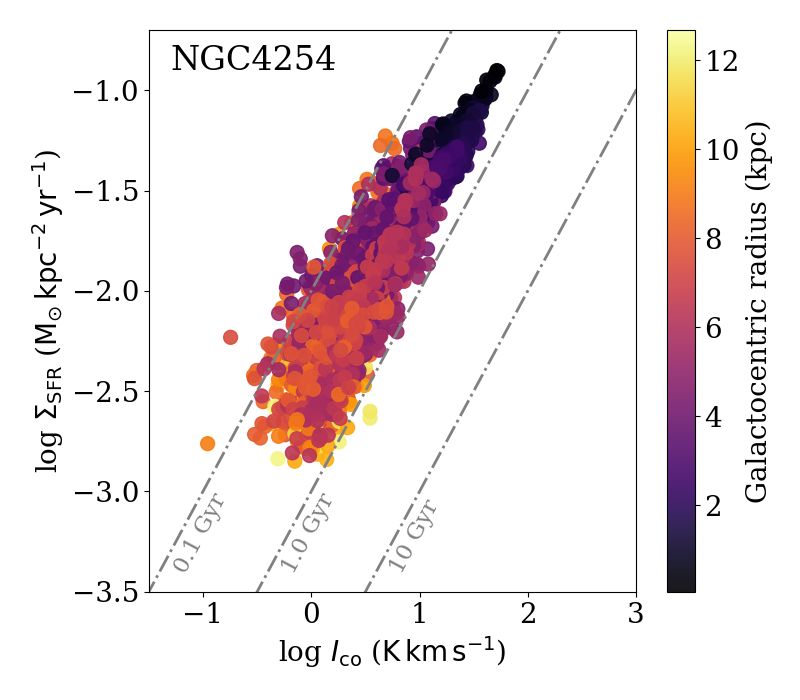}\,
\includegraphics[scale=0.30]{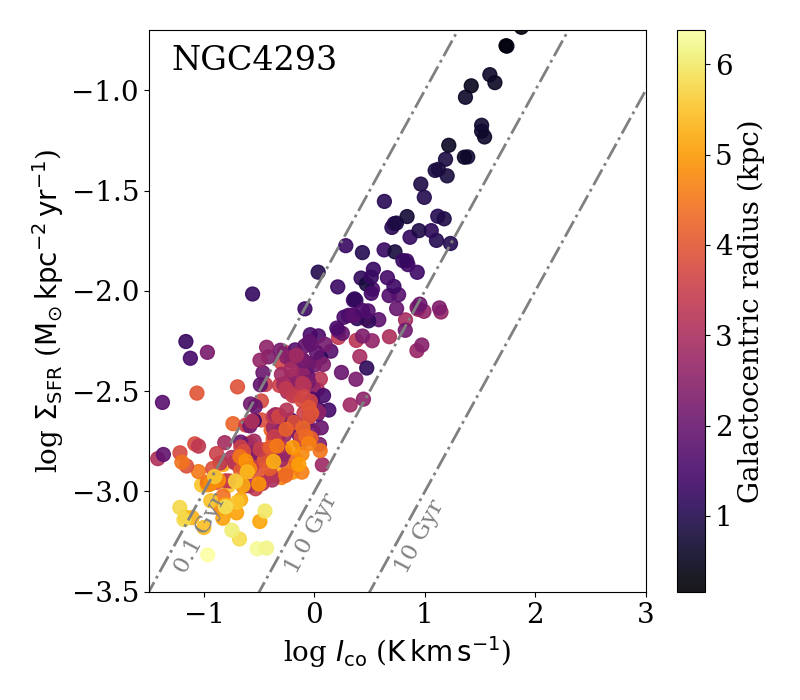}\\

\includegraphics[scale=0.30]{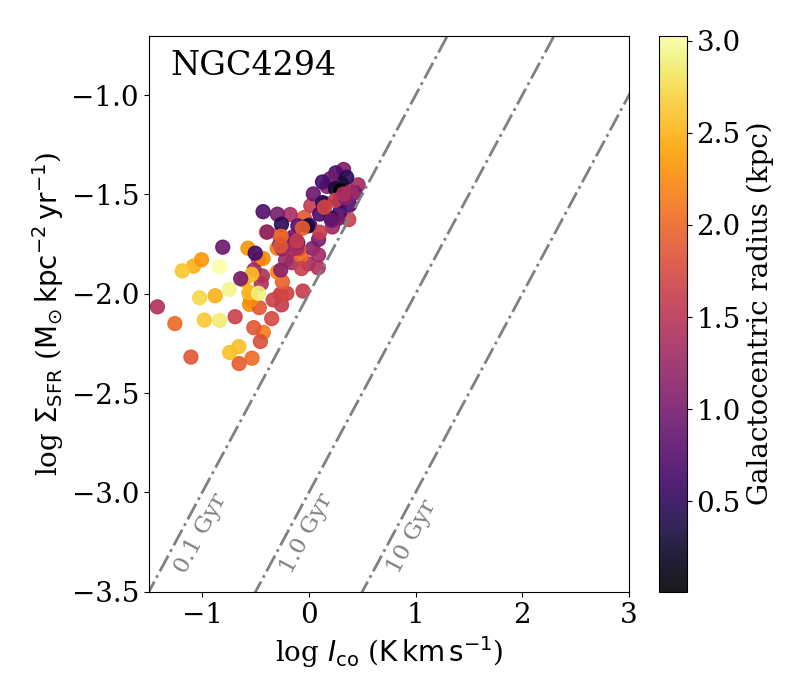}\,
\includegraphics[scale=0.30]{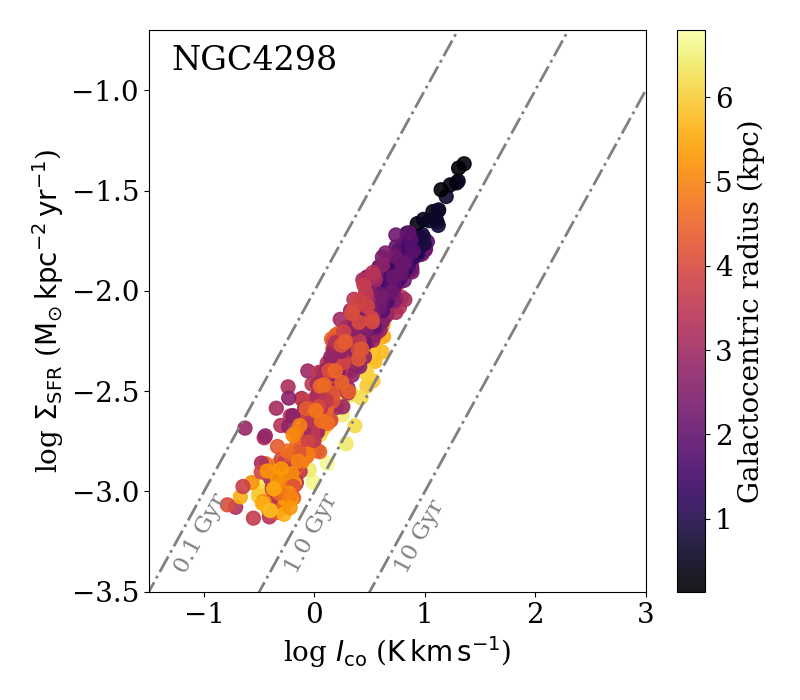}\,
\includegraphics[scale=0.30]{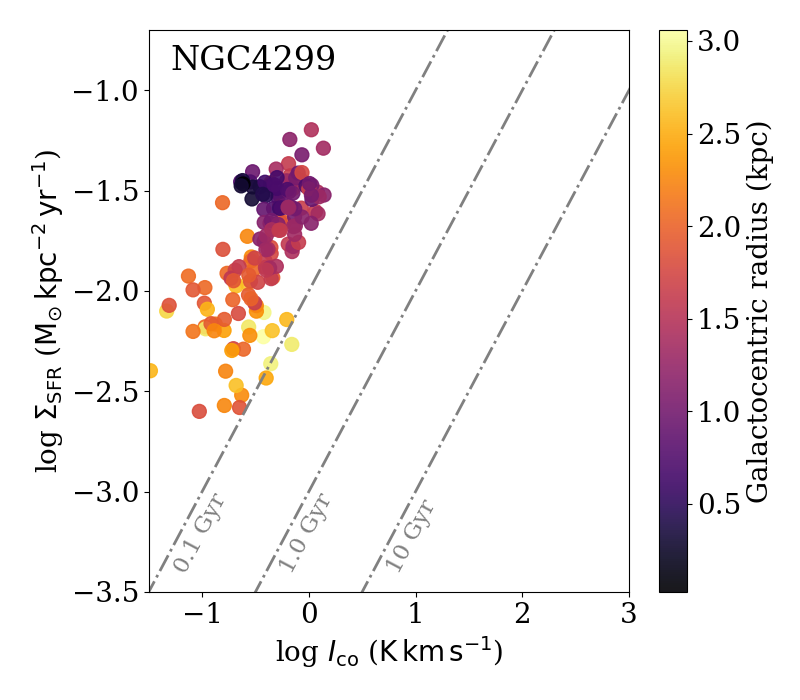}\\

\caption{The resolved Kennicutt-Schmidt relation for all galaxies in the VERTICO sample, using directly observed CO integrated intensities as a proxy for $\Sigma_\textrm{mol}$ as well as the observed $\Sigma_\textrm{SFR}$ (e.g., without a correction for $\textrm{cos}\,i$). All data points are convolved at a common working resolution of 720\,pc. Each data point is color-coded by the distance to the galaxy center. The diagonal dashed, gray lines show constant depletion times of 0.1, 1 and 10\,Gyr, respectively.}
\label{fig:obs_ks}
\end{figure*}

\renewcommand\thefigure{\thesection.\arabic{figure}}
\setcounter{figure}{0}

\begin{figure*}
\includegraphics[scale=0.30]{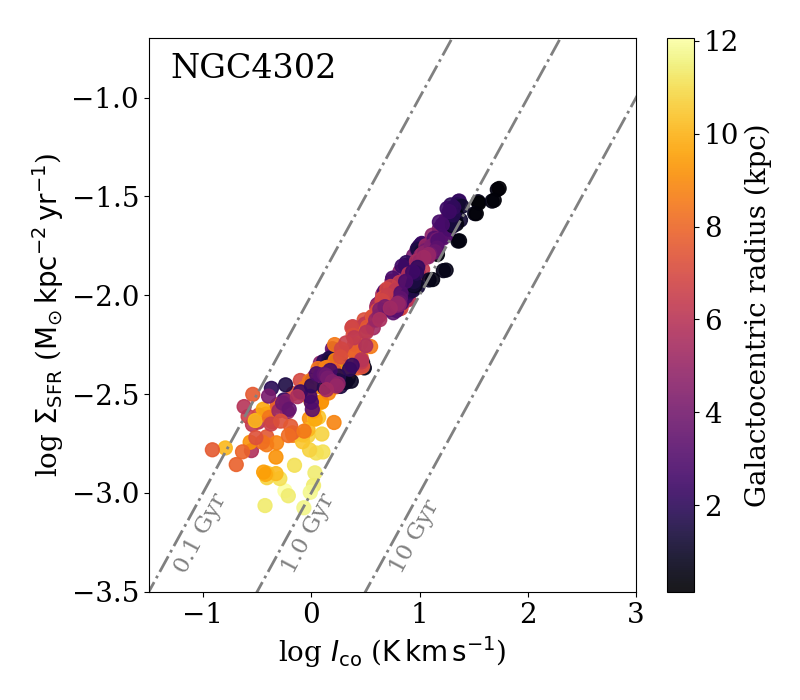}\,
\includegraphics[scale=0.30]{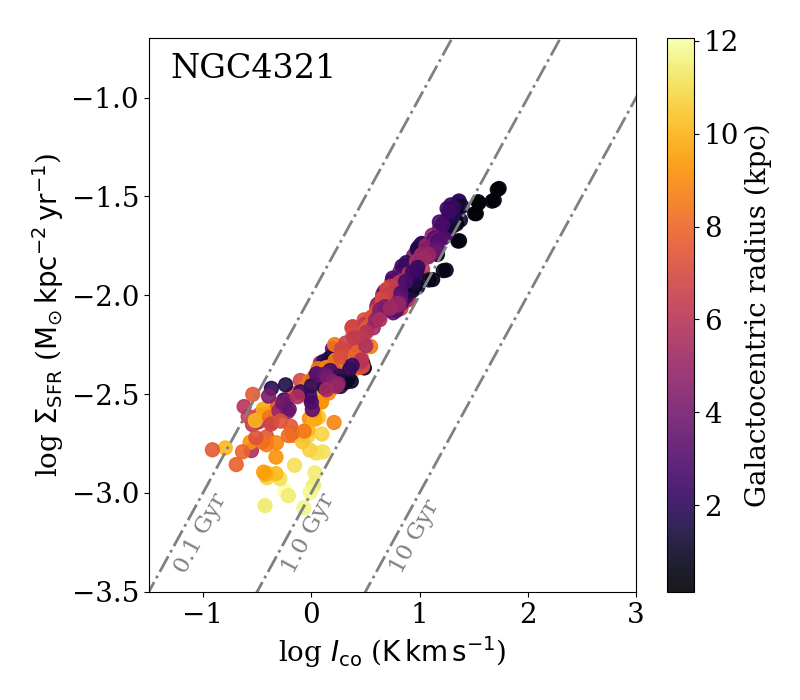}\,
\includegraphics[scale=0.30]{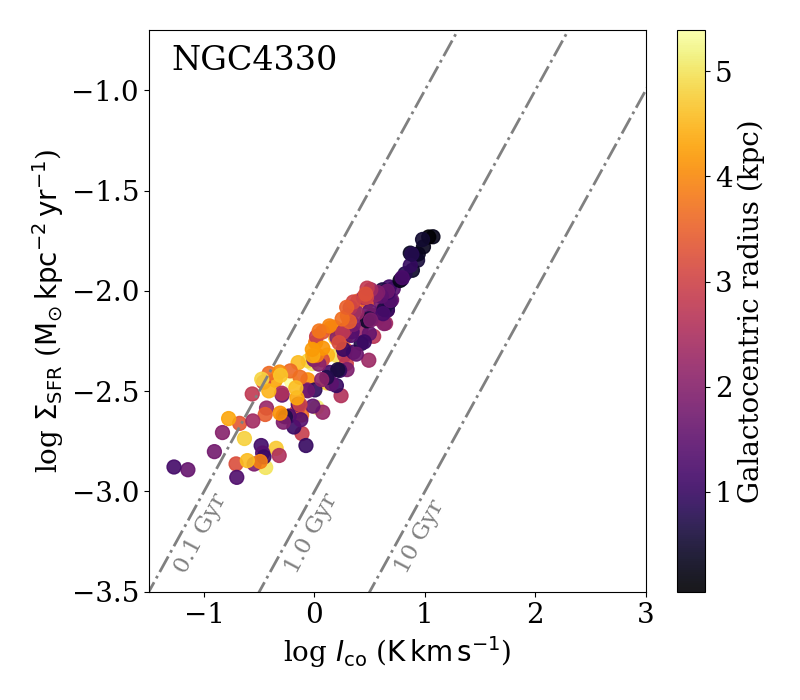}\\

\includegraphics[scale=0.30]{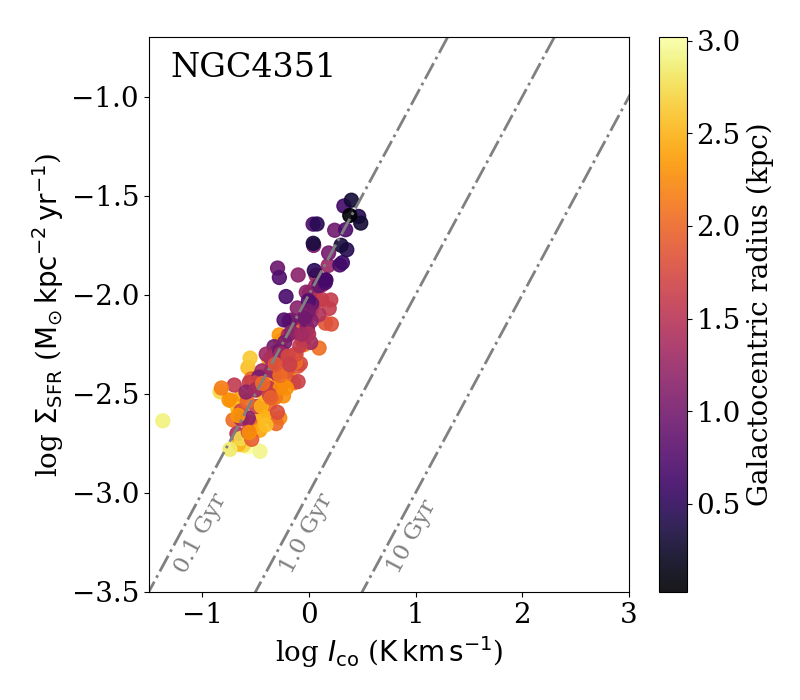}\,
\includegraphics[scale=0.30]{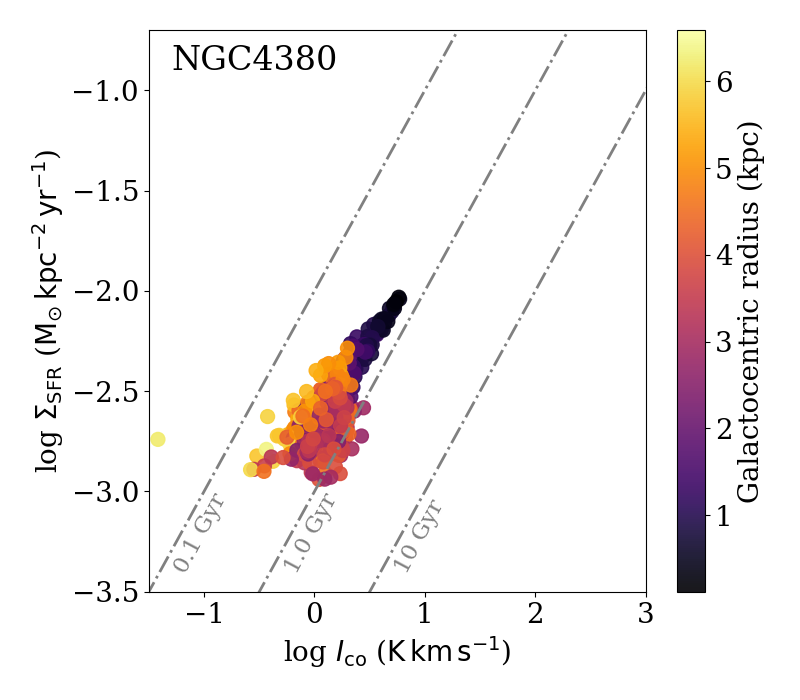}\,
\includegraphics[scale=0.30]{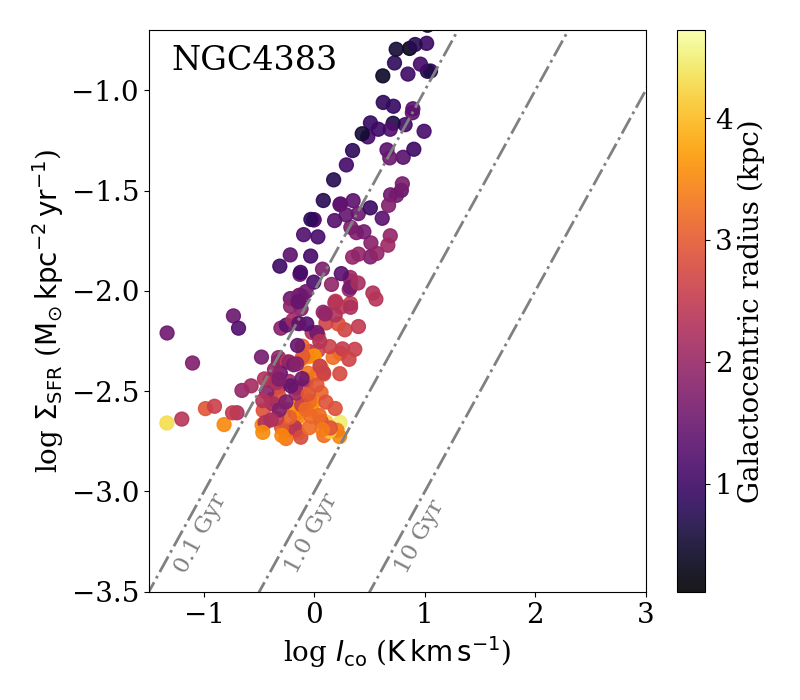}\\

\includegraphics[scale=0.30]{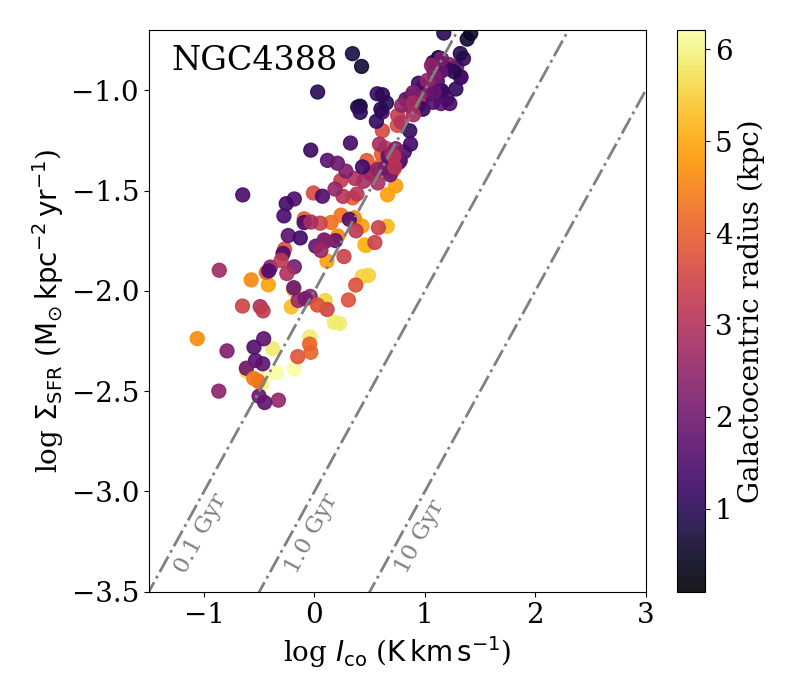}\,
\includegraphics[scale=0.30]{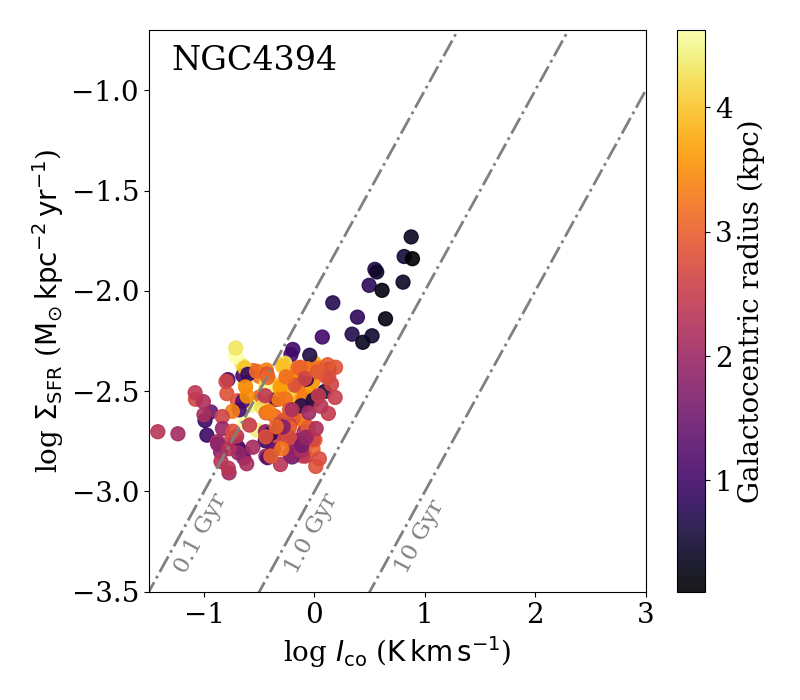}\,
\includegraphics[scale=0.30]{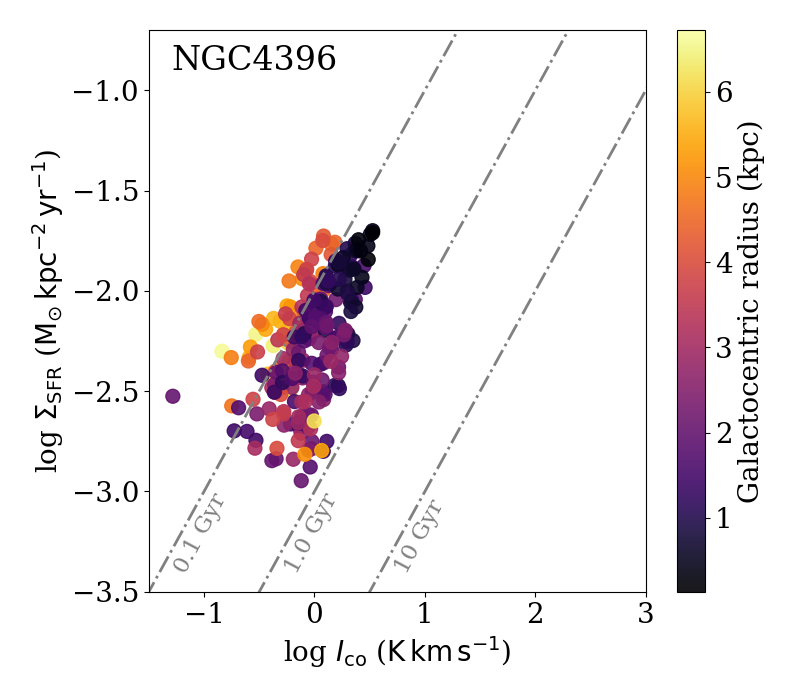}\\

\includegraphics[scale=0.30]{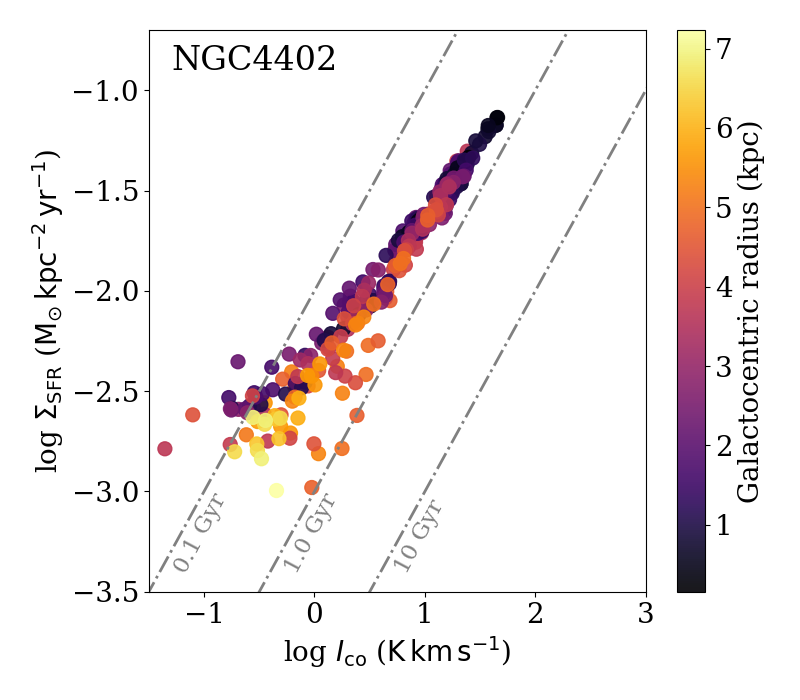}\,
\includegraphics[scale=0.30]{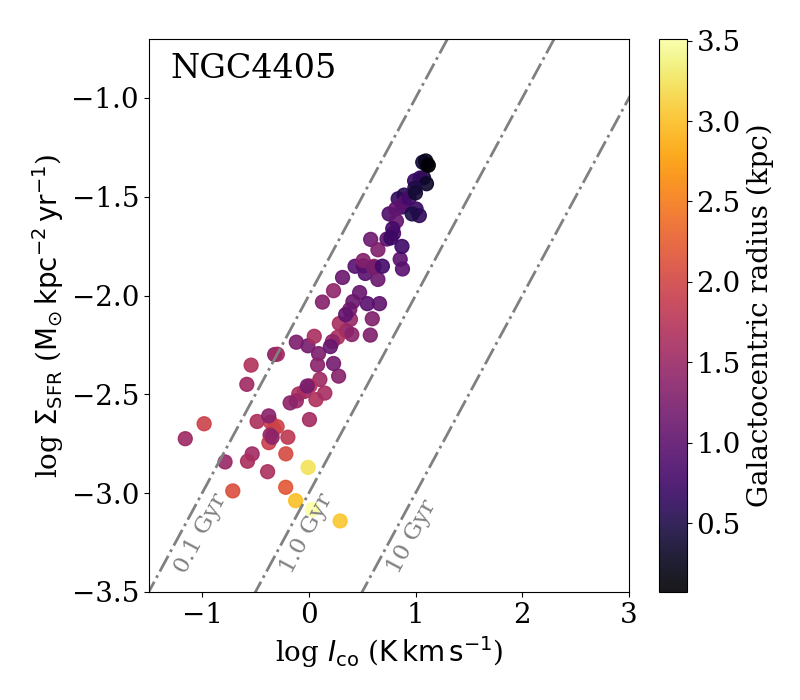}\,
\includegraphics[scale=0.30]{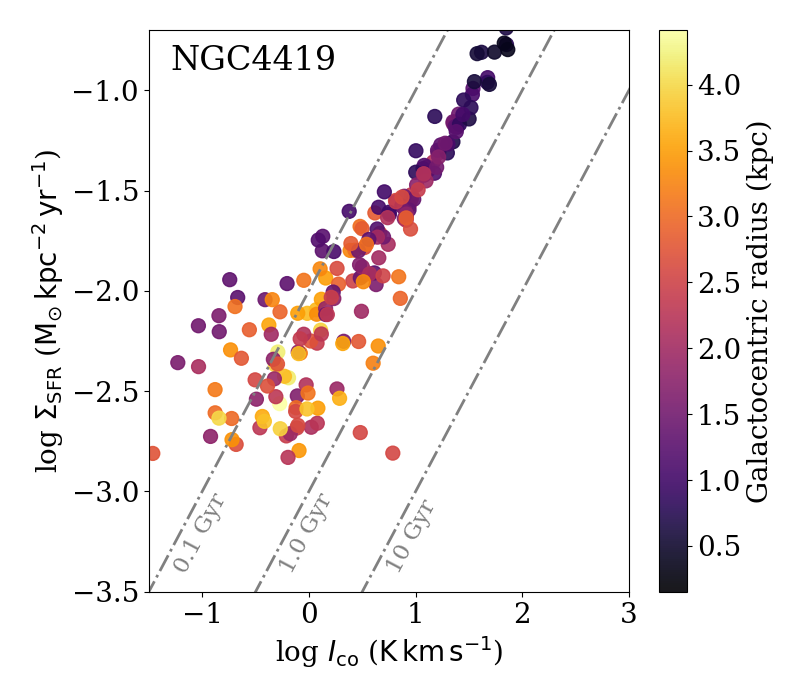}\\

\caption{continued.}
\end{figure*}

\renewcommand\thefigure{\thesection.\arabic{figure}}
\setcounter{figure}{0}

\begin{figure*}
\includegraphics[scale=0.30]{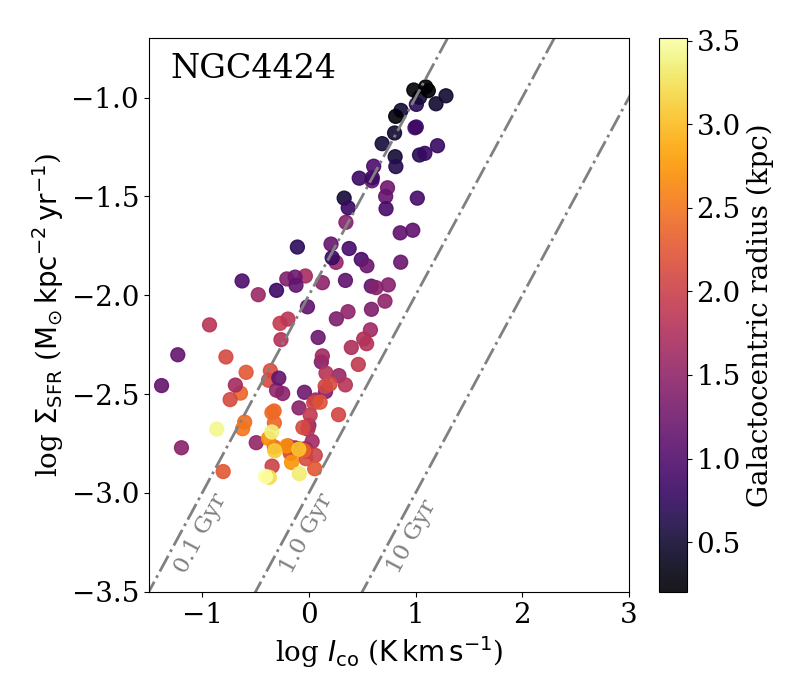}\,
\includegraphics[scale=0.30]{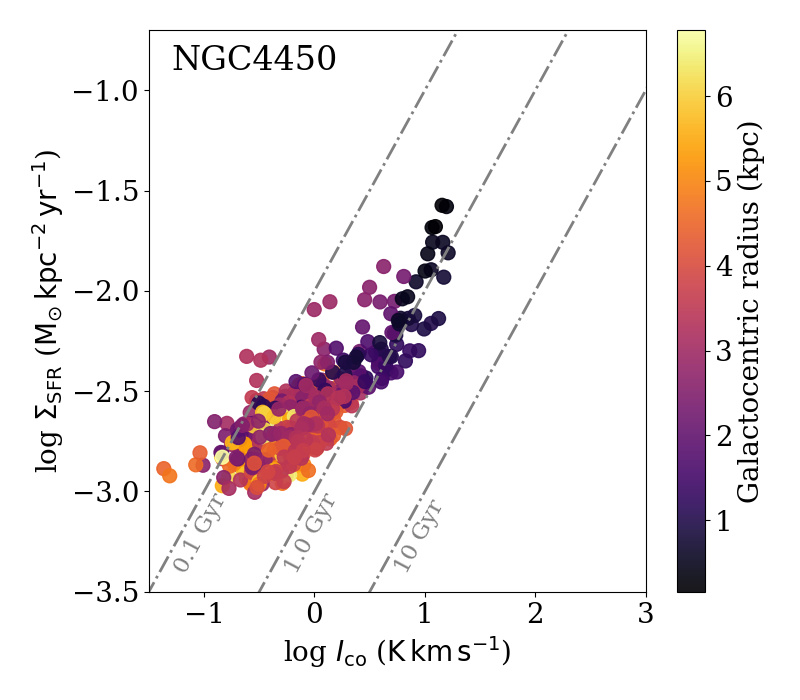}\,
\includegraphics[scale=0.30]{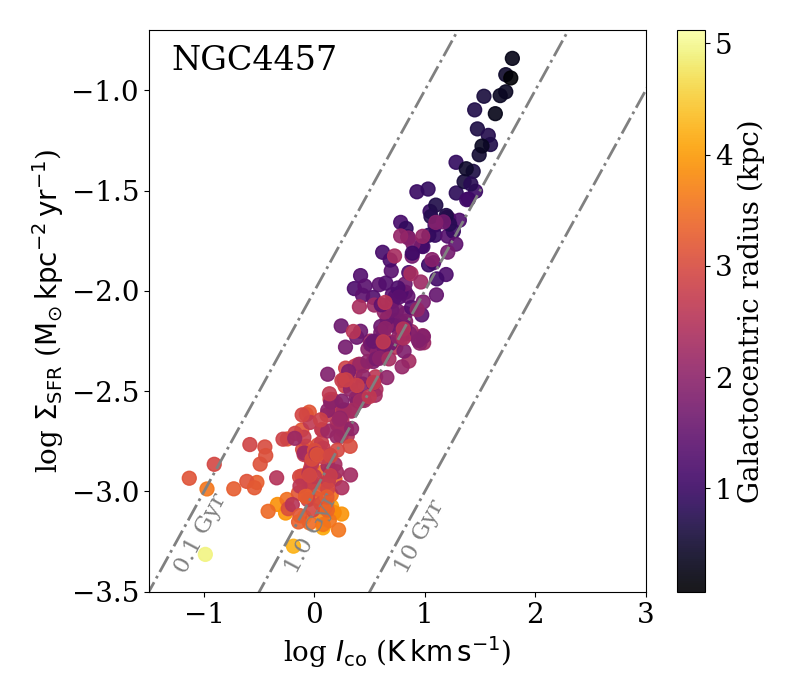}\\

\includegraphics[scale=0.30]{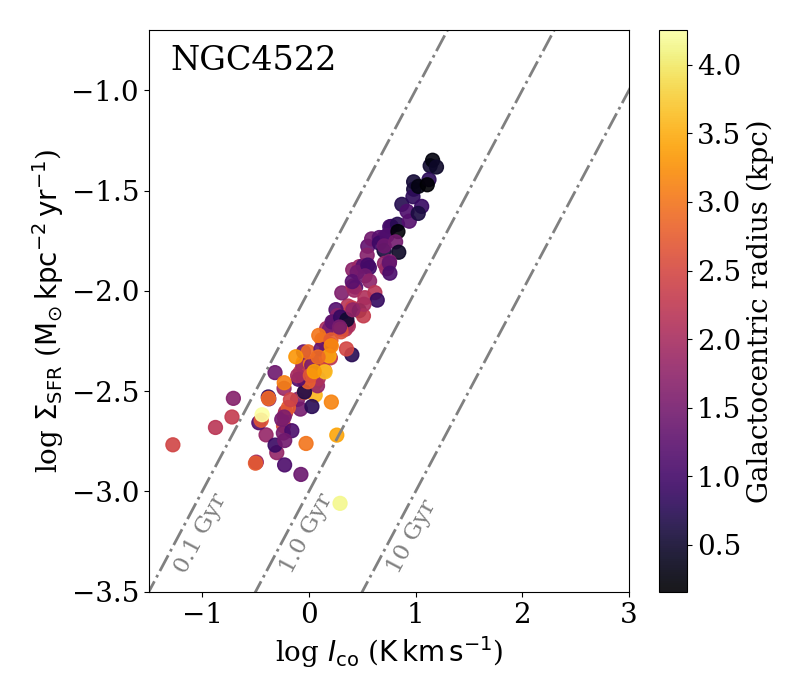}\,
\includegraphics[scale=0.30]{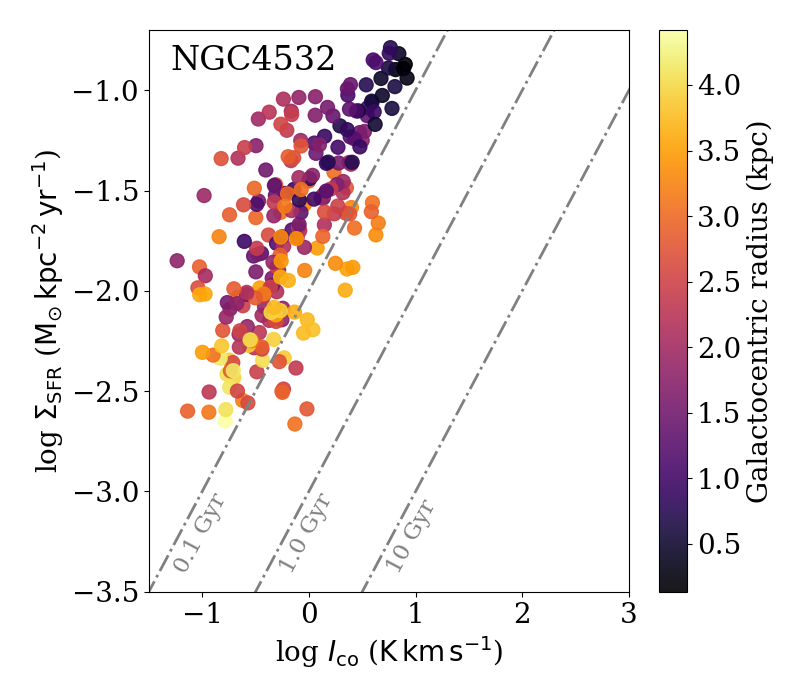}\,
\includegraphics[scale=0.30]{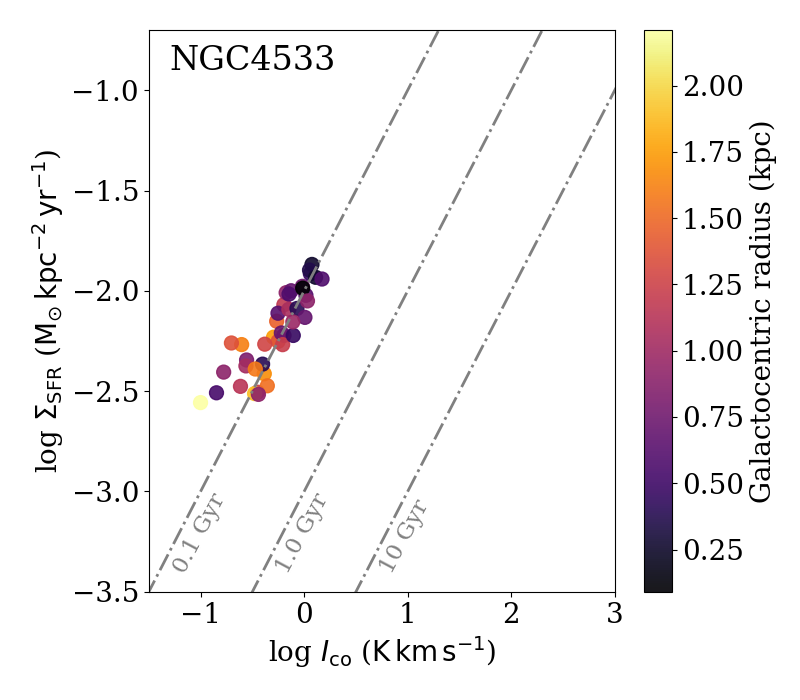}\\

\includegraphics[scale=0.30]{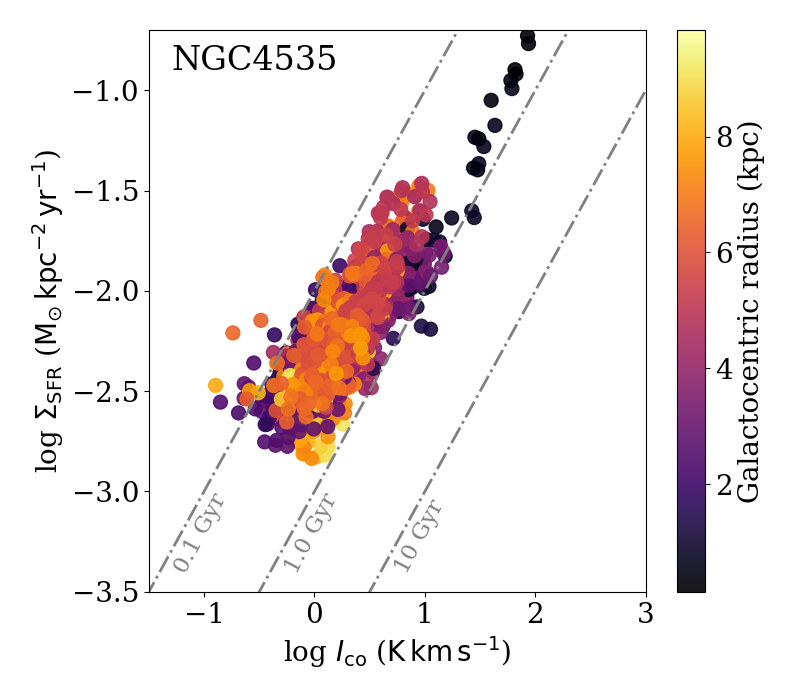}\,
\includegraphics[scale=0.30]{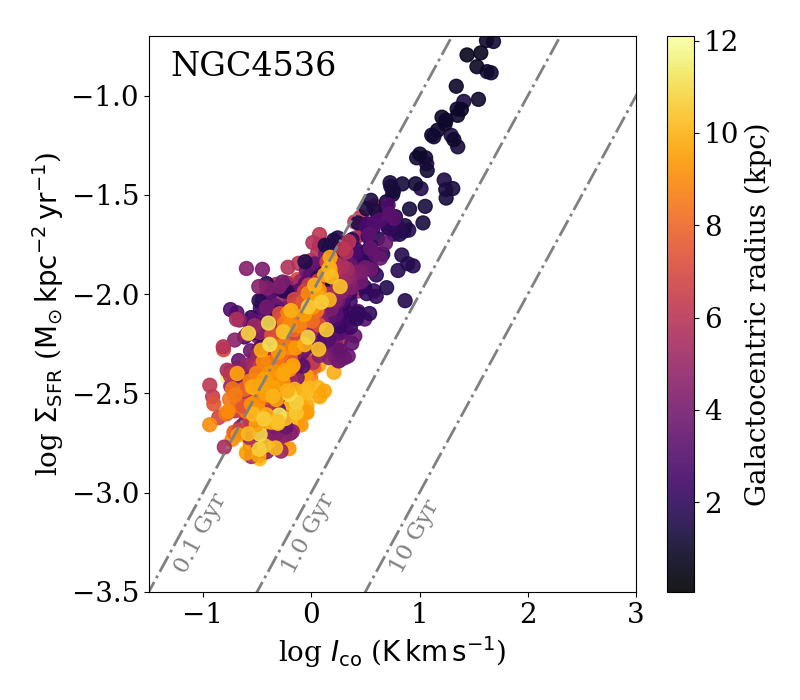}\,
\includegraphics[scale=0.30]{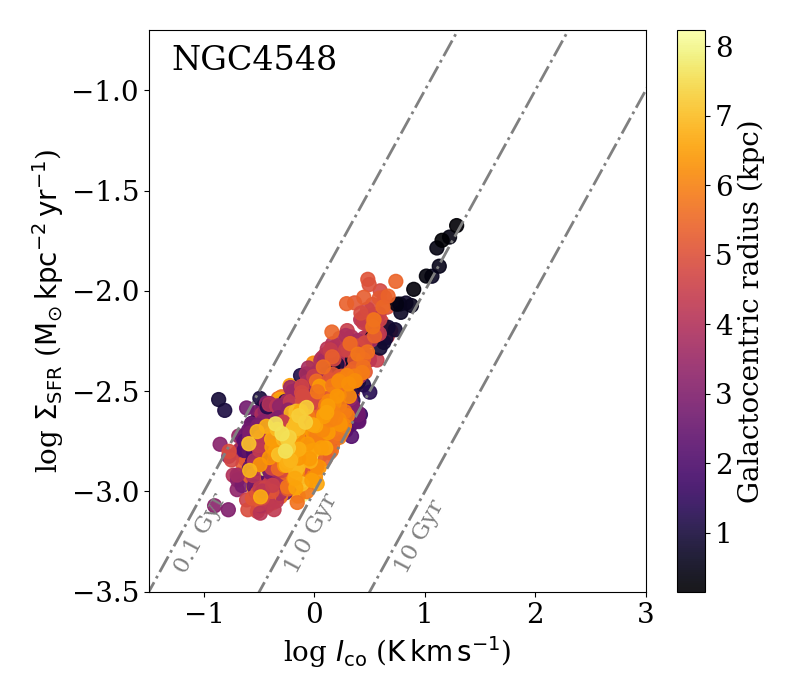}\\

\includegraphics[scale=0.30]{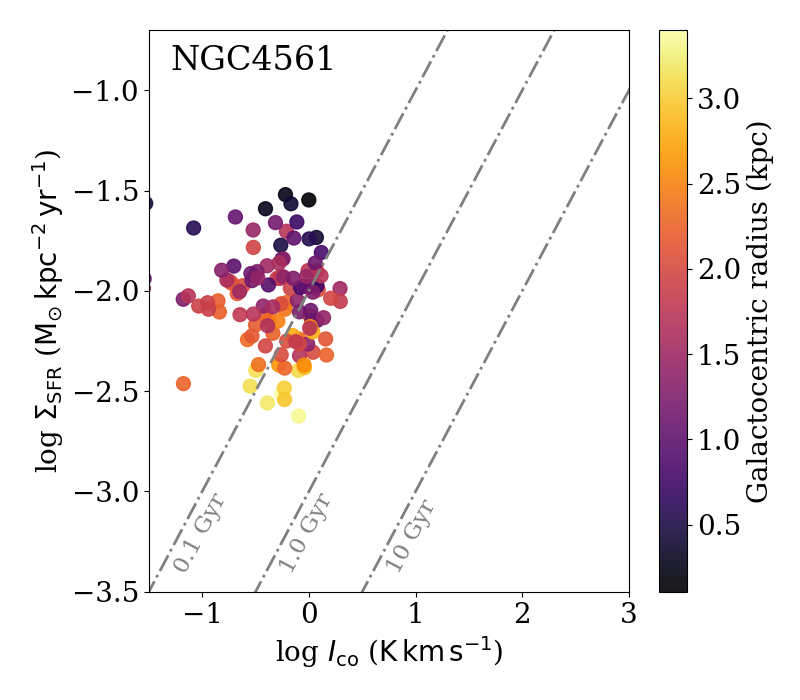}\,
\includegraphics[scale=0.30]{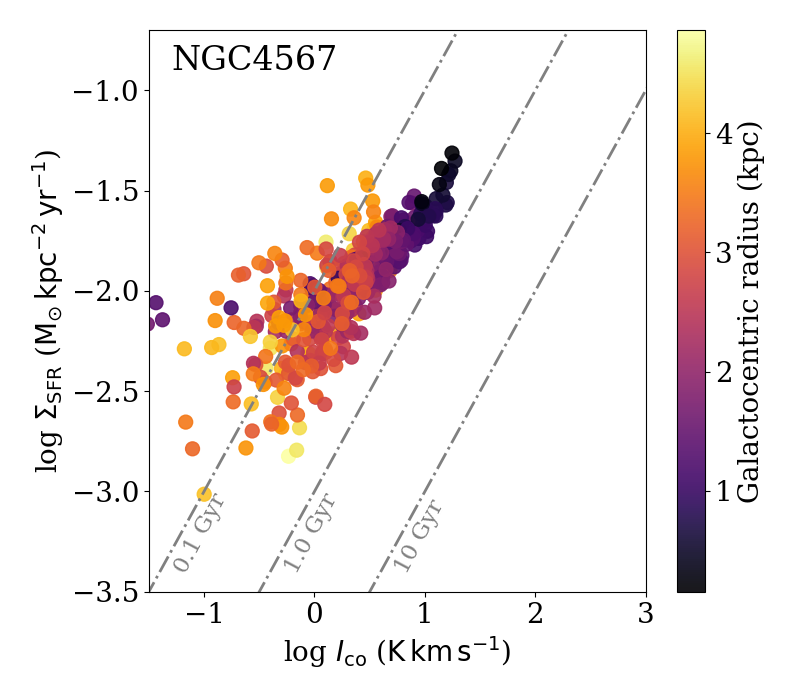}\,
\includegraphics[scale=0.30]{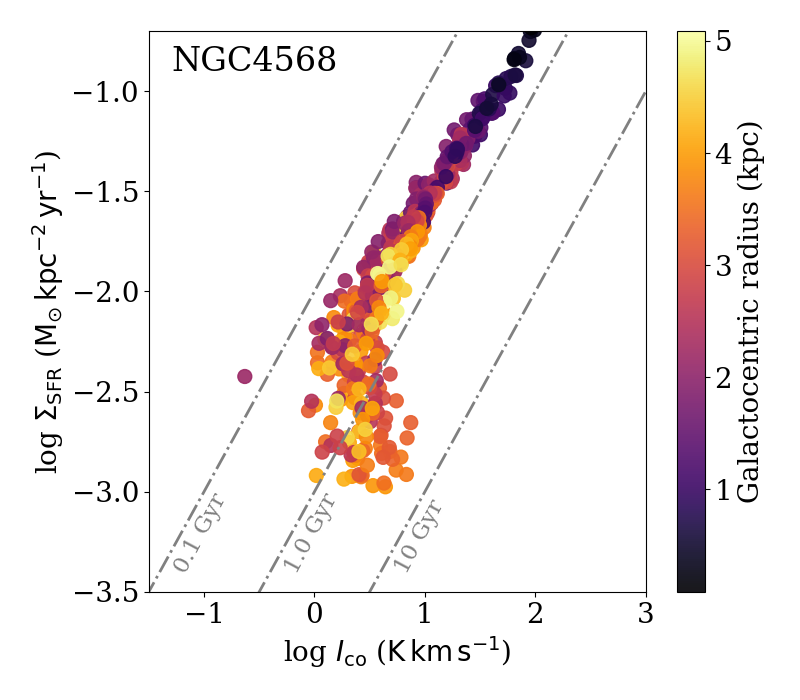}\\

\caption{continued.}
\end{figure*}

\renewcommand\thefigure{\thesection.\arabic{figure}}
\setcounter{figure}{0}

\begin{figure*}
\includegraphics[scale=0.30]{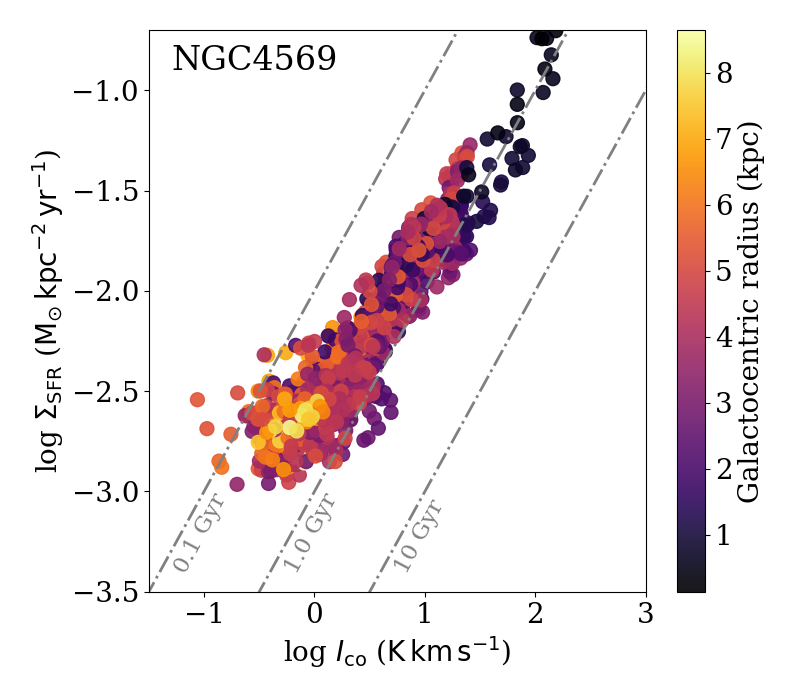}\,
\includegraphics[scale=0.30]{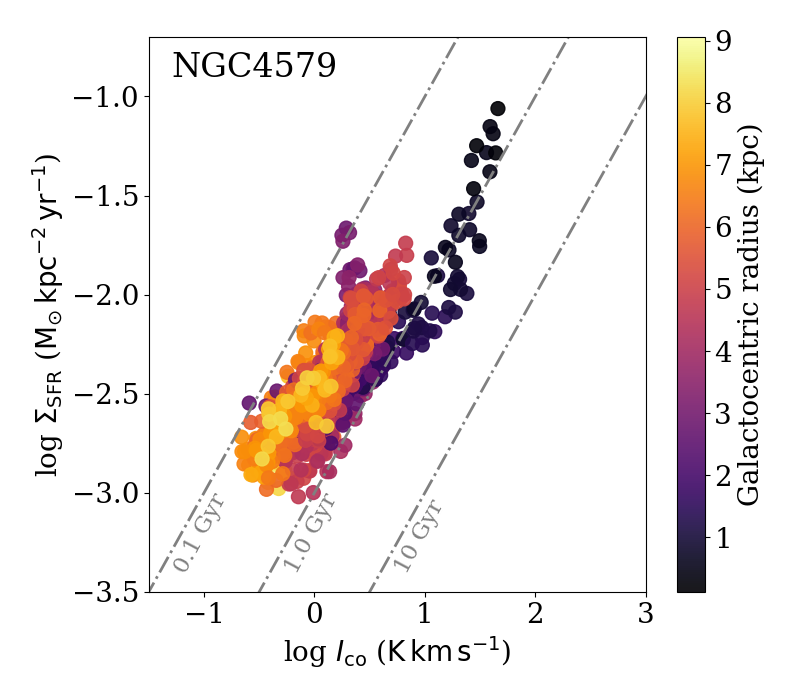}\,
\includegraphics[scale=0.30]{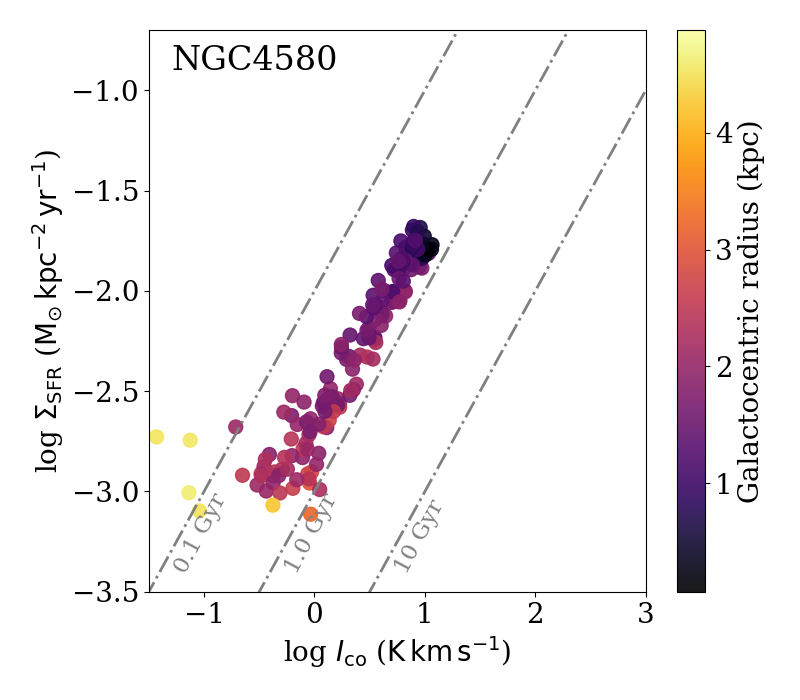}\\

\includegraphics[scale=0.30]{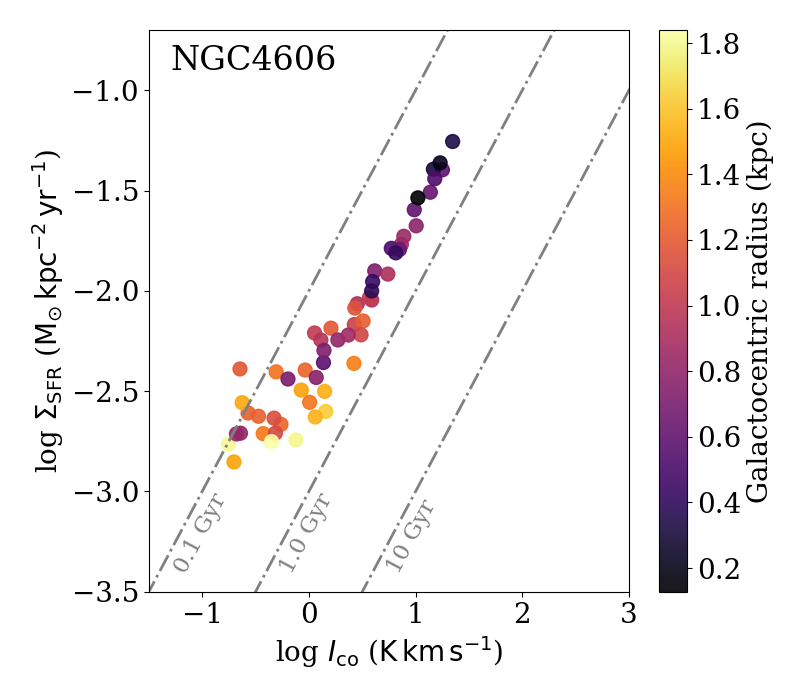}\,
\includegraphics[scale=0.30]{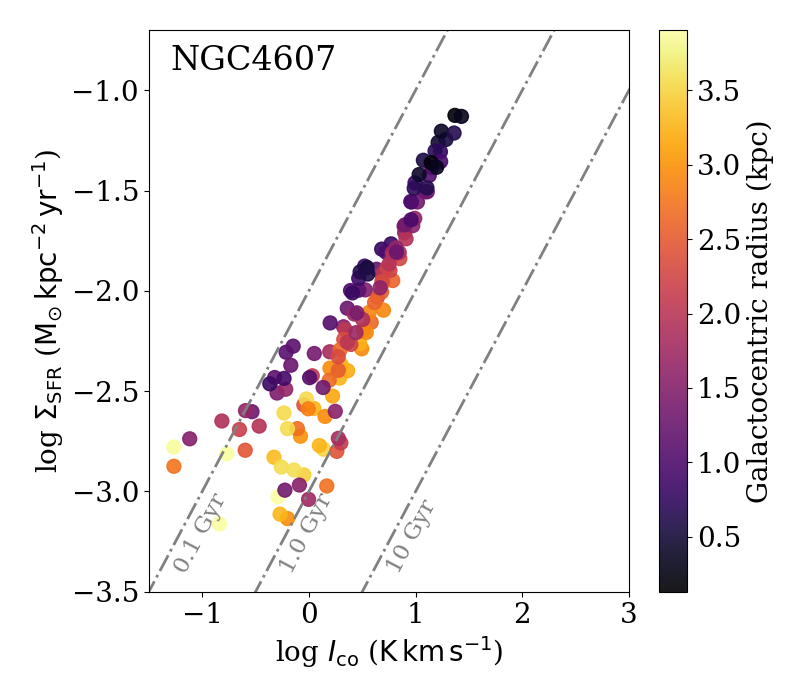}\,
\includegraphics[scale=0.30]{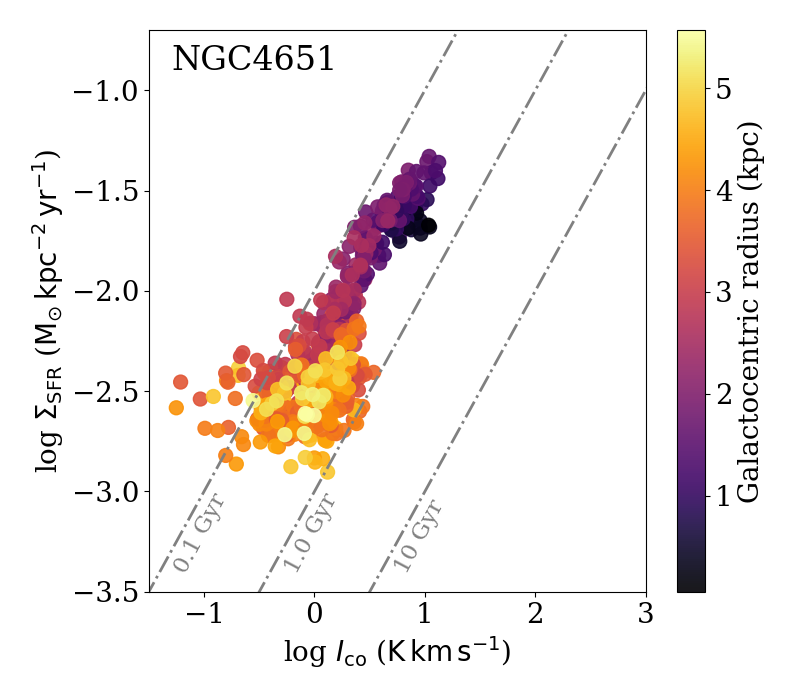}\\

\includegraphics[scale=0.30]{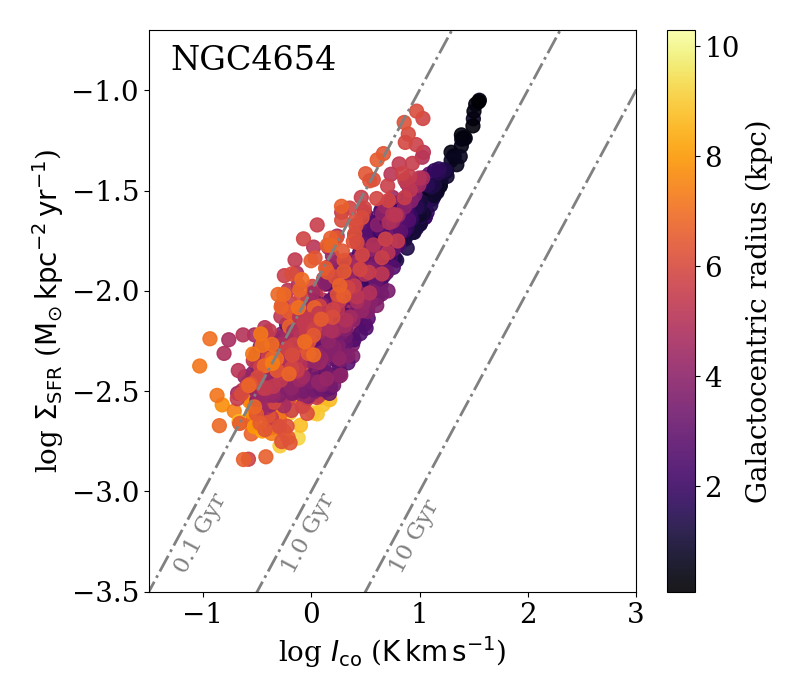}\,
\includegraphics[scale=0.30]{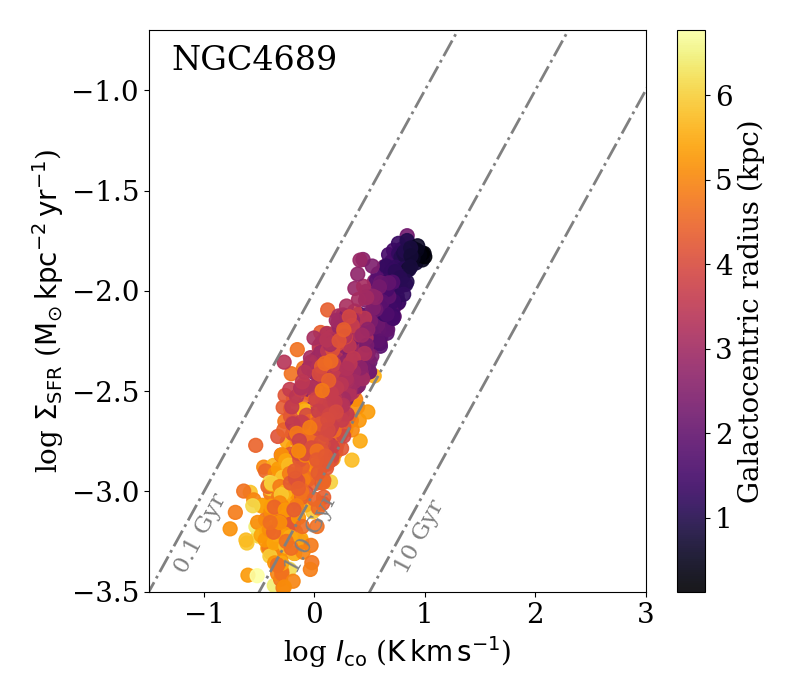}\,
\includegraphics[scale=0.30]{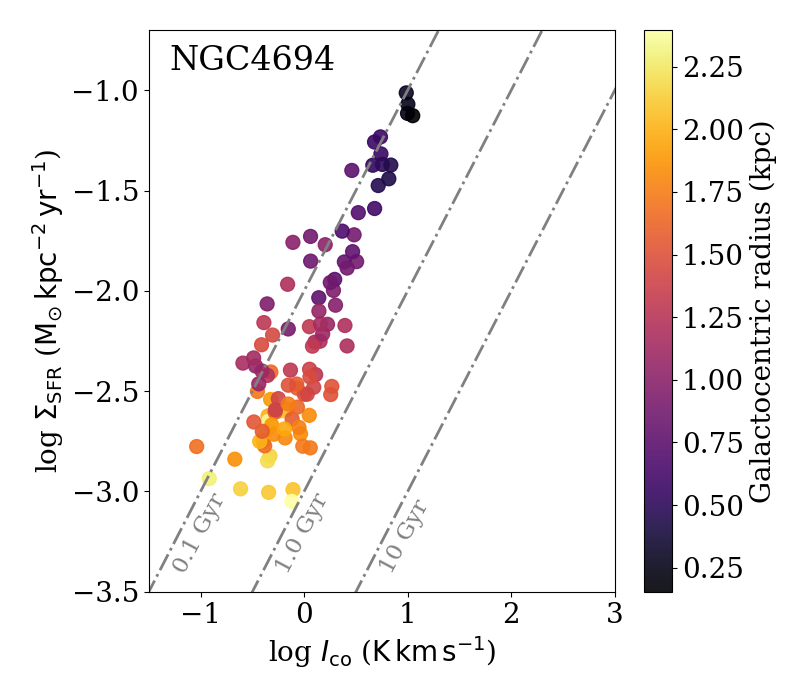}\\

\includegraphics[scale=0.30]{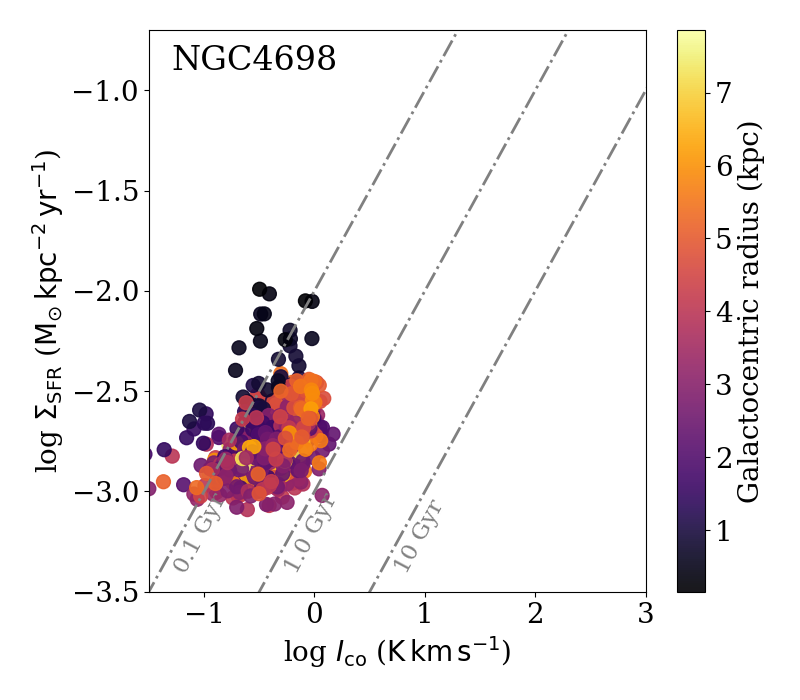}\,
\includegraphics[scale=0.30]{observed_ks_NGC4698-9as.png}\,
\includegraphics[scale=0.30]{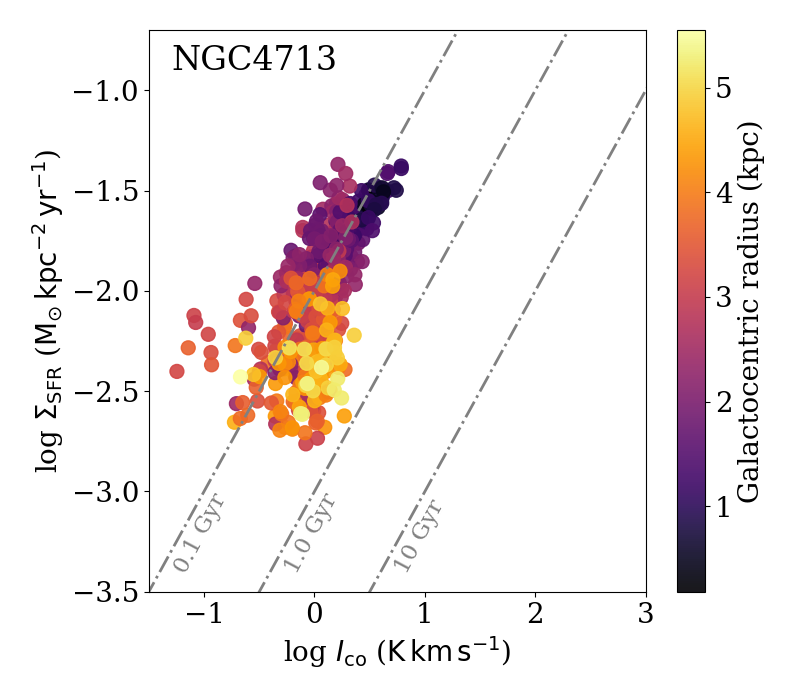}\\

\caption{continued.}
\end{figure*}

\renewcommand\thefigure{\thesection.\arabic{figure}}
\setcounter{figure}{0}

\begin{figure*}
\includegraphics[scale=0.30]{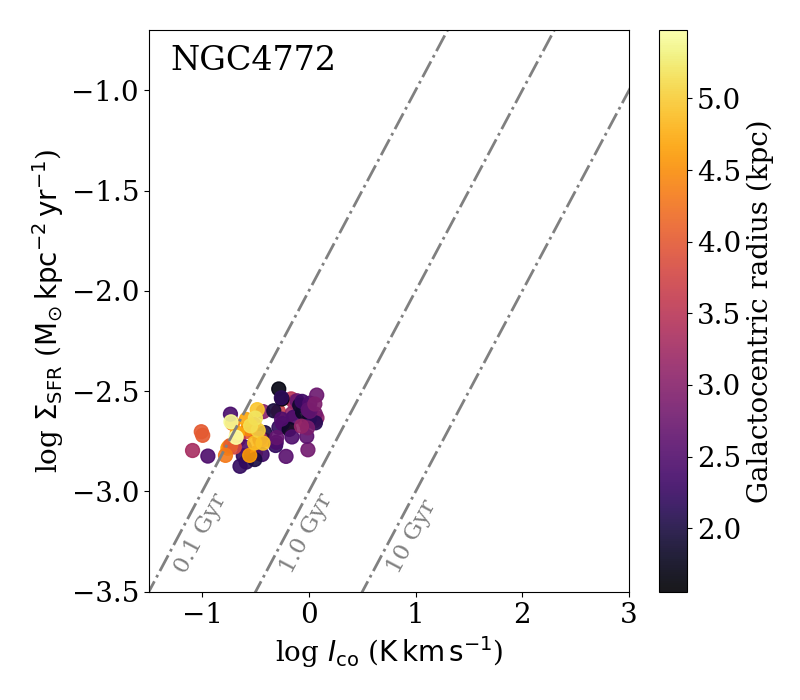}\,
\includegraphics[scale=0.30]{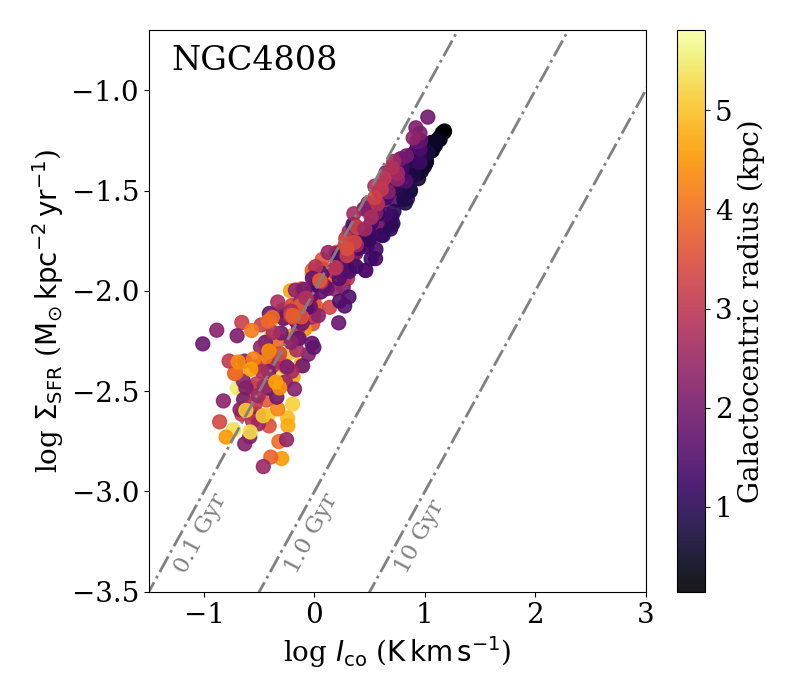}\\
\caption{continued.}
\end{figure*}

\end{appendix}


\bibliographystyle{aa}
\bibliography{biblio}

\end{document}